\documentclass{aastex62}
\stepcounter{section}
\usepackage{natbib}
\usepackage{subfloat}

\accepted{for publication in the Astrophyisical Journal, September 5, 2020}

\shorttitle{The T Tauri Triple System}
\shortauthors{ Beck et al.}

\begin{document}

\title{On the Nature of the T Tauri Triple System}

\correspondingauthor{Tracy L. Beck}

\author{Tracy L. Beck}
\affiliation{The Space Telescope Science Institute, 3700 San Martin Drive, Baltimore, MD 21218, USA 
\email{tbeck@stsci.edu} \\
}
\author{G. H. Schaefer}
\affil{The CHARA Array of Georgia State University, Mount Wilson Observatory, Mount Wilson, CA 91023, USA
}
\author{S. Guilloteau}
\affil{Laboratoire dÕAstrophysique de Bordeaux, UniversitŽ de Bordeaux, CNRS, B18N, AllŽe Geoffroy Saint-Hilaire, 33615 Pessac, France
}
\author{M. Simon}
\affil{Department of Physics and Astronomy, Stony Brook University, Stony Brook, NY 11794, USA
}
\author{A. Dutrey}
\affil{Laboratoire dÕAstrophysique de Bordeaux, UniversitŽ de Bordeaux, CNRS, B18N, AllŽe Geoffroy Saint-Hilaire, 33615 Pessac, France
}
\author{E. Di Folco}
\affil{Laboratoire dÕAstrophysique de Bordeaux, UniversitŽ de Bordeaux, CNRS, B18N, AllŽe Geoffroy Saint-Hilaire, 33615 Pessac, France
}
\author{E. Chapillon$^{3,}$}
\affil{IRAM, 300 Rue de la Piscine, 38406 Saint Martin dÕHres Cedex, France
}





\begin{abstract}
We present a multi-wavelength analysis to reveal the nature of the enigmatic T~Tauri triple star system.  New optical and infrared measurements are coupled with archival X-ray, UV and mm datasets to show morphologies of disk material and outflow kinematics.  A dark lane of obscuring material is seen in silhouette in several emission lines and in model-subtracted ALMA mm continuum dust residuals near the position of T~Tau~Sa+Sb, revealing the attenuating circumbinary ring around T~Tau~S.  The flux variability of T~Tau~S is linked in part to the binary orbit; T~Tau~Sb brightens near orbital apastron as it emerges from behind circumbinary material.  Outflow diagnostics confirm that T~Tau~N powers the blue-shifted western outflow, and the T~Tau~S binary drives the northwest-southeastern flow.  Analysis of the southern outflow shows periodic arcs ejected from the T~Tau system. Correlation of these arc locations and tangential kinematics with the orbit timing suggests that launch of the last four southern outflow ejections is contemporaneous with, and perhaps triggered by, the T~Tau~Sa+Sb binary periastron passage.  We present a geometry of the T~Tau triple that has the southern components foreground to T~Tau~N, obscured by a circumbinary ring, with mis-aligned disks and interacting outflows.  Particularly, a wind from T~Tauri~Sa that is perpendicular to its circumstellar disk might interact with the circumbinary material, which may explain conflicting high contrast measurements of the system outflows in the literature.  T~Tauri is an important laboratory to understand early dynamical processes in young multiple systems. We discuss the historical and future characteristics of the system in this context. 
\end{abstract}


\setcounter{section}{0}
\section{Introduction} \label{sec:intro}

\begin{figure}
\includegraphics[scale=0.6,angle=0]{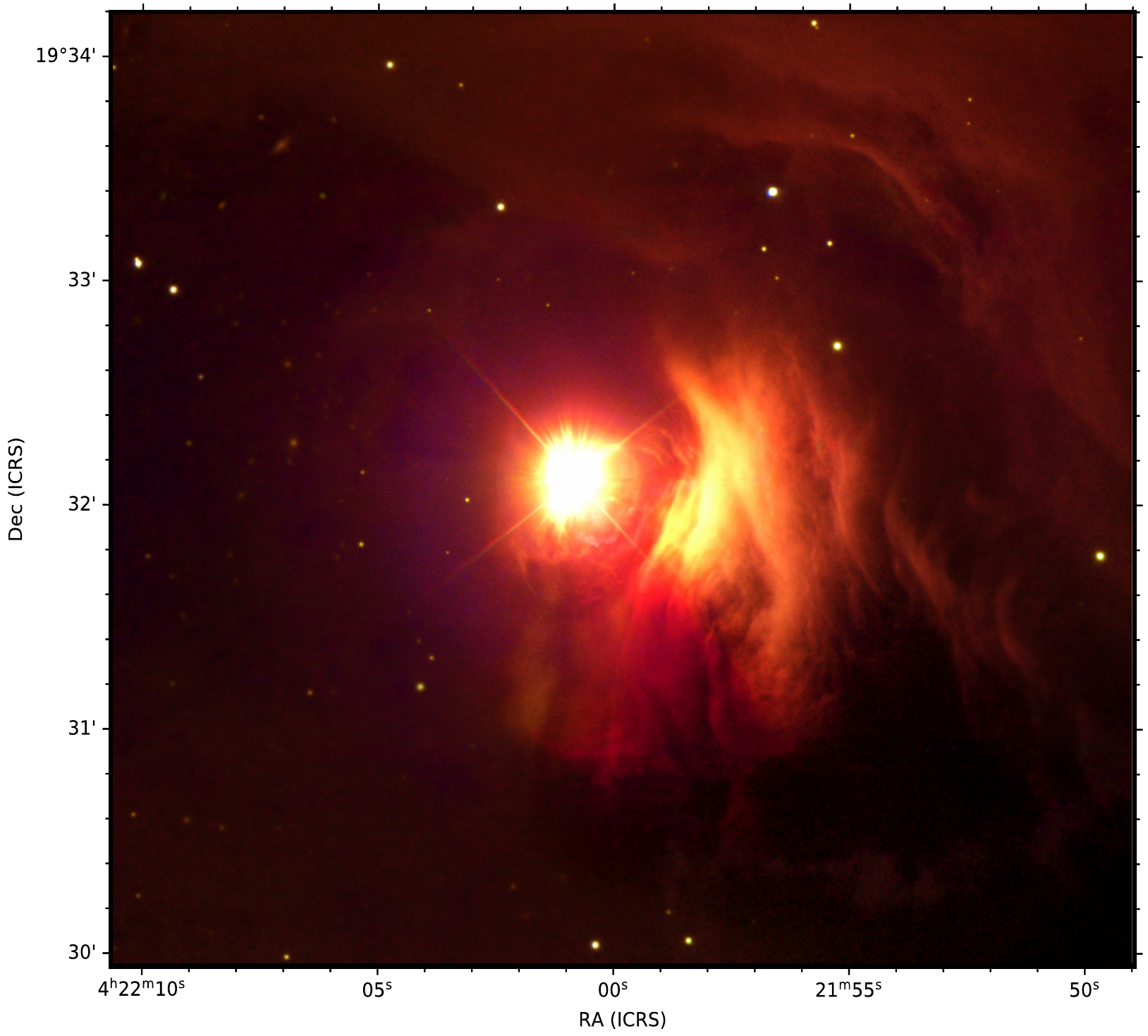}
\caption{An optical broadband g, i and z three-color image of T Tauri and its surrounding environment, including Hind's variable nebula, NGC 1555, to the west.  The image display scale enhances the low level extended emission in the environment of T Tau, but as a result the southern arc-like extensions of Burnham's nebula are overwhelmed by central stellar flux.  Observations were acquired using the GMOS imaging camera at the Gemini North Observatory (Table~\ref{tab.obs}). \label{fig.opt_img}}
\end{figure}

T Tauri is the eponymous pre-main sequence sun-like star \citep{joy45}, but investigation over the past 30 years has revealed that it is a remarkable member of this class of young stars.  T Tau was identified as an interesting variable star in the 1800's, particularly when the associated extended nebula 20-30$\arcsec$ to the west disappeared from detection in the 1860's \citep{hind64}.  Hind's variable nebula underwent multiple dimming events throughout the late 1800s and early 1900s, and was the first astronomical nebula confirmed to vary in brightness level \citep{burn90}.   In 1890, Burnham reported on low level nebulous emission surrounding T Tau, with a prominent $\sim$4$\arcsec$ extension at a position angle (PA) of $\sim$151$^{\circ}$.  This nebula, now designated Burnham's nebula, has also varied in morphology and brightness over its observation history \citep{burn90,stru62,robb95}.  T Tau itself was historically seen to flicker in brightness in optical wavelengths, from its peak level at V$_{mag}\sim$9.5 to fainter than the detection limit on photographic plates ($<$13.5-14th magnitude;\cite{lozi49, beck01b}).  Figure~\ref{fig.opt_img} presents a three color optical view of T Tau, revealing the optical star and its surrounding nebulosities.

The T Tau triple star system consists of the optical star, T Tau North (T Tau N), and the infrared binary, T Tau South (T Tau S), at a separation of 0.$\arcsec$7 to the south \citep{dyck82}. T Tau S has never been detected in the optical, to a limiting V magnitude of 19.6 \citep{stap98}. \cite{kore00} discovered that T Tau S is itself a binary with a projected spatial separation of $\sim$7 AU ($\sim$0.$\arcsec$05) at the time of its discovery observation in 1997.  The stars of the T Tau S binary, designated Sa and Sb, have been monitored for the past two decades for orbital motion using high resolution near-infrared (IR) adaptive optics imaging \citep{scha06, kohl08, scha14, csep15, kasp16, kohl16, scha20}.  In this time, the T Tau S binary has been mapped through more than two thirds of its 27.2$\pm$0.7 year orbit \citep{scha20}.  The total mass of T Tau S is $\sim$2.5-2.7M$_{\odot}$, T Tau Sb is 0.4-0.5M$_{\odot}$ and Sa is 2.0-2.3M$_{\odot}$, based on the most recent published dynamical model of the orbit \citep{scha20}.   The range of the derived mass values depend on the distance adopted for the T Tau system \citep{scha20}; 143.7pc from Gaia DR2 \citep{bail18}, or 148.7pc from VLBI measurements \citep{xu19}.  Recent IR spectral measurements find evidence for a cooler photospheric temperature and later spectral type for T Tau N;  \cite{flor20} postulate that it is lower mass and young, while \cite{scha20} argue for variable star spot activity.  Although T Tau N is one of the most luminous T Tauri stars within 200pc of the sun and T Tau S is optically undetected, the orbital models reveal that T Tau Sa is the most massive star in this young triple system.  Our reference to "T Tau S" throughout this document is to the combined T Tau Sa+Sb binary system.

T Tau N is among one the brightest Classical T Tauri Stars (CTTSs) in the Taurus-Auriga association of young stars.  The T Tau system is known to have natal envelope material in its surrounding environment \citep{momo96, schu97}, which suggests that this triple system is on the young side for T Tauri Stars (TTSs; \cite{adam87,lada87}).  HR diagram analysis with modern stellar models span a wide range: T Tau N is either at an older state with star spot activity (to 4 Myr; \cite{scha20}), or younger with a cooler photospheric temperature and hence lower mass \citep{flor20}.  T Tau N also exhibited rampant historical optical variability with more than 3 magnitudes of brightness variation, a defining characteristic of the T Tauri class of stars \citep{joy45}, which ceased abruptly in the 1920s \citep{lozi49, beck01b}.   The near-IR K magnitudes of T Tau N were found to be stable with a low level of variability in monitoring measurements \citep{beck04, scha20}.  By contrast, the IR flux of the combined T Tau S binary system has varied wildly.  \cite{ghez91} discovered a $\sim$2 magnitude flare in the brightness of T Tau S from 2 through 10$\mu$m, and attributed the increase to an accretion outburst in the system.  \cite{beck04} found that the near-IR flux of the combined T Tau S binary can change drastically on timescales of a week with a redder when faint character, suggesting that variable obscuration along the line of sight may play a role in the fluctuation.  \cite{vanb10} also found that the mid-IR flux of T Tau S can change measurably over several nights, and they present a model that incorporates rapid changes in accretion to explain the mid-IR emission heating of dust in the inner disk.  Moreover, \cite{vanb10} further postulated that IR variability characteristics of the combined T Tau S system may be linked to the binary orbital motion, perhaps from pulsed accretion onto T Tau Sa as Sb passes at periastron.

\begin{figure}
\includegraphics[scale=0.5,angle=0]{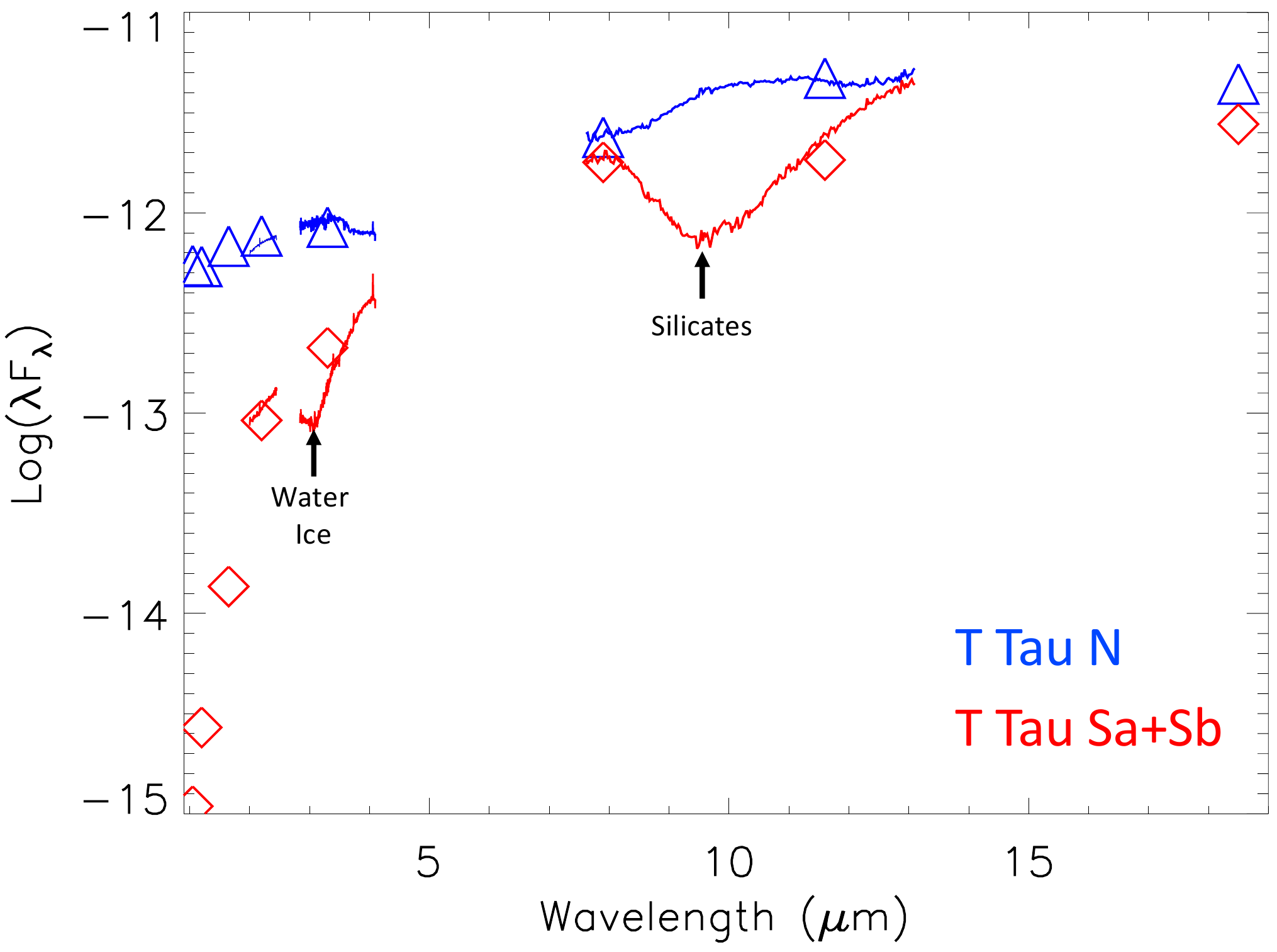}
\caption{The 1.0 to 18$\mu$m spectral energy distribution of T Tau N (blue) and T Tau S (red); the Sa and Sb binary components were unresolved.  Symbols plot the broadband fluxes in the near-IR Z through L band and 7.9, 11.6 and 18.3$\mu$m mid-IR bands (T Tau N triangles, T Tau S, diamonds).  2-4$\mu$m (IRTF SpeX) and 7.5-13.2$\mu$m (Gemini North Michelle) longslit spectra are overplotted to reveal broad features of 3$\mu$m water ice and $\sim$10$\mu$m silicates.  The Z and J-band measurements are from 2005, the near-IR and mid-IR measurements are from Fall 2003, see Table~\ref{tab.obs}.}  The shortest wavelength measurement in the Z-band presents the upper bound to the flux of T Tau S for this non-detection. \label{fig.sed}
\end{figure}

The T Tau triple system has evidence for non-coplanar circumstellar disks and two nearly perpendicular outflows \citep{bohm94, hoge97, akes98, solf99, beck04, duch05, mana19}.  Each of the three stars has a circumstellar disk, with mass accretion measured by atomic hydrogen emission transitions that trace accretion flows from a disk onto a young star \citep{kasp02, duch02, beck04, duch05}. The circumstellar disk around T Tau N has been studied extensively at mm through infrared wavelengths and is viewed $\sim$20-30$^{\circ}$ from face-on \citep{akes98, gust08, guil11, guil13, podi14, mana19}.   A small ($<$4~AU) circumstellar disk around T Tau Sa is detected with a near-edge on orientation ($>70^{\circ}$; \cite{ratz09, mana19}).  Material along the line of sight toward T Tau S that causes A$_v\sim$20 magnitudes of foreground attenuation is detected in solid state absorption features of water ice, silicates and CO$_2$ ice \citep{ghez91, beck01a, beck04, skem08, ratz09, vand99}.  Figure~\ref{fig.sed} shows the 1 - 18$\mu$m spectral energy distribution (SED) of T Tau N and S (Sa+Sb unresolved), including 2-4$\mu$m and 8-13$\mu$m spectral measurements that trace the ice and silicates.  The SED of T Tau N is typical of young sun-like stars with silicates in emission tracing hot dust in its inner heated disk \citep{ghez91, skem08, ratz09}.   Figure~\ref{fig.sed} demonstrates the strong flux attenuation and solid state absorption features in the SED of T Tau S.  Because of the significant obscuration toward the T Tau S binary and small extinction toward T Tau N, the attenuating material that causes the ice and silicate absorption has been best described as a circumbinary disk encircling the T Tau Sa+Sb stars.  Indirect evidence for a circumbinary distribution of material encircling T Tau S has been detected in near IR polarimetric maps \citep{yang18}, and as a dimming at its position in the background ultraviolet (UV) molecular hydrogen emission \citep{walt03}.  However, spatially resolved IR measurements suggest that the flux attenuation of Sb may be significantly less than Sa \citep{ratz09}.  The nature of the foreground obscuring material along the line of sight toward T Tau S has been elusive. 



 
 
 
The T Tau triple system hosts three cataloged Herbig-Haro flows.  The HH~155 outflow is oriented in the east-west direction, arises from T Tau N and extends to Hind's variable nebulosity 20-30$\arcsec$ to the west \cite{bohm94, solf99}.   The extended HH 255 outflow is believed to arise from T Tau S, extends $\sim$40$\arcsec$ in the North-South direction, terminating in the HH 355 red and blue shifted lobes which are $\sim$20 arc minutes distant \citep{reip97, solf99}.   This outflow represents one of the few extensive, $\sim$1.55 parsec giant Herbig-Haro flows in Taurus \citep{reip97}.  Optical spectral measurements of Burnham's nebula (HH 255) reveal an extensive array of over 80 optical emission lines, attributed to shocks in the inner outflows \citep{bohm88}.  Resolved spectral imaging has found multiple loops and arcs of emission from H$\alpha$, [S~II] and near-IR H$_2$ associated with the inner HH~255 flow and Burnham's nebula \citep{herb97, robb95, herb07, beck08, gust10}.  However, a new generation of extremely sensitive high contrast adaptive optics imaging has challenged the outflow geometry found from the past kinematic measurements.   Particularly, \cite{herb07} and \cite{yang18} measured high contrast near-IR H$_2$ and H-band polarized continuum, respectively, and found evidence that T Tau Sa or Sb may be driving a western component of the outflows, previously attributed to T Tau~N.  As observations of the inner environment of T Tau have improved and become more sensitive, confusion in the orientation and nature of the outflows from the system has followed.

T Tau exhibits intriguing evidence for mis-aligned circumstellar disks and outflows, which has made it a demonstration case for orbital dynamical evolution during the formation stages of multiple star systems \citep{reip00}.  In this paper, we present a new multi-wavelength investigation of the T Tauri triple to reveal the new observable characteristics of the system and its geometry.  We analyze archival ALMA maps of the 1.3mm dust continuum, and present multi-wavelength UV through optical emission maps to reveal the circumstellar and circumbinary material in the system.  We present and analyze previously unpublished optical and infrared measurements of outflow diagnostics and relate the flows to characteristics of this young triple system.  We emphasize in this study that a multi-wavelength understanding of the characteristics of young star systems is important, and that the T Tau triple system is a fundamental laboratory for understanding dynamical effects in the early evolutionary stages of triple and high order multiple systems.  

\subsection{A Historical Summary of T Tauri}\label{sec:history}

As the prototypical member of its class, T Tau has become a laboratory to reveal intriguing astrophysical phenomenon on the formation of across the electromagnetic spectrum.  The T Tau system has a number of historical observational firsts.  To study variability in the nearby Hind's nebula, T Tau was the first astrophysical source to have low level extended nebular morphology analyzed by computer digitized subtraction of the bright stellar core (first electronic PSF subtraction; \cite{lorr75}).  T Tau was the first clear detection of the bright 2.12$\mu$m H$_2$ feature arising from a young star \citep{beck78}, although ro-vibrational molecular hydrogen in the Orion star forming region had been identified a few years earlier by \cite{gaut76}.  It was also the first detection of the UV electronic H$_2$ emissions from hot gas in the environments of a TTS \citep{brow81}, and discovery of T Tau S was the first identified companion seen only in the infrared \citep{dyck82, kore97}.  The system has exhibited strong and highly variable circular polarization and non-thermal radio emission, serving as a demonstration for radio properties of inner magnetized outflows \citep{phil93, skin94, ray97, john03, smit03}.  \cite{schn18} used T Tau as an accretion laboratory to reveal that strong mass accretion affects the hot 10$^6$K coronal emission, resulting in an observed anti-correlation between mass accretion and stellar X-Ray activity. 

\begin{figure}
\plotone{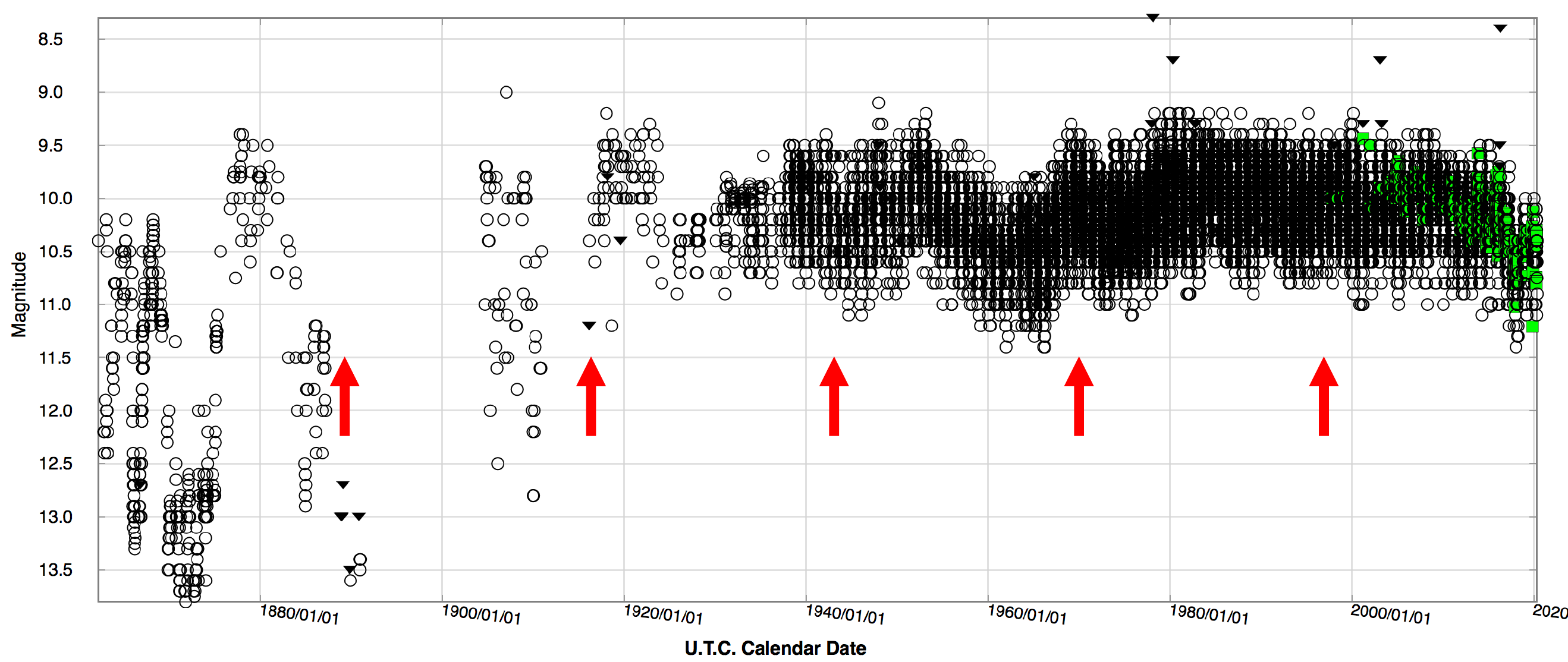}
\caption{The historical optical variability of T Tau from over 120 years of monitoring observations, as reported by the American Association of Variable Star Observers.  The historical measurements are in an optical visible bandpass equivalent, the green data points since 2000 are in a v-band filter.  The times highlighted by red arrows show the approximate orbital periastron passage date of the T Tau S binary for the past 5 orbits. \label{fig.figaavso}}
\end{figure}

Prior to $\sim$1900, T Tau was highlighted in the literature as an interesting system because of significant optical brightness variability.  T Tau N varied in optical brightness randomly and by more than 3 magnitudes \citep{lozi49, beck01b}.  The American Association of Variable Star Observers (AAVSO) has collected over 28,000 optical brightness measurements of T Tau from the last 150 years, including historical literature reports and contributions from amateur astronomers.  Figure~\ref{fig.figaavso} presents the AAVSO optical V-band light curve of the T Tau system from the early 1860s to 2020 \citep{kafk20}.   As also found by \cite{lozi49} and \cite{beck01b}, Figure~\ref{fig.figaavso} shows that the optical brightness of T Tau varied by over 3 magnitudes, and then the rampant variability abruptly ceased before the 1930's.  Regular measurements by the AAVSO since that time show T Tau to be much more stable, with annual variations on at the $\pm$0.5-0.7 magnitude level.  This is also consistent with variability monitoring observations of \cite{herb94} and the available PanStarrs measurements \citep{cham16}.  \cite{herb94} found that the $\sim$1 magnitude optical photometric brightness variations of T Tau indicated a likely origin caused by accretion variability and possible variations in flux attenuation, such as from dust structures passing through the line of sight.  From the AAVSO light curve in Figure~\ref{fig.figaavso} we see that a slow, extended $\sim$0.5 mag decrease in the average brightness occurred between 1950 to 1965.  More recent measurements in the past $\sim$5-7 years suggest another slight decrease in the optical flux of T Tau may again be occurring.  However, for the most part, the optical flux of T Tau has been much more stable in the past 8 to 9 decades compared to its historical rampant variability.   Variations in line of sight obscuration to the system were postulated as likely cause of the large and random changes in optical brightness of T Tau prior to $\sim$1930 \citep{beck01b}, but the abrupt cessation of this variability has never been explained.

\section{Observations} \label{sec:obs}

\begin{figure}
\plotone{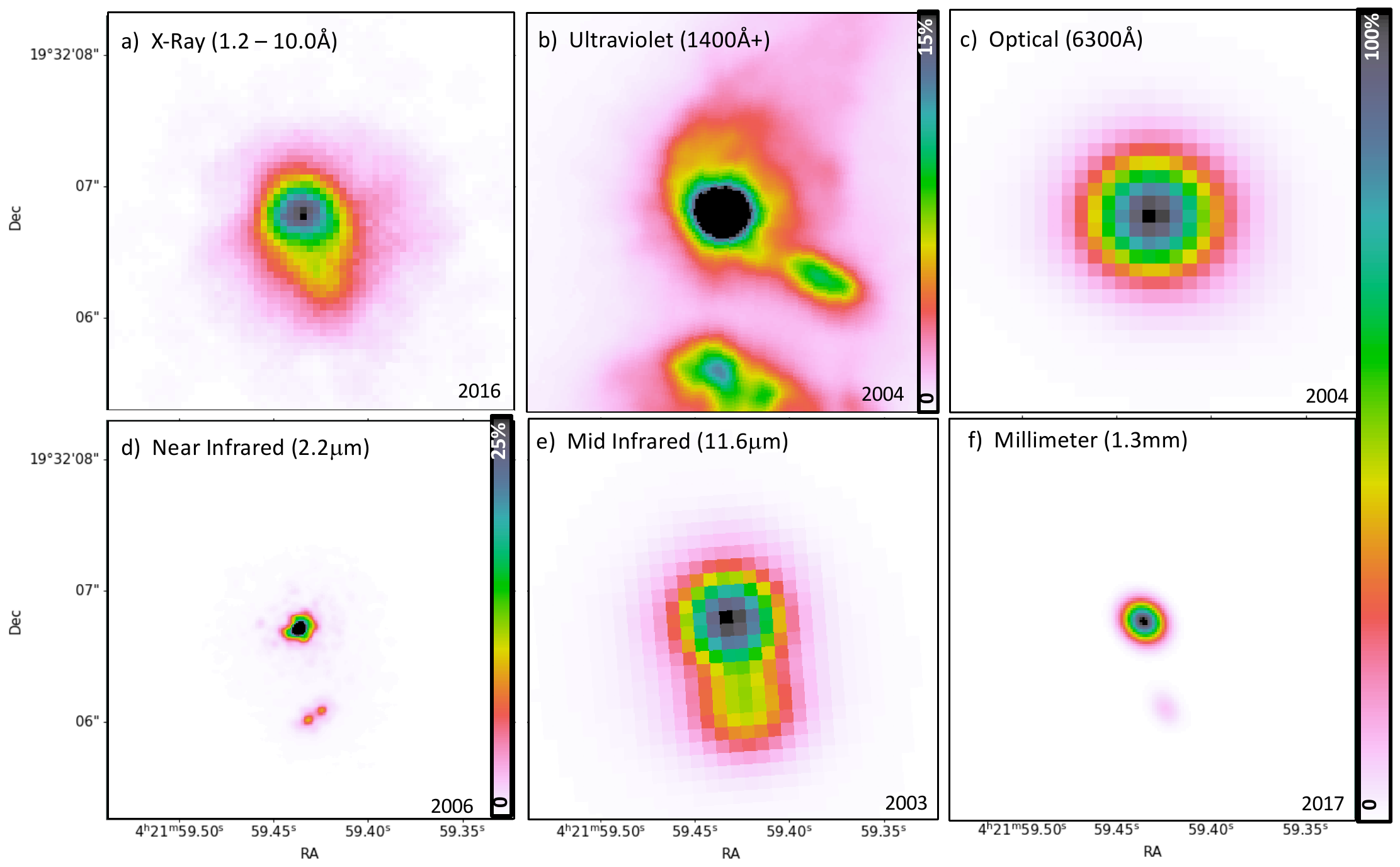}
\caption{The inner 3$\arcsec\times$3$\arcsec$ region around T Tau from X-ray through mm wavelengths.  All images are scaled logarithmically from 0 to 100\% of the peak flux (of T Tau N), except for panels b) and d), which are scaled from 0 to 15\% and 25\% of the peak flux, respectively, to enhance the significantly fainter extended emission structure and the fainter stars, T Tau Sa+Sb.\label{fig.fig4cont}}
\end{figure}

In this paper we present previously unpublished imaging, spectroscopy and imaging spectroscopy at optical, near-IR and mid-IR wavelengths.  We also present and reanalyze archival datasets at X-Ray, UV and mm wavelengths.   Table~\ref{tab.obs} summarizes the observations presented in this study (top) and the archival datasets from the Chandra X-Ray Observatory, {\it Hubble Space Telescope} (HST) and {\it Atacama Large Mm Array} (ALMA) downloaded and presented here (bottom).  The archival X-ray dataset from Chandra was presented and analyzed in detail by \cite{schn18} and the ALMA observations of T Tau were included in the 1.3mm dust continuum survey paper of \cite{mana19}.  To the best of our knowledge, this is the first publication of the archival UV imaging dataset from the HST.  Figure~\ref{fig.fig4cont} presents the continuum flux of the inner 3$\arcsec\times$3$\arcsec$ region around the T Tau triple system from X-ray through 1.3mm from the datasets summarized in Table~\ref{tab.obs}.  The three stars in the T Tau triple system are spatially resolved and detectable only in the near-IR continuum imaging (Figure~\ref{fig.fig4cont}d).  In all other images the T Tau S binary is either obscured to non-detection (UV and optical; Figure~\ref{fig.fig4cont}b \& c), or below the sensitivity or spatial resolution limit of the measurements without specialized processing (X-Ray, mid-IR, mm; Figure~\ref{fig.fig4cont}a, e \& f).  The UV image presented here is from the HST Advanced Camera for Surveys (ACS) F140LP long pass filter, and in addition to continuum flux, strong extended UV H$_2$ emission is also measured.  This is the only image in Figure~\ref{fig.fig4cont} that exhibits strong line emission above the continuum. 

\begin{deluxetable*}{l|ccccc|c|c}[b!]
\tablecaption{Summary of Observations \label{tab.obs}}
\rotate
\tablewidth{0pt}
\tablehead{
\colhead{UT Date}  & \colhead{Telescope}  & \colhead{Instrument } & \colhead{Configuration} & \colhead{Exposure Time (s)} & \colhead{Total Exposure} 
& \colhead{Program ID} & \colhead{Figure}  \\
\colhead{}  & \colhead{}  & \colhead{ } & \colhead{} & \colhead{/ \# Exp.} & \colhead{Time} & \colhead{/ PI} & \colhead{} 
}
\startdata
\hline
 &  &  &  Previously Unpublished Data &  &  &  &   \\
\hline
\hline
2003 Sep. 03 & Gemini North & Michelle & 7.9, 11.6, 18.3$\mu$m &  30 / 20 & 600  & GN-2003B-SV-81 /  & 2, 4  \\
 &  &  &  Imaging &  &  & T. Beck &   \\
 \hline
2003 Sep. 03 & Gemini North & Michelle & 9-13$\mu$m  &  30 / 20 & 600  & GN-2003B-SV-81 /  &  2 \\
&  & Michelle &  Spectroscopy &  &   &  T. Beck &   \\
\hline
2003 Dec. 05 & NASA IRTF & SpeX & LXD1.9, 2-4$\mu$m &  30 / 20 & 600  & -- / T. Beck &  2 \\
\hline
2004 Oct. 04 & Gemini North  & GMOS  & g, i, z, H$\alpha$ &  30 (180 & 120  & GN-2004B-Q-80 / & 1, 16 \\
 &   &  Imaging &  &  @ H$\alpha$) / 4 & (720 @ H$\alpha$) & T. Beck & 17 \\
\hline
 2019 Oct. 01 & Gemini North  & GMOS  & H$\alpha$ &  180 / 5 & 900  & GN-2019B-Q-204 / & 17 \\
 &   &  Imaging &  &  &  & T. Beck &  \\
\hline
2004 Oct. 04 & Gemini North  & GMOS IFU & R831 Grating &  600 / 1 & 600 & GN-2004B-Q-80 / &  4, 8, \\
 &  & &  @768nm &  8 positions & &  T. Beck &  10-13 \\
\hline
2005 Oct. 19 & Gemini North  & NIFS & Z, J &  10 (20 @ Z) / 12 & 120 (240) & Engineering Time & 2, 8 \\
\hline
2005 Oct. 25 & Gemini North  & NIFS & K &  127.2 (5.3$\times$24) / 36 & 4580 & Engineering Time & 15, 20 \\
\hline
2006 Dec. 25 & Gemini North & NIFS & H and K w/  & 120 / 16 & 1920  & GN-2006B-DD-6  / & 14, 15\\
 &  &  &  0.$\arcsec$5 OD &  &  & T. Beck & \\
\hline
\hline
 &  &  &  Archival Data &  &  &  &   \\
\hline
\hline
2015 Jan. 1-3 & Chandra X-Ray  & ACIS & HETG 0th  &  - & 127.4ks &  16672 /  & 4 \\
 & Observatory &  &  Order Image &  - & &   P. Schneider &  \\
\hline
2004 Nov. 05 & Hubble Space  & ACS & F140LP \& F165LP &  990 / 4 &  & 10489 /  & 4, 9, \\
2004 Nov. 05 &  Telescope &  & Filters &   &  &  A. Brown &  16, 20\\
\hline
2017 Aug. 18 & ALMA & Band 6 & 233Ghz &  - & 526 & 2016.1.01164.S / & 4, 7, \\
 &  &  &   &  &  &  G. Herczeg & 8, 20 \\
 \enddata
\end{deluxetable*}

\subsection{Optical Imaging}\label{sec:op}

To study the nebulosity and extended outflow structure, optical imaging of T Tau with the Gemini Multi-Object Spectrograph (GMOS) imaging camera was carried out in October 2004 in the g, i, z broadband filters and the H$\alpha$ narrow band filter (Table~\ref{tab.obs}; Gemini program ID: GN-2004B-Q-80, PI=T. Beck).  The observations were executed using two exposure times to increase dynamic range with a straightforward 4 point dither pattern.  These data were reduced into combined images using the Gemini IRAF package.  The GMOS imager has a $\sim$6$\arcmin\times$6$\arcmin$ field of view sampled in these observations at a 0.$\arcsec$0727 plate scale.  GMOS on-instrument wavefront sensor (OIWFS) delivered seeing enhanced (tip-tilt corrected) images with a FWHM of $\sim$0.$\arcsec$5.  The data processing software combined individual images in the 4 point dither pattern into a single image, and binned by a factor of 2 for an 0.$\arcsec$1458/pixel dominant WCS axis.  Field distortions were preserved in the WCS header matrix.  In all of the broad-band images, the central $\sim$0.$\arcsec$5-1$\arcsec$ region of T Tau N was saturated in the long exposure images.  The saturated pixels were replaced with unsaturated data from shorter exposure images, and inner saturation spikes on the detector were interpolated using a linear fit.  The three color g, i, z image mosaic is presented in Figure~\ref{fig.opt_img}, and the H$\alpha$ image of extended outflow structure is analyzed in \S6.

A second epoch of optical imaging through the H$\alpha$ filter was acquired with the GMOS imaging camera in October 2019 (Table~\ref{tab.obs}; Gemini program ID: GN-2019B-Q-204, PI=T. Beck).  The GMOS imaging mode had a detector change between these two observations, with the updated camera having a 0.$\arcsec$080 plate scale. The data were acquired with a 180s individual exposure time with a 5 point dither pattern.  The DRAGONS data reduction package offered by the Gemini Observatory was used to process the images, which were then also binned by a factor of 2 for a WCS with a 0.$\arcsec$1619/pixel dominant axis.  Field distortions were preserved in the WCS header matrix.  For these images, the central region was saturated and the inner spikes on the detector were not interpolated.  This imaging was used in conjunction with the 2004 epoch to measure outflow proper motions, see \S6.

\subsection{High Resolution Near-IR Imaging}\label{sec:irim}

Adaptive optics (AO) imaging provides precise measurements of the orbital motion and relative flux ratios of the three components in the T Tau system.  Near-IR AO measurements of T Tau at or near the diffraction limit were made at Gemini and Keck Observatories from 2015 - 2019 by G. Schaefer (PI).  These observations are presented and discussed in detail in a companion paper, \cite{scha20}. Many of the images are used here for reference positions of the T Tau N and Sa+Sb stars.  See \cite{scha20} for the details and a log of the photometric and astrometric results for the near-IR AO imaging observations of components in the T Tau system.  Figure~\ref{fig.fig4cont}d presents the December 2006 near IR (K-band) AO image of T Tau with the N, Sa and Sb stars identified.  

\subsection{IR Longslit Spectroscopy and Mid-IR Imaging}\label{sec:irspec}

Observations of T Tau with the SpeX instrument \citep{rayn03} on the NASA Infrared Telescope Facility (IRTF) were acquired using the LXD1.9 2-4$\mu$m grating setting (R$\sim$1000 spectra) with an 0.$\arcsec$5 slit width during scheduled visiting observations (PI=T. Beck).  Weather conditions were photometric and dry, with infrared seeing of $\sim$0.$\arcsec$55.  The spatial point spread functions (PSFs) of T Tau N and S were blended in the cross dispersion direction.  Accurate extraction of the spectra of T Tau N and S was carried out by scaling the spatial PSF measured using the telluric standard observation, and scaling, shifting and adding profiles to fit the binary using our $\chi^2$ minimization routines created for this purpose \citep{prat03, beck04}.  The component spectra were extracted and formatted into a structure for subsequent processing using the SpexTool software package provided by the IRTF \citep{cush04}.  Spatially resolved 2-4$\mu$m spectra of T Tau N and S are presented in Figure~\ref{fig.sed}.

Mid-IR imaging and spectral measurements of T Tau were acquired with the Michelle instrument at Gemini North Observatory for system verification observations (Gemini Program ID: GN-2003B-SV-81, PI = T. Beck).  The mid-IR diffraction-limited images were observed through the 7.9, 11.6 and 18.3$\mu$m medium band filters and low resolution spectroscopy was acquired from 7.6-13.5$\mu$m at resolution R$\sim$100.  Data were observed using standard chop-nod observation strategies for mid-IR wavelengths to produce near diffraction limited images. Packages in the Gemini IRAF suite of reduction routines were constructed to reduce and analyze this data.  T Tau N and S were easily resolved in the images, but the close Sa+Sb binary system was well below the resolution limit.  Flux calibration of the images was carried out using observations of Vega ($\alpha$ Lyr) as the mid-IR standard reference source, observations of which were well matched in airmass.   The spectral extraction was carried out by scaling, shifting and summing the PSF observations of Vega into a model binary using routines adapted from the near-IR processing \citep{prat03, beck04}.  The flux calibration of the spectral observations was accomplished by referencing and scaling the broadband mid-IR images at 7.9 and 11.6$\mu$m to the appropriate wavelength range in the spectra.  The the SED of Figure 2 includes the photometry and spectra, and the 11.6$\mu$m image of T Tau is presented in Figure~\ref{fig.fig4cont}e. 

\subsection{Optical Integral Field Spectroscopy}\label{sec:opspec}

High resolution optical spectroscopy with 2-D spatial resolution provides the unique opportunity to detect and isolate the inner HH emission to probe the nature of outflows from the T Tau system. The GMOS Integral Field Unit (IFU) was used with the blue slit (3.$\arcsec$5$\times$5$\arcsec$ field of view), the R831 grating, RG610 long pass filter with a 768nm central wavelength setting to acquire simultaneous R$\sim$6000 imaging spectra from 620nm - 850nm (Gemini program ID: GN-2004B-Q-80, PI=T. Beck).  Within this bandpass lie the common nebular emission features of [OI], H$\alpha$, [N II] and [S II] from 630nm to 673nm.  The fiber-fed GMOS IFU has an 0.$\arcsec$1 sampling in hexagonal spatial elements in its single slit setting, providing a spatial view that spans 3.5$"$$\times$5$\arcsec$ on the sky.  The OIWFS seeing-enhanced spectral images had an 0.$\arcsec$45 FWHM.  Two exposures of 600s were acquired at each of eight positions to map a 5$\arcsec$$\times$10$\arcsec$ region around T Tau.  The optical IFU data were processed using the standard routines in the Gemini GMOS IRAF package, and the dithered IFU positions were stitched together using specific IDL routines created for this purpose.  The optical continuum image (extracted at 630nm) is shown in Figure~\ref{fig.fig4cont}c, and the line emission maps are presented and analyzed in detail in the outflow discussion of \S5.

\subsection{Near-IR Integral Field Spectroscopy}\label{sec:irifu}

We include here and summarize in Table~\ref{tab.obs} the three different sets of near-IR integral field spectroscopy from Gemini North Observatory's Near-IR Integral Field Spectrograph (NIFS) that have been used for this project.  All NIFS observations have a two pixel spectral resolving power of $\sim$5300 and provide the spatially resolved imaging spectroscopy over a 3$\arcsec\times$3$\arcsec$ field of view with $\sim$0.$\arcsec$1 spatial FWHMs ($\sim$0.$\arcsec$1$\times\sim$0.$\arcsec$04 spatial pixels).  In all cases, observations were acquired using the Gemini North Facility adaptive optics system, Altair, using T Tau N as the optical wavefront reference star (Rmag = 9.6) with 1000Hz guiding for typical diffraction limited imaging at 1.6-2.2$\mu$m.  This provided AO-corrected imaging spectroscopy with spatial FWHM of $\sim.\arcsec$1.  IFS data were reduced and formatted into datacubes with 0.$\arcsec$04$\times$0.$\arcsec$04 square spatial elements using the tasks in the Gemini NIFS IRAF package; the reduction process is described in detail in \cite{beck08}. 

The first dataset consists of a short set of exposures through all four of the NIFS grating settings at the Z-band (0.95-1.12$\mu$m) and J-band (1.17-1.34$\mu$m) wavelengths.  These data were acquired during engineering time as a test of NIFS observing tool (OT) sequence execution, on the night that NIFS had first light at Gemini (2005 Oct. 19).  The Z and J-band continuum photometry are included in Figure~\ref{fig.sed}, and continuum subtracted map of the extended and resolved 1.08$\mu$m He~I line emission is presented and discussed in \S4.

The second dataset was from long duration flexure tests acquired during NIFS engineering commissioning time to investigate the flexure movement of astrophysical targets versus time within the NIFS imaging field.  This test used the K-band grating setting and had a long clock duration of over 4 hrs (including overheads).   T Tau is bright in the near IR, the short individual observations had to be co-added to avoid saturation, resulting in integrated IFS observations of T Tau with a combined time of 4580s.  The resulting maps of near-IR ro-vibrational H$_2$ from this dataset were previously published and analyzed in \cite{beck08}.  The H$_2$ line emission map from these observations is discussed in detail in \S5.

The third NIFS dataset presented here consists of H and K-band observations of T Tau using the 0.$\arcsec$5 diameter occulting disk (OD) to block the bright flux from T Tau N (Table~\ref{tab.obs}; program ID GN-2006B-DD-6, PI=T.Beck).   The occulting spot was used for this observation to block the bright central flux of T Tau N and to optimize the dynamic range and sensitivity to low level line emission in a shorter amount of total exposure time.  The goal of this project was to measure the extended line emission from 1.644$\mu$m [Fe II] and acquire a second epoch of near-IR H$_2$ emission maps to compare with the prior observations described above.  In \S5, the [Fe II] Line emission map from these observations is presented in detail in \S5, and the H$_2$ maps are used to analyze tangential outflow motions.

\subsection{Archival {\it Chandra} X-Ray Imaging}\label{sec:xr}

T Tau was observed with the Chandra X-ray Observatory from 01-03 January 2015 under program number 16672 (PI = P. C. Schneider).   The ACIS instrument with the High energy transmission grating (HETG) acquired spectra and the 0th order HETG map of X-ray emission from T Tau over a 127.4~ks exposure time.   The 0th order image has a 0.4 - 10kEV bandpass (31 - 1.2\AA).  The image and spectra from this observation are presented and analyzed in detail in \cite{schn18}.  Particularly, the X-ray emission character from T Tau N, and its observed anti-correlation with optical tracers of mass accretion activity, are analyzed and modeled in that study.  For wavelength completeness in our view on the T Tauri triple system, the Chandra ACIS HETG 0th order image of T Tau is included in the Figure~\ref{fig.fig4cont}a.  This super-sampled image was shared by P. C. Schneider (private communication).  The X-ray map of T Tau shows appreciable emission to the south of T Tau N, at a position angle that seems consistent with high energy photons near the position of T Tau Sa+Sb passing through the foreground attenuating material \citep{schn18}.

\subsection{Archival {\it HST} Ultraviolet Imaging}\label{sec:uv}

We re-analyzed archival data from the HST ACS ultraviolet imaging from project 10489 ( PI = A. Brown).    The HST is the only available option for characterizing astrophysical emission at far-ultraviolet (FUV) wavelengths.   Images of CTTSs acquired through the ACS Solar Blind Channel (SBC) camera using the F140LP filter include stellar continuum and chromospheric flux at the position of young stars, but also within this band-pass is spatially extended emission from the H$_2$ electronic transitions of the Lyman bands. The UV H$_2$ luminosity correlates directly with central accretion luminosity (e.g., from CIV emission \cite{fran12}).  Images acquired with the ACS SBC F165LP filter span a bandpass that is longward of the UV H$_2$ emission, only central stellar, chromospheric flux and scattered light is seen.  Hence, the F140LP and F165LP filters effectively provide a ÒUV H$_2$ onÓ and ÒUV H$_2$ offÓ filter combination to study extended UV H$_2$ emission in the environments of young stars. The photospheres of young $\sim$K spectral type CTTSs are intrinsically faint at these FUV wavelengths, but the UV H$_2$ emission can be strong and circumstellar dust attenuation and scattering is stronger than at longer wavelengths.  Models of X-ray and UV heating of H$_2$ in disks predict that emission should be compact and not extended beyond 10-30AU from the star \citep{nomu05, nomu07}.  The vast majority of the disk H$_2$ arises from the hot gas in the inner 10AU, and the extended emission traces low density surface layers in the upper disk \citep{fran12}.  Here we present the ACS F140LP image of T Tau in Figure~\ref{fig.fig4cont}b, and it is discussed in detail in sections \S4 and \S6.

\subsection{Archival 1.3mm ALMA Continuum Imaging}\label{sec:mm}

To reveal and understand the disks in the T Tau triple system, we re-analyzed archival data from the Atacama Large MM Array (ALMA) from project 2016.1.01164.S (Band 6 at 0.$\arcsec$12 resolution; PI= Herczeg; published in \cite{long19} and \cite{mana19}).  The observations of T Tau were obtained on 18 August 2017 using 47 12-m antennas on baselines of 21$\sim$3697m.  The general program observations are described in \cite{mana19}.  The pipeline science data products in the archive had visible artifacts, so we improved the image processing in the 233GHz (1.33mm) data using phase self-calibration techniques.  The phase self calibration with 6sec smoothing time improved the dynamic range of the default calibrations from 300 to 1600.  An amplitude self calibration with 12sec smoothing time resulted in no further gain.  These new processed images show two obvious sources, T Tau N and the combined S+Sb system.  This dust map is included in continuum images of Figure~\ref{fig.fig4cont}f and analyzed in greater detail in the disk discusion of \S4.  Unfortunately, the $^{13}$CO and C$^{18}$O data included in the ALMA setting was difficult to interpret because the measured signal traces only the low level fluctuations on an underlying distribution of material that is resolved out; these emission line images seem to lack information from the short interferometric spacings.

\section{Infrared Variability of T Tau South} \label{sec:ir_var}

\begin{figure}
\includegraphics[scale=0.6,angle=0]{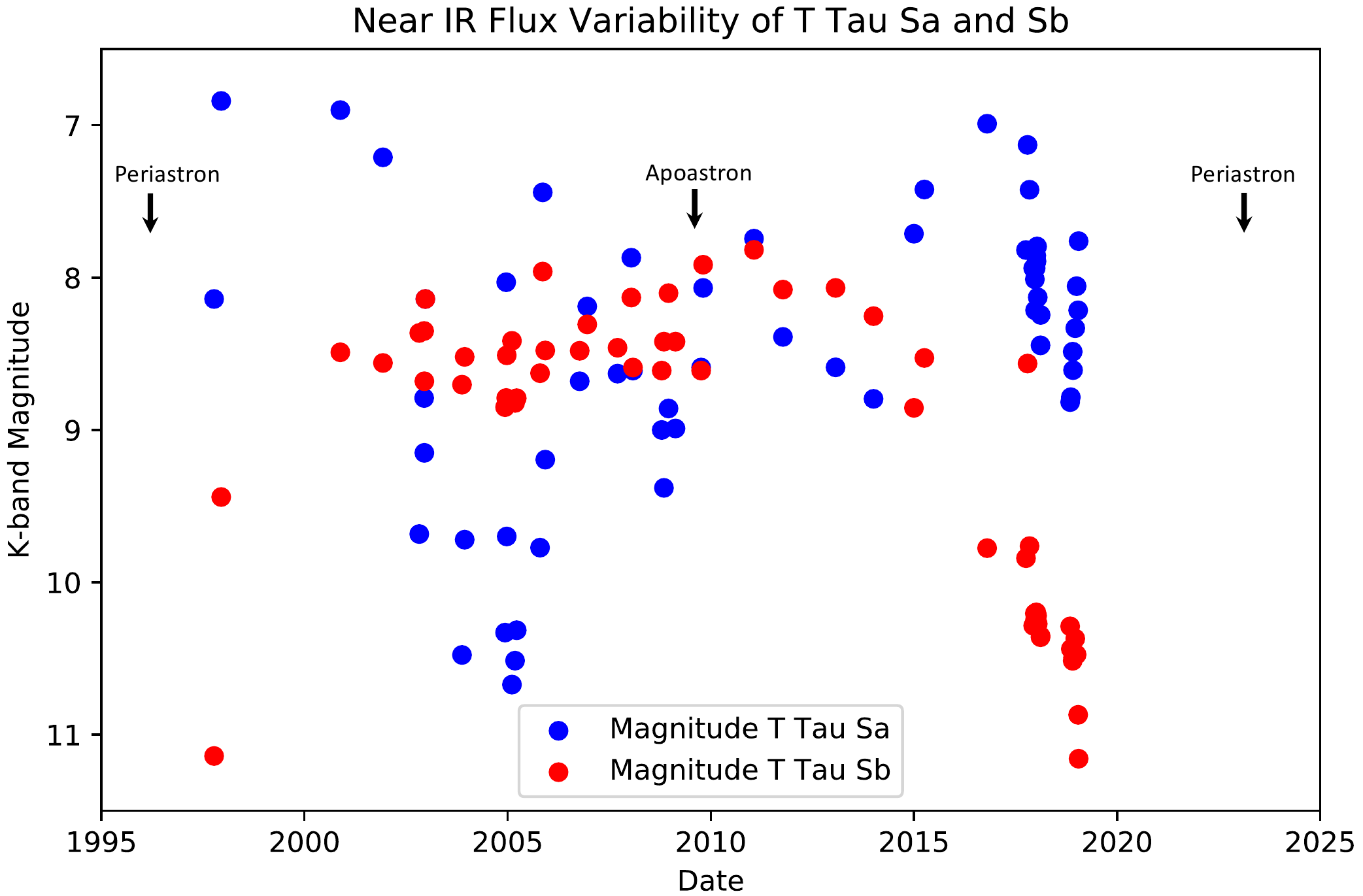}
\caption{The infrared variability of T Tau Sa (blue) and Sb (red) from spatially resolved measurements of the system made from 1997 to 2019.  Typical uncertainties on the measurements of T Tau Sa are $\sim$0.1 - 0.3mag, and Sb are 0.3 - 0.5mag in the absolute photometry (see \cite{scha20}). 
\label{fig.ir_var}}
\end{figure}

The combined light from Tau South has been known to vary significantly since the discovery of the system as a young multiple star \citep{dyck82, ghez91, beck04}. \cite{scha20} compiled and presented 24 years of spatially resolved brightness measurements of T Tau Sa and Sb, showing rampant variations in the near IR (K-band).  The compiled light curve of T Tau Sa and Sb is presented in Figure~\ref{fig.ir_var} and includes data from \cite{kore00, duch02, furl03, duch05, duch06, vanb10, scha14, scha20}.  Highlighted by arrows in Figure~\ref{fig.ir_var} are the temporal locations of the prior periastron passage date (1996.10) the most recent orbital apastron (2009.69) and the next periastron passage (2023.28) in the 27.2$\pm$0.7 year orbital period \citep{scha20}.  T Tau Sb was discovered in late 1997, nearly two years after the last periastron passage.  Hence, to-date there are no resolved measurements of the brightness of T Tau Sa and Sb at the time of periastron.  Indeed, these measurements are likely difficult to make, as shown in Figure~\ref{fig.ir_var} and emphasized by \cite{scha20}.  T Tau Sb has declined in brightness by over 3 magnitudes since $\sim$2015 and it seems to be increasingly faint as it approaches periastron.  At the greatest stellar separation during apastron in mid 2009, the brightness of both T Tau Sa and Sb seemed to stabilize.  Within $\sim$2-3 years of apastron, from 2006 to 2012, the IR flux measurements vary by about 1 magnitude rather than the 3+ magnitudes seen outside of this time frame.  This indicates that properties that drive the variability, such as rampant stellar accretion or variations in line-of-sight material, are less pronounced. This might be expected during a more quiescent system state during the time frame around apastron passage \citep{vanb10}.

\begin{figure}
\includegraphics[scale=0.6,angle=0]{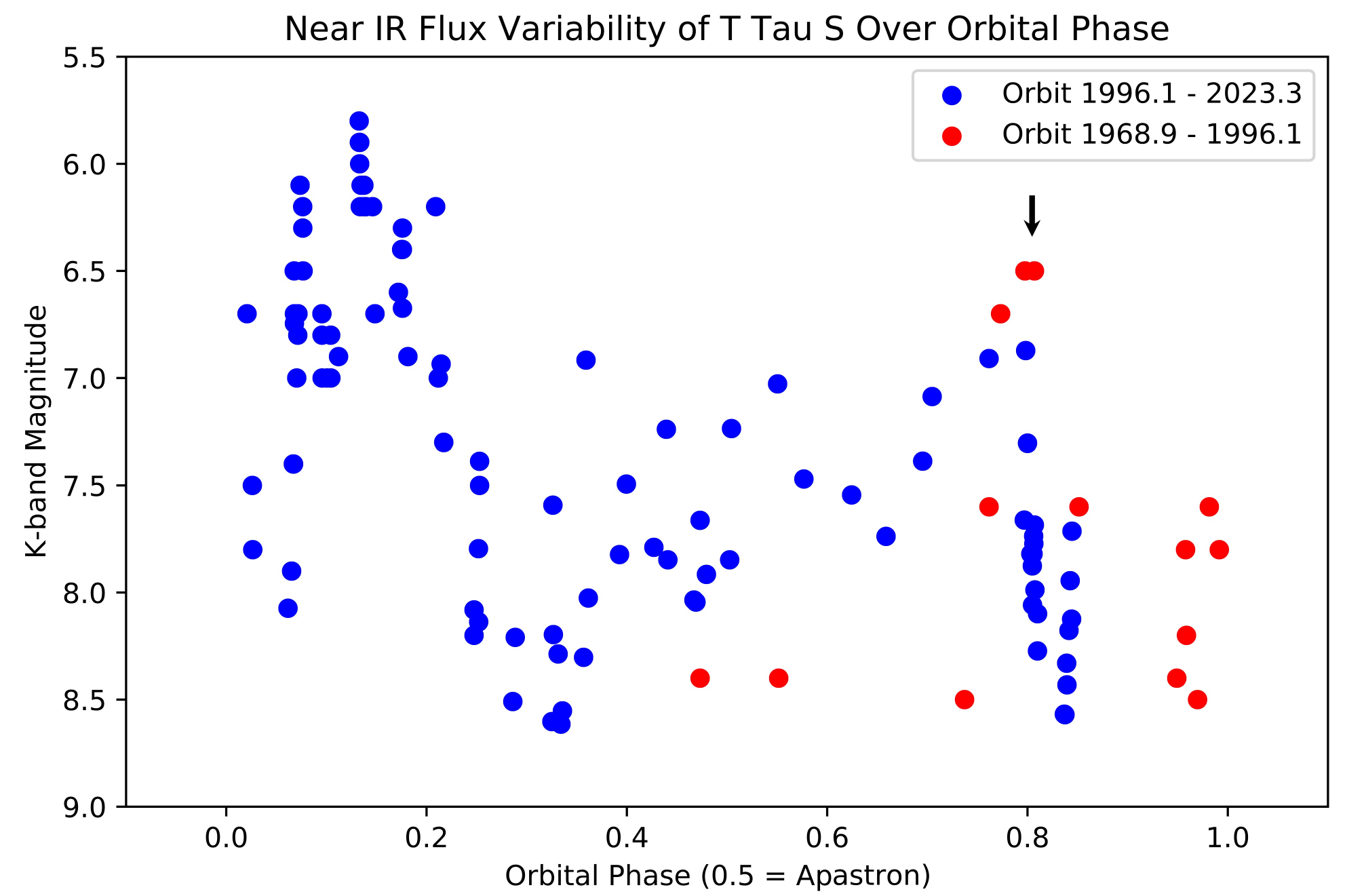}
\caption{The phase-wrapped light curve of the combined flux of T Tau Sa+Sb from all measurements that resolved T Tau S from N. Blue points show the data from 1996.10 to 2019, and red points are from the prior orbit before the last periastron at passage at epoch 1996.10.  Phase 0.5 corresponds to the orbital apastron.  The black arrow presents the phase location of the $\sim$2 magnitude grey flare discovered by \cite{ghez91}.  Typical uncertainties on the combined magnitudes of T Tau S are $\sim$0.2 - 0.3mag.  \label{fig.ir_var2}}
\end{figure}

\cite{beck04} compiled all of the unresolved brightness measurements since the discovery observation of T Tau S in 1981 \citep{dyck82}.  When coupled with the summed component fluxes of T Tau Sa+Sb collected by \cite{scha20}, 120 measurements of the brightness of T Tau S are available that span nearly 40 years.  Figure~\ref{fig.ir_var2} presents the phase-wrapped orbital light curve of the T Tau S system, based on this combined dataset.  The data points in blue are from T Tau S system magnitudes during the 1996.10 to 2023.28 orbit of the Sa+Sb binary, and the data points in red are from the prior orbit from 1968.93 to 1996.10.  T Tau S was fainter in the measurements made in the early 1980s compared to those during the same phase in the current orbit.  Interestingly, a brightness increase of the combined system in late 2016 near orbital phase ~ 0.8, driven primarily by the nearly $\sim$2 magnitude increase of the flux of T Tau Sa, was contemporaneous in orbital phase with the first gray flare of T Tau S discovered by \cite{ghez91}.  This coincidence suggests that the combined flux variations of T Tau S may be linked to the orbital phase of the binary. 

\section{Disks in the T Tau System} \label{sec:disks}
 
\begin{figure}
\plotone{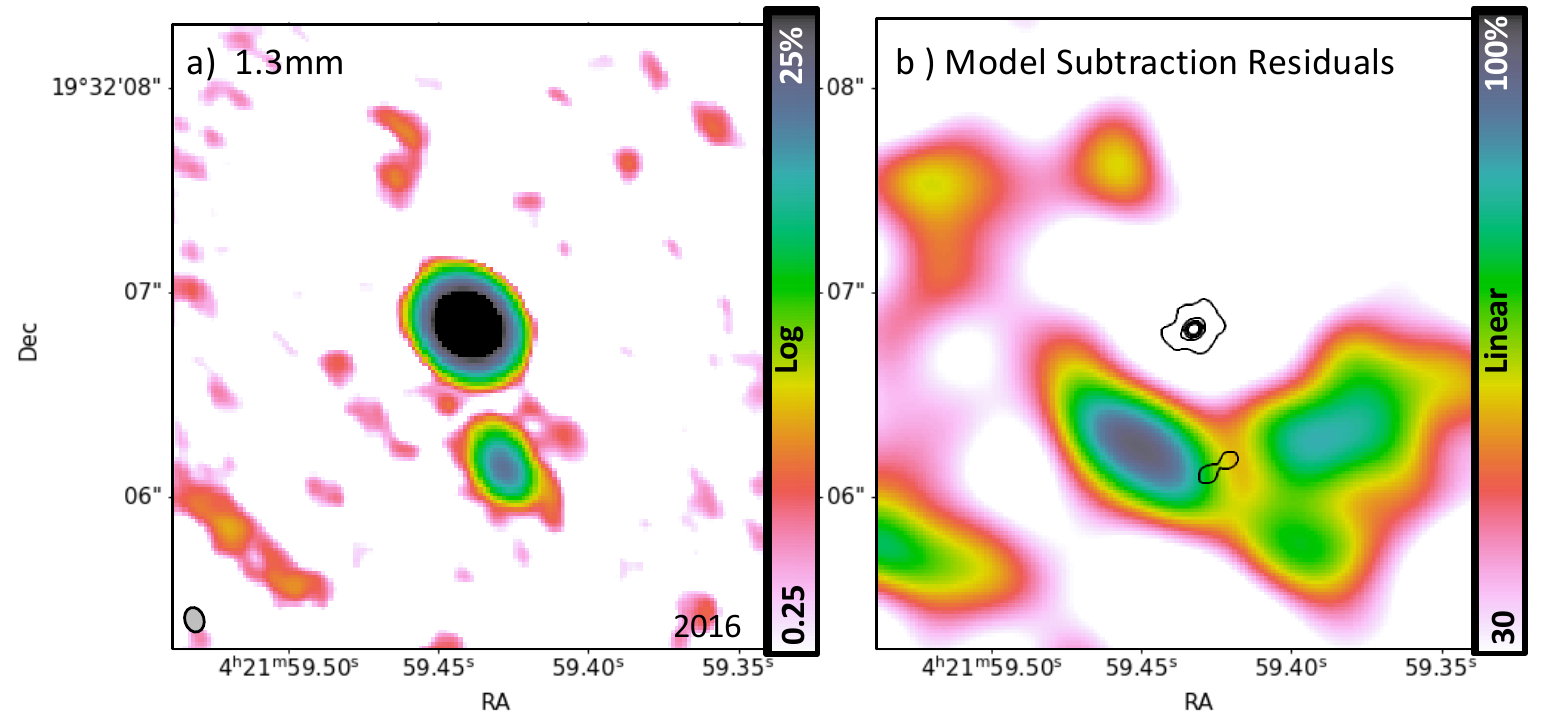}
\caption{The ALMA 1.3mm image of the T Tau system (a) and the residual dust emission after model subtraction (c).  Overplotted on the residual image are contours of near IR continuum flux showing the positions of the stars (epoch 2006), with levels of 5, 10, 40, 70 and 90\% of the peak flux.  The epoch 2006 contours are used to present positions of Sa and Sb here because T Tau Sb had become too faint to easily appear in the contour maps from the contemporaneous epoch 2017 near IR measurement. \label{fig.mm_emission}}
\end{figure}

We used the archival Band 6 233GHz ALMA data to reveal and understand the disks in the T Tau triple system \cite{long19, mana19}.   Figure~\ref{fig.fig4cont}f presents the 1.3mm continuum image that shows two obvious sources; T Tau N and the unresolved Sa+Sb system.  A detailed analysis to fit disks in the UV plane suggests that the 1.3mm dust continuum emission from T Tau is best described by four components: resolved CS disks around both T Tau N and Sa, emission from an unresolved CS disk around Sb, and elongated emission to the south of T Tau N, extended around Sa+Sb.  We modeled the emission from T Tau N and Sa each by an elliptical Gaussian, and that from T Tau Sb by a point source.  A global fit was performed with this overall model to the measured visibilities, keeping the position, intensity and, where relevant, sizes and orientation of the model sources as free parameters. We used the UV{\_}FIT command in the 'IMAGER' program of the GILDAS software tool \citep{gild13}.  The fits optimally found three components, with the northern component resolved, and the southern-most resolved in one axis.  The position of the mm fits to the resolved and unresolved southern components correspond with the IR positions of Sa and Sb, respectively.   Table~\ref{tab.cs_disks} summarizes our fit parameters for the 1.3mm continuum circumstellar dust disk emission around T Tau N, Sa and Sb, and Figure~\ref{fig.mm_emission} presents a graphical depiction of the results.  Numerical model uncertainties on the disk flux fits are at a level of 0.2mJy or less, disk radii are $\sim\pm$2AU.  After removing the best fit model from the data, the residual image clearly reveals a spatially resolved emission slightly South of T Tau N, extending around Sa+Sb. Figure~\ref{fig.mm_emission}a shows the continuum map and Figure~\ref{fig.mm_emission}b presents the residual after the best-fit model was subtracted from the measured image.  For all three of the stars, the measured disk positions in the 1.3mm dust observations are consistent with the IR continuum positions from AO measurements at the 2017 epoch of the observations \citep{scha20}.  In Figure~\ref{fig.mm_emission} the IR continuum positions of the stars are presented using epoch 2006 observations because in the near contemporaneous epoch 2017 IR measurements T Tau Sb was too faint relative to T Tau Sa to show up in the flux contours.

\begin{deluxetable*}{l|cccc}[b!]
\tablecaption{Summary of Circumstellar Disk Properties Derived from mm Dust \label{tab.cs_disks}}
\tablewidth{0pt}
\tablehead{
\colhead{Star}  & \colhead{Integrated 1.3mm Flux (mJy) } & \colhead{Disk Radius (AU)} & \colhead{Disk Inclination}  & \colhead{Disk PA} 
}
\startdata
T Tau N & 190.0$\pm$0.2 & 18 &  28$\pm$1$^{\circ}$ & 89$\pm$1$^{\circ}$ \\
T Tau Sa & 8.4$\pm$0.06 & 3.5 &  $>$70$^{\circ}$ & 4$\pm$2$^{\circ}$  \\
T Tau Sb & 1.17$\pm$0.05 &  $<$2 &  -- & --  \\
\enddata
\end{deluxetable*}

\subsection{The Circumstellar Disks}\label{sec:csdisks}

The circumstellar disk around T Tau N has been previously measured in mm dust and is known to be viewed nearly face-on \citep{akes98, long19, mana19}.  In our reanalysis of the ALMA maps, the relatively bright ($\sim$190 mJy) 1.3mm dust emission around T Tau N is well fit by a uniform disk of 0.$\arcsec$25 (35 AU) diameter.  The disk is slightly elongated in the east-west direction at a position angle of 89 degrees, and the inclination of 28$\pm$1 reveals that the disk is nearly coplanar with the orbital plane of T Tau Sb around T Tau Sa \citep{scha20}.  The results for our disk fit for T Tau N are similar to those derived by \cite{mana19} from the same dataset.

The measured disk around Sa has a 1.3mm emission level of $\sim$8.4$\pm$0.06 mJy with a $>$70$^{\circ}$ inclination at a position angle of $\sim$4$^{\circ}$.  The elliptical Gaussian disk fit is unresolved in the east-west dimension and has a FWHM of $\sim$50 mas (7 AU diameter) in the north-south direction.  Hence, the fit to the Sa emission is consistent with a highly inclined disk that is not coplanar with either the disk of T Tau N or the Sa-Sb orbital motion.  In fact, the orientation and character of the disk might be consistent with a perpendicular $\sim90^{\circ}$ orientation with respect to the T Tau Sa-Sb orbital motion, as predicted by the dynamical models of \cite{lubo18}.  The results for our disk fit for T Tau Sa are similar to those derived by \cite{mana19}.  Additionally, the size and orientation of the disk around T Tau Sa that we find in the ALMA maps is similar to the measurements at 10$\mu$m from the VLTI \citep{ratz09}. 

\cite{mana19} fit the 1.3mm ALMA continuum maps of emission from T Tau with a two component model for disks around T Tau N and Sa.  Although our respective circumstellar disk position angles and elongations are similar, they find a slightly lower disk emission level for T Tau N (179mJy versus our 190mJy), and a slightly higher 1.3mm continuum disk level for T Tau Sa (9.7mJy versus our 8.4mJy).  Our reanalysis finds a best model fit with disk emission from all three of the stars.  Incorporating the third component, T Tau Sb, decreases the modeled flux from T Tau Sa and increases the flux from N.  Our fit to the 1.3mm emission around T Tau Sb is at a level of 1.17$\pm$0.05 mJy and is unresolved, suggesting a small and compact ($<$2 AU radius) CS disk.  The small physical sizes of the CS disks around both the Sa and Sb stars can be naturally explained by tidal truncation effects, as the binary periastron distance is only $\sim$5AU \citep{arty94}.  The combined T Tau S system is known to be variable in cm emission of non-thermal origin at a level of $\sim$1.3 mJy \citep{ray97}; the variable radio emission may also contribute at a low level to the measured flux at 1.3 mm.

\subsection{A Circumbinary Ring Around T Tau Sa and Sb}\label{sec:cbdisks}

The T Tau S binary is known to have a moderate distribution of material along the line of sight to the central stars from the strong reddening of the stellar flux, and the broad spectral absorption features of solid-state 3$\mu$m water ice (Figure~\ref{fig.sed}; \cite{beck01a, beck04}) and 10$\mu$m silicates (also Figure~\ref{fig.sed}; \cite{ghez91, skem08, ratz09}).   Here, we show evidence for direct measurement of weak mm flux from the extended dust along the line of sight to T Tau S.  We also present silhouette images of the attenuating circumbinary ring material encircling T Tau Sa and Sb, as seen from multi-wavelength emission line species.  

After fitting and subtracting the model of 3 circumstellar disks described in the previous section (with the disk of Sb compact and spatially unresolved), we detect faint extended emission at $\sim$1$\arcsec$ southwards of T Tau N, over about $\sim$2$\arcsec$ with a northwest to southeast elongation at PA$\sim$60${^{\circ}}$ east of north.  Figure~\ref{fig.mm_emission}b shows this emission in the 1.3mm residual map after the three component disk model has been subtracted.  We interpret this as possible detection of the circumbinary dust around T Tau Sa+Sb.  The total integrated flux of this emission across its spatial extent is 17$\pm$2 mJy.  The summed flux of this feature is more than 10 times greater than the compact disk from T Tau Sb, and is an integrated detection with a moderate level significance.   Taken by itself, this direct detection of dust in a circumbinary distribution is unconvincing.  However, when viewed side-by-side with multi-wavelength emission line maps, the origin of this residual dust distribution in the circumbinary ring becomes more compelling.

\begin{figure}
\plotone{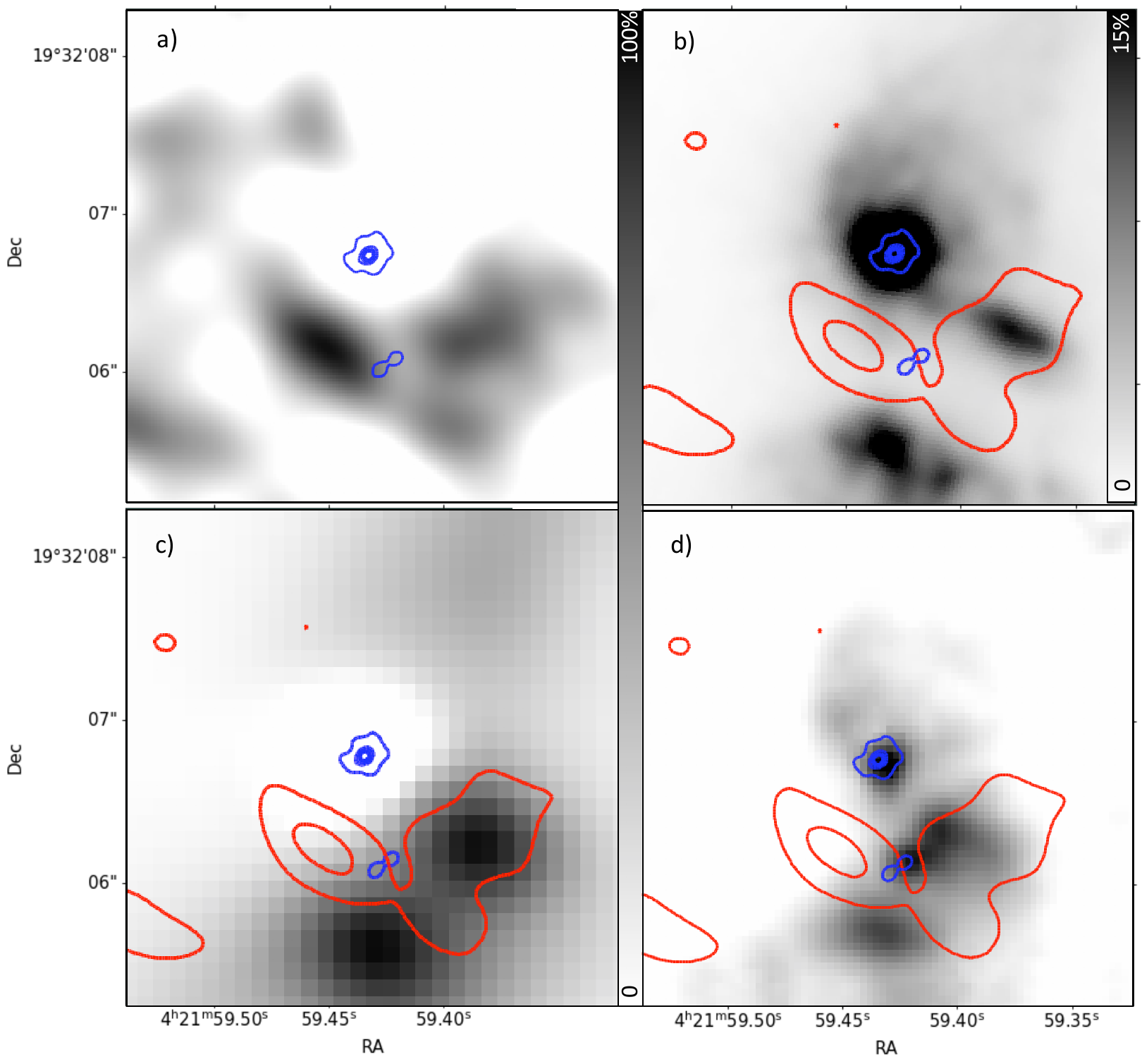}
\caption{A 3$\arcsec\times$3$\arcsec$ view of the ALMA 1.3mm residual dust image after removal of the three component disk model of T Tau (a), the UV H$_2$ emission image (b), the 6300\AA [OI] image (c) and the 1.08$\mu$m [He I] image (d).  Panels c) and d) have had continuum emission and the point-source line emission from T Tau N removed.  Overplotted in blue in each panel are five equally spaced contours from 10 to 90\% of the near-IR 2$\mu$m continuum showing the positions of the stars.  Red contours on b) to d) show the 4$\sigma$ and 6$\sigma$ RMS residual 1.3mm dust, tracing the location of the circumbinary dust ring. \label{fig.cbring}}
\end{figure}

Figure~\ref{fig.cbring}a presents the 1.3mm dust residual image after the three star disk model has been subtracted.  Figures~\ref{fig.cbring}b,~\ref{fig.cbring}c and~\ref{fig.cbring}d show images of three emission line species that exhibit a dark silhouette at the approximate position and orientation traced by the residual mm dust.  The Hubble ACS ultraviolet image in the F140LP filter (Figure~\ref{fig.cbring}b) traces bright chromospheric flux from the central position of T Tau N, and strong extended emissions from the numerous electronic transitions of H$_2$ that exist within this bandpass \citep{sauc03, walt03, fran12}.  A "dark lane" of material is seen at the position of the mm dust (traced by red contours) and encompassing the T Tau S binary (which is shown in blue).  Bright lobes of UV H$_2$ are measured to the northwest and southeast of the position of T Tau S.  This UV H$_2$ emission structure showing a lack of line emission at the position of T Tau S was first noted in the spatially resolved slit maps of \cite{walt03}.   The position angle of this dark silhouette seen in UV H$_2$ is also $\sim$50-60${^{\circ}}$ east of north.  The spatial extent of the silhouette region in between the bright lobes of emission is $\sim$0.$\arcsec$4, or nearly 60~AU at the distance of T Tau \citep{xu19}.  Figure~\ref{fig.cbring}c shows the $\lambda$6300\AA [O~I] image that has been subtracted for continuum emission, and then had a point-source image scaled and subtracted for the bright line emission from T Tau N.  The [O~I] line emission has two lobes of bright flux to the northwest and southeast of the position of T Tau Sa+Sb, with a dark silhouette of material in between.  The dark lane silhouette in the $\lambda$6300\AA [O~I] image is at approximately the same position angle as the UV silhouette.  Figure~\ref{fig.cbring}d similarly shows the map of continuum and T Tau N point-source subtracted line emission from the 1.08$\mu$m He~I feature, which also exhibits the dark lane and bright emission lobe structure of the circumbinary ring silhouette.  The He~I image shows detection of T Tau Sb through the foreground attenuating material, T Tau Sa is not seen as a point-source in He~I.   

The measured extended lobes of UV H$_2$ and [O~I] emission revealed in Figure~\ref{fig.cbring}c likely have a component of scattered line flux from strong central emission from the stars and inner disk region, but this species can also arise from shock excitation of gas in the extended outflows from T Tau S.  The He~I gas traces hot dense material with an excitation temp of $\sim$10$^5$ and is believed to primarily from the stellar accretion zone and the innermost regions of a wide angle wind or outflow \citep{beri01}.  Hence, the extended lobes of He~I are seen likely in silhouette after being scattered off of the circumbinary dust from strong emission in the central accretion+outflow region.  The symmetry of the dark-lane silhouettes in the UV, [OI] and He~I maps suggests that the circumbinary ring is viewed within $<$10$^{\circ}$ of edge-on, which is inclined by $\sim$70$^{\circ}$ with respect to the orbit of T Tau Sa+Sb.  Taken in combination and plotted side-by-side, the emission line maps clearly reveal the silhouette of the circumbinary ring material and lend credence to the valid direct detection of dust in the 1.3mm residual map.
 
We interpret the significant residual 1.3mm dust emission presented in Figures~\ref{fig.mm_emission}b and~\ref{fig.cbring}a as direct signature of the long suspected circumbinary disk around T Tau S. The peak brightness of the dust measured at 1.3mm is of order 60mK, which corresponds to A$_V$ $\sim$40, assuming T$_{dust}$ = 20 K and $\kappa_{dust}$ = 0.02 cm$^2$/g.   This circumbinary ring is responsible for the foreground attenuation of $\sim$15-20 A$_V$ toward Sa and Sb, as only the front-side of the dust measured at mm wavelengths is along the line of sight to T Tau Sa+Sb absorbs the stellar flux.   Hence, the measured 1.3mm residual dust is in strong agreement with the magnitude of the attenuating line-of-sight circumbinary disk material, and its location and orientation is well matched to the dark silhouette disk seen in emission line maps from UV through near-IR wavelengths.  The strongest emission in the 1.3mm residual dust map is offset from the position of T Tau Sa+Sb to the northeast, this may be because of over subtraction of the circumbinary ring material in the circumstellar disk models for the binary.   The extent of the disk is difficult to determine, though the mm residual extents $\sim$2$\arcsec$ or over 250~AU to the northeast they are at a low level over the background fluctuations. Higher spatial resolution mm observations to better constrain inner circumstellar disk extents and orientations for T Tau Sa and Sb would provide the necessary basis to improve the direct measurement of the circumbinary ring material.

\begin{figure}
\plotone{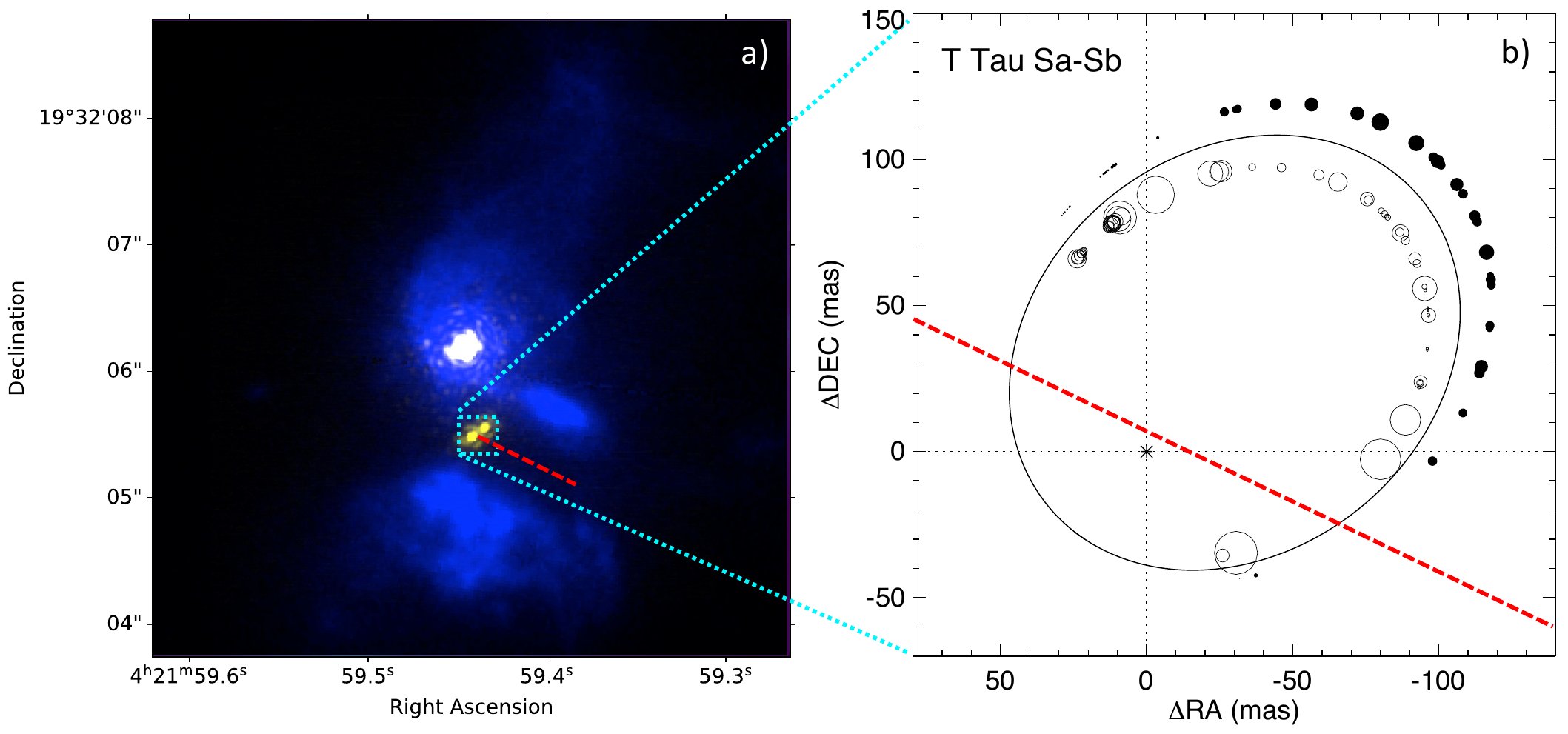}
\caption{The F140LP UV image tracing extended H$_2$ emission (blue; 2004) with the near IR continuum image (yellow; 2006) showing the circumbinary ring silhouette and stellar positions, respectively.  The red line traces the approximate mid-point and angle of the dark lane of material seen in the UV image. The box is expanded to the right showing the best-fit orbit model of the T Tau Sa+Sb binary (solid line).  The position of T Tau Sa is marked by an asterisk.  The approximate position and orientation of the mid-point in the UV-dark lane is also over plotted as a red dashed line in (b). The near IR (K-band) brightness of T Tau Sa (open circles) and T Tau Sb (filled circles) is shown near the orbital position of Sb.  The size of the symbols represents the brightness of each star in linear flux units at a corresponding time relative to the orbital position of Sb around Sa.  \label{fig.orbit_ir}}
\end{figure}

Figure~\ref{fig.orbit_ir}a presents the HST F140LP UV image (blue) which traces extended emission in the electronic transitions of H$_2$, with the epoch 2006 near IR image of the stars (yellow).    We interpret the dark lane seen in the UV image to be the silhouette of dust emission in the circumbinary ring around T Tau S.  The dashed red line shows the approximate central position of this dark lane.  T Tau Sa and Sb are offset to the northeast of the central position of the UV-dark lane, however the stars, particularly T Tau Sa, lie close to the observed center of the UV dark silhoutte (red line).  The cyan box is expanded in Figure~\ref{fig.orbit_ir}b to show the orbit of T Tau Sb around Sa.  Plotted in Figure~\ref{fig.orbit_ir}b are the near IR brightnesses of T Tau Sa (open circles) and Sb (filled circles) over their respective positions in the orbit.  T Tau Sa's orbit around the system center of mass is of coursenear the asterisk, it is shown here near the position of Sb to clarify the magnitude of the IR variability.   The brightnesses are compiled from \cite{scha20} and shown in Figure~\ref{fig.ir_var}.  T Tau Sb is faint near periastron (filled points to the left in Figure~\ref{fig.orbit_ir}b) and at its brightest near apastron (filled points to the right).  T Tau Sa seems to have an opposite character, though with much more significant variability in brightness over its orbit.  Interestingly, T Tau Sb is faintest when it is closest to the UV flux silhouette mid-point as traced by the red line in Figure~\ref{fig.orbit_ir}a, and it brightens as it reaches apastron at its greatest distance from this line.  We interpret this to mean that a substantial measure of the brightness variation of T Tau Sb arises from changes in line-of-sight obscuration as it emerges from behind the densest circumbinary dust during its apastron passage.  This interpretation must clearly mean that the orbit of Sa+Sb is non-coplanar with the distribution of circumbinary ring material, and that T Tau Sb will continue to be extremely faint relative to Sa through its upcoming periastron passage in early 2023. \cite{kohl20} analyzed IR brightness measurements of T Tau S through late 2016 and also concluded that T Tau Sb emerges from behind the circumstellar or circumsystem material as it goes through its orbital apastron.  From Figure~\ref{fig.orbit_ir}a we find that the north western UV-bright lobe around T Tau S is offset from the apastron position of Sb by $\sim$0.$\arcsec$15 ($\sim$20AU).  This means that: (1) low density UV-absorbing material (A$_v\sim$1-2) that causes the shape of the dark silhouette revealed by the UV image must exist to large distances ($\sim$30-40AU) beyond the circumbinary ring mid-plane and (2) when T Tau Sb 'emerges' from behind the circumbinary ring at apastron it is not to a line-of-sight at A$_v$=0. 

The 4$\sigma$ lower contour in the residual dust emission map shown in Figures~\ref{fig.mm_emission}b-d has an extension to the northwest.  If this structure is real and foreground to the T Tau S system, then it should effectively block the detection of any UV H$_2$ emission in the northwest emission lobe (Figure~\ref{fig.cbring}b).  This dust structure might be in the background to the UV emission, or it might be tracing a time variable change in dust morphology.  The ALMA dust maps were acquired almost 13 years after the UV H$_2$ and [O~I] emission images; it is conceivable that the northwestern dust extension is a time variable feature in the system.  Additionally, the 1.3mm wavelength might be affected by non-thermal emission from the known active magnetic radio outflows from the system \citep{phil93, skin94, ray97, john03, smit03}.  Measurements of the residual dust emission at additional mm wavelengths can determine the spectral energy to clarify the origin of this extended low level emission.

\section{Inner Outflows in the T Tau System} \label{sec:outflows}

The T Tauri system hosts three cataloged Herbig-Haro objects: the north-south HH 255 outflow from T Tau S terminates in the giant HH 355 lobes which are $\sim$20 arc minutes away, and HH 155 from T Tau N in the east-west direction extends to Hinds variable nebula 20-30$\arcsec$ distant \citep{bohm88, bohm94, reip97}.  In the past, five inner knots associated with HH 155 and 255 were located within ~3$\arcsec$ of the central stars, and longslit mapping spectroscopy kinematically traced the east-west flow to T Tau N and the northwest-southeast flow to T Tau S \citep{bohm88, bohm94}.  \cite{reip00} first postulated that the T Tauri triple system is in the process of dynamical reorganization from a non-hierarchical to a hierarchical configuration. In collimated jets, subtle changes in direction of the outflow may indicate a change in the orbital motion of an underlying binary, and regular, periodic knots of ejecta have been interpreted simply to reflect an average orbital period \citep{angl07, louv18, hao16}.  Although the inner regions of HH 155 and 255 may not be well collimated, structure observed in the innermost knots can provide a record of the kinematics which drive the flows.  High spatial resolution adaptive optics imaging investigations of the morphologies in extended emission features within $\sim$2$\arcsec$ of T Tauri have provided some conflicting results on the origins of the inner outflows, with postulation that T Tau S drives the strong western flow \citep{herb07, yang18}.  However, imaging alone does not provide a complete picture; understanding the gas velocities is especially important to identify the sources of outflows.  Spectral imaging measurements of optical and near infrared emission line species of H$\alpha$, [O~I], [S~II], [Fe~II] and H$_2$ provide sensitive tracers to disentangle the inner outflow morphologies and kinematics in T Tau.

\subsection{The Optical Outflows: [SII], [OI] and H$\alpha$}\label{sec:outflows.op}

\begin{figure}
\plotone{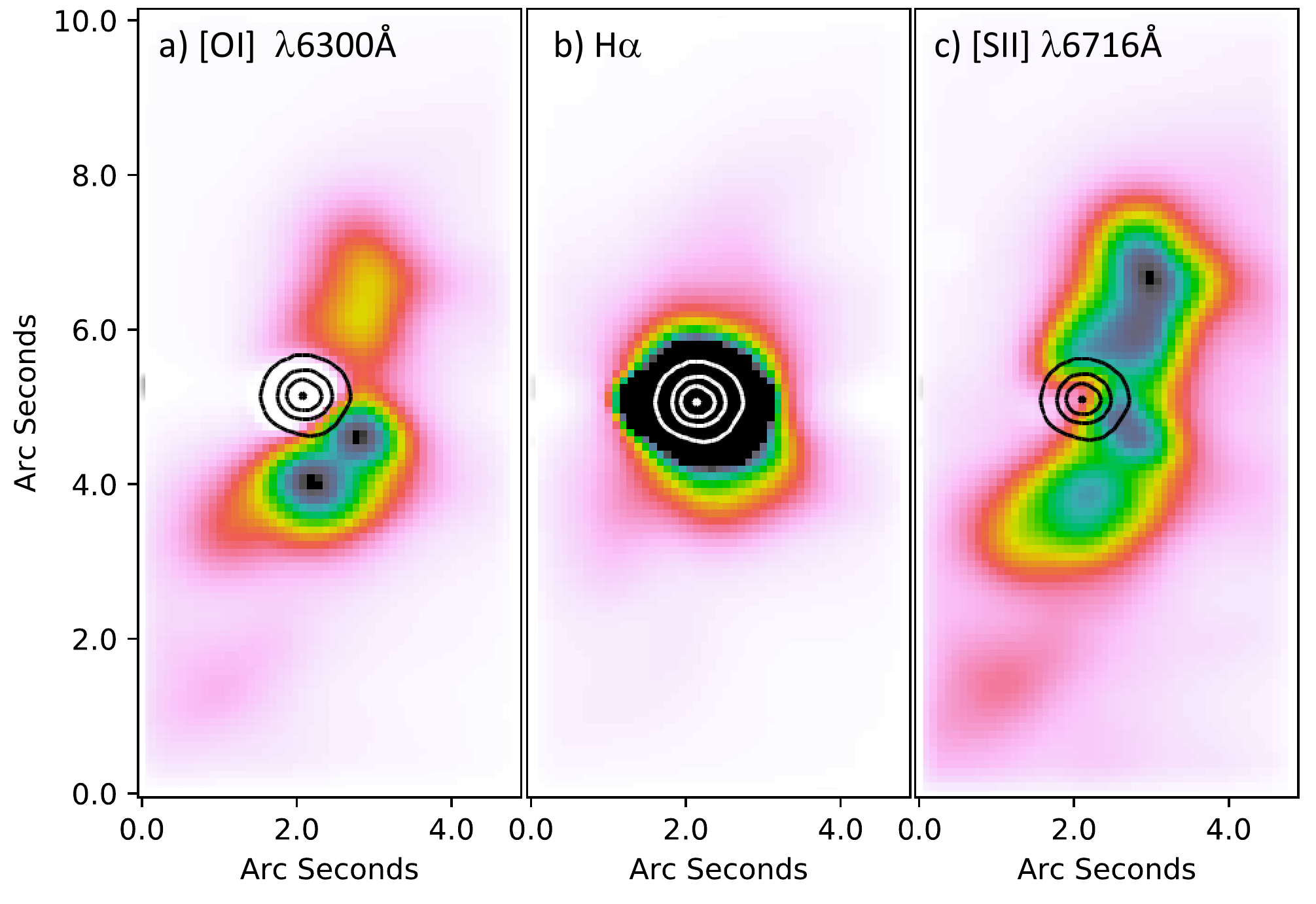}
\caption{Integrated maps of extended optical emission line species of [OI] ($\lambda$6300\AA),  H$\alpha$ ($\lambda$6563\AA) and [SII] ($\lambda$6716\AA).  The three maps have been subtracted for the continuum emission and had a point-source model for T Tau N scaled to its peak emission and subtracted, to enhance the extended structures.  Contours trace the $\sim$10, 40, 70, and 90\% emission levels of optical continuum flux, showing the position of T Tau N in these measurements.  [OI] and [SII] maps are scaled logarithmically from 0.0 to 1.0 of the peak emission line flux.  The H$\alpha$ map is scaled logarithmically from 0.0 to 8.0 times the peak flux to improve sensitivity to the bright central emission structure. \label{fig.opt_morph}}
\end{figure}

Figure~\ref{fig.opt_morph} shows the 5$\arcsec \times$ 10$\arcsec$ mosaiced IFU maps of the optical emission lines of [O~I] ($\lambda$6300.304; Figure~\ref{fig.opt_morph}a), H$\alpha$ ($\lambda$6562.83; Figure~\ref{fig.opt_morph}b), and [S~II] ($\lambda$6716.440; Figure~\ref{fig.opt_morph}c).  The H$\alpha$ map is saturated in the inner 0.$\arcsec$5 and displayed to show some of the lower level structure.  Although the optical IFU measurements include other transitions, these three emission lines are the main species we present to focus on high signal-to-noise measures of outflow morphology and kinematics.  The spectrally integrated line maps (Figure~\ref{fig.opt_morph}) have been subtracted for the continuum emission, and the continuum PSF model was scaled to the central line emission from T Tau N and subtracted to enhance detection of low level extended spatial structure.  The map of [OI] emission shows a strong bi-lobed structure encompassing the position of T Tau S.  As discussed in the previous section, we interpret the dark-lane morphology as indirect detection of the significant circumbinary ring material obscuring the T Tau S binary.  The bright [OI] lobes may arise from the line emission from a central photoevaporating disk flow \citep{rigl13}, which is then scattered off of the circumbinary dust distribution seen in silhouette.  The [O~I] map also exhibits extended emission to the southeast and north of the strong circumbinary emission lobes, this emission traces extended outflows in the T Tau environment.  The H$\alpha$ map was heavily saturated at the position of T Tau N, and the residual central flux after PSF subtraction is strong.  The H$\alpha$ line emission has multiple bright emission extensions within $\sim$1 from the position of T Tau N.  An obvious arc shaped structure extends to the southeast in H$\alpha$ emission, with a corresponding arc to the north west.  Lobes of emission similar to those seen in [OI] that trace the circumbinary ring from T Tau S might be present in the H$\alpha$ map, but clear detection is swamped by the strong emission from T Tau N.

[S~II] is an efficient tracer of extended shocks in young star outflows.  The [S~II] emission from jets is not significant near to the star because it is not emitted in the central accretion region, and the gas conditions in the central disk or outflow material are typically too dense \citep{hirt94, hirt94b, solf99}.  The [S II] emission in our spatially resolved map of T Tau shows a bi-lobed encompassing the position of the central circumbinary ring, but the relative strength of the [S~II] emission lobes is much weaker than the [O~I].  The [S~II] emission to the northwest of the silhouette in Figure~\ref{fig.opt_morph} has a clear collimated linear extension oriented directly to the north of T Tau S.  The brightest [S~II] line emission is along this linear jet extension, $\sim$2.4$\arcsec$ due north of the circumbinary silhouette.  The [S~II] map also shows an arc of emission that starts at the position of T Tau N and extends north and west.  The morphology in our spatially resolved [S~II] line emission map suggests that multiple outflow components could be superimposed along the line of sight, consistent with the past optical measurements of the outflow kinematics from T Tau \citep{bohm94, solf99}.  For these reasons, [S~II] is the primary emission species that we analyze in the following section to disentangle the multiple outflow kinematics.  

\begin{figure}
\plotone{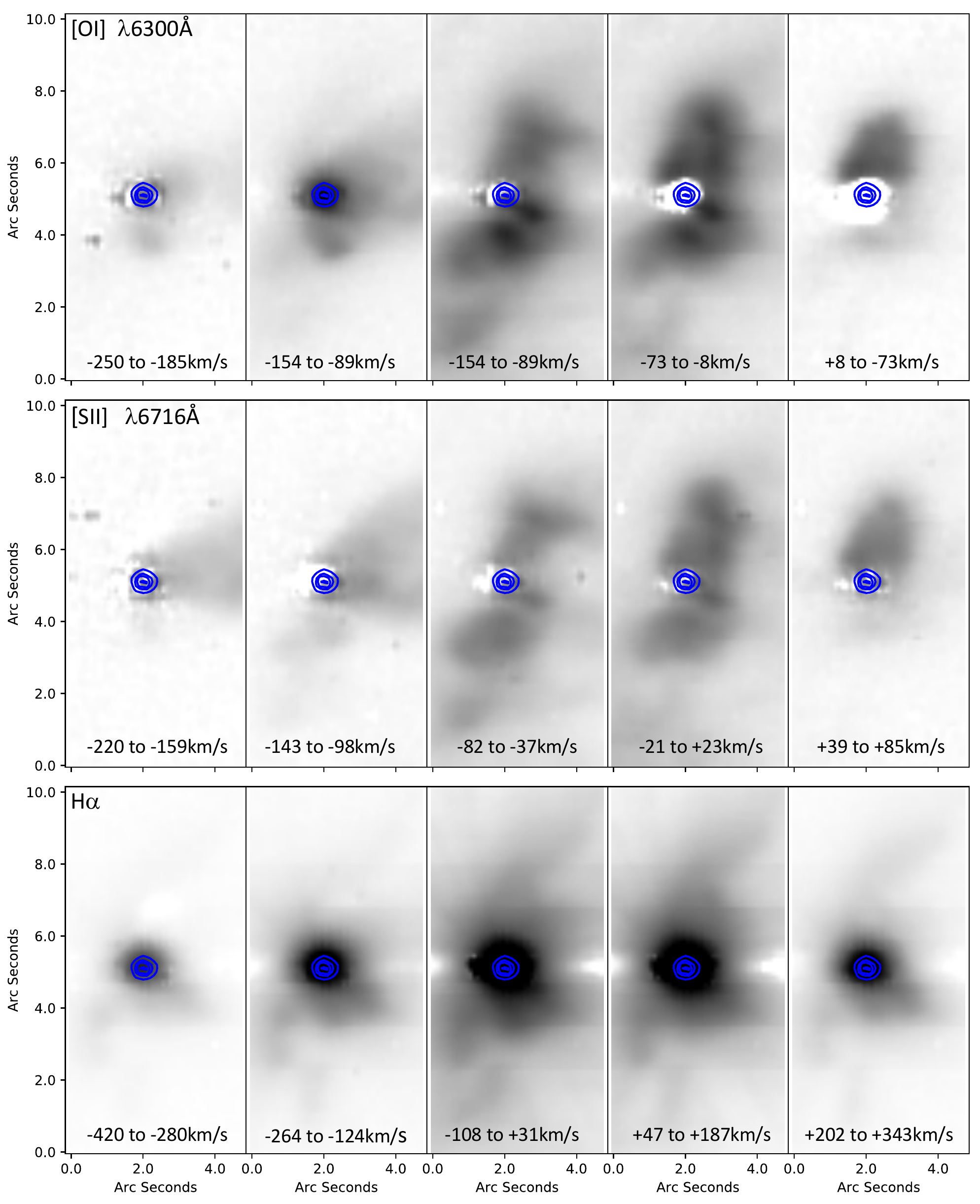}
\caption{Velocity channel maps of continuum and T Tau N point-source subtracted optical emission line species of [OI] ($\lambda$6300\AA; top), [SII] ($\lambda$6716\AA; middle) and H$\alpha$ ($\lambda$6563\AA; bottom).  The integrated velocity ranges used to construct each channel map are presented at the bottom of each panel.  Blue contours trace the 10, 40, 70 and 90\% flux contours of the optical continuum emission showing the position of T Tau N.  [OI] and [SII] maps are scaled logarithmically from 0.0 to 1.0 of the peak emission line channel flux.  The H$\alpha$ map is scaled logarithmically from 0.0 to 8.0 times the peak channel line flux to improve sensitivity to central emission structure. \label{fig.velocity_channel}}
\end{figure}

The integrated optical emission line maps of [O~I], H$\alpha$ and[S~II] shown in Figure~\ref{fig.opt_morph} were made by subtracting the continuum flux and summing the IFU data cube over the spectral axis to create narrow-band emission line images.  Figure~\ref{fig.velocity_channel} shows the kinematically resolved velocity channel maps of these three emission species, with [O~I] at the top, [S~II] middle and H$\alpha$ at the bottom.  Each panel shows velocity integrated segments of the emission, from blue-shifted at left to red-shifted emission in the right panels.   The [OI] and [S~II] line emission is seen from $\sim$-250km/s in blueward channels to $\sim$+80 in the red. The strong H$\alpha$ emission extends over 700km/s in the spectral dimension.  The morphological structures in these emission channel maps differ.  In the channel maps of [OI], the line emission shows stronger flux from the position of the bi-lobes of the circumbinary ring silhouette.  The integrated centroid velocity of this emission is $\sim$-50km/s, it is bright with high dispersion so it is seen throughout the velocity channels.  Beyond the central lobes of the circumbinary ring, the other extended [OI] emission from the outflows is at a lower flux level but similar in morphology as seen in the extended [S~II].

In the blueshifted channels of [S~II], the line emission shows strong flux arising from the position of T Tau N and extending to the west.  The southern flow material becomes more apparent in the central velocity channels, and the fourth channel (-21 to +23~km~s$^{-1}$) strongly shows the northwestern collimated jet arising directly from the upper lobe of the circumbinary ring silhouette.  The redshifted channel of [S~II] emission reveals the strongest receding emission toward the northwest.  Investigation of the [S~II] emission channel maps suggests that multiple outflow components from the T Tau triple system are overlapping along our line of sight.  The H$\alpha$ channel maps show nearly identical structure through all of the wavelength slices, with a morphology that almost exactly mimics that seen in the integrated emission image (Figure~\ref{fig.opt_morph}b).  As such, it seems that the majority of the H$\alpha$ emission from the T Tau environment is from strong central line emission that is scattered off of dust structures in the inner region.  Subtraction of a scaled average H$\alpha$ emission image from each channel (not shown) reveals extended line emission structure similar to the extended  [S~II].

\begin{figure}
\includegraphics[scale=0.5,angle=0]{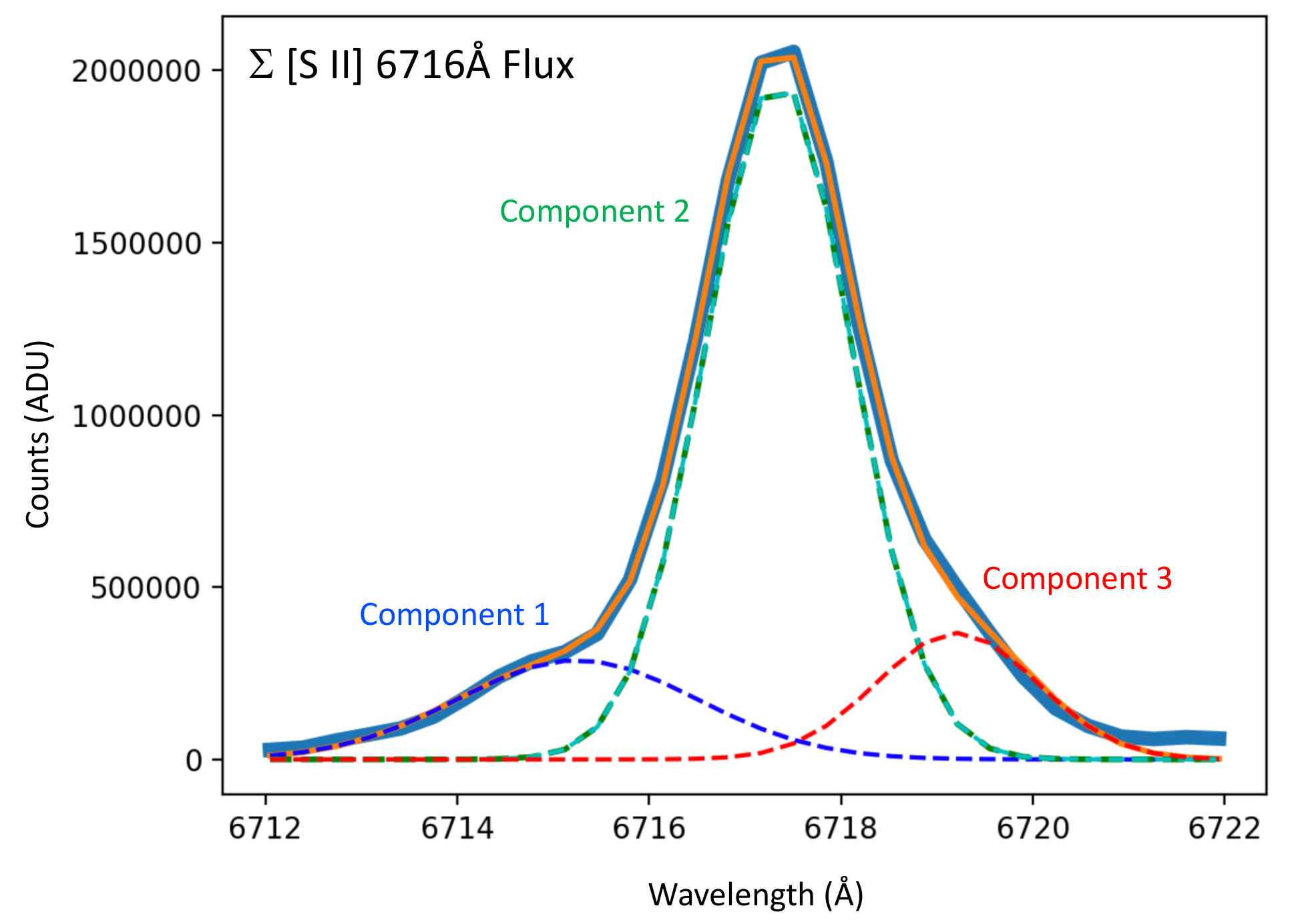}
\caption{The $\lambda$6716\AA [S~II] line emission profile integrated over the full IFU spatial field.  The integrated emission profile is well fit by a blend of three Gaussians, a blue-shifted component (Component 1), a component near the LSR velocity of T Tau (Component 2) and a red shifted component (Component 3). \label{fig.sii_profile}}
\end{figure}

Figure~\ref{fig.sii_profile} shows the line profile of [S~II] ($\lambda$6716\AA) emission that has been integrated over every spatial position in the IFU field.  This emission profile is well fit by three Gaussian components:  Component 1 is somewhat broadened and blueshifted by -78~km~s$^{-1}$,  Component 2 is near the systemic velocity and shifted overall by +12~km~s$^{-1}$, and Component 3 is narrower and redshifted by +98~km~s$^{-1}$ (typical uncertainties are $\pm$8~km/s).  Investigation and fitting of the [S~II] line emission profile at several positions within the IFU field revealed that this three component fit was successful at nearly all spatial locations.  Hence, to disentangle the line-of-sight outflow components in the T Tau system, the line profile fit shown in Figure~\ref{fig.sii_profile} was extended and carried out at all locations that had [S~II] line emission of greater than 10\% of the peak channel emission line flux.

\begin{figure}
\includegraphics[scale=0.75,angle=0]{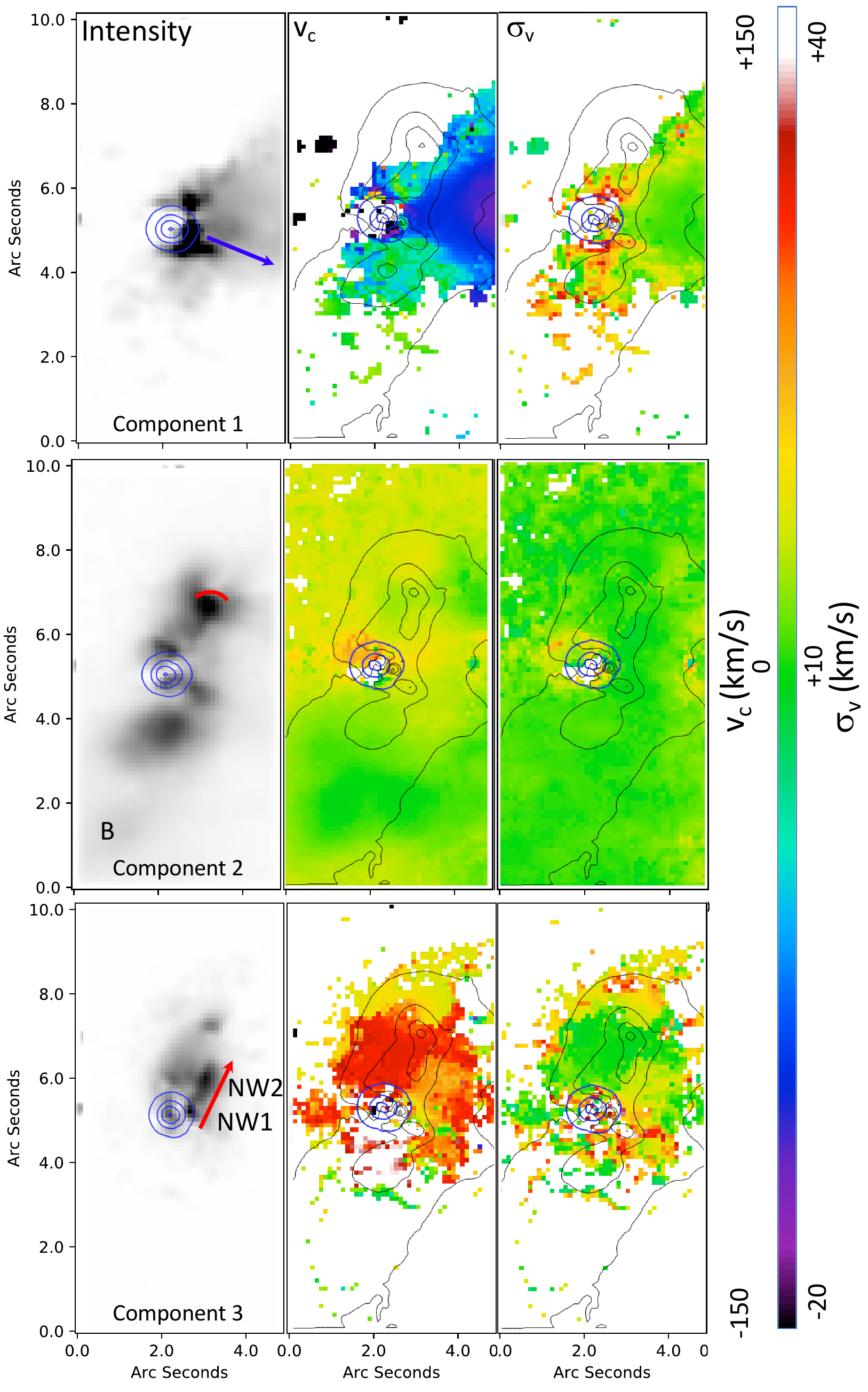}
\caption{The Gaussian emission line profile fits for $\lambda$6716\AA [S~II] shown in Figure~\ref{fig.sii_profile} were applied to every spatial position in the IFU field (with line flux$>$10\% of the profile peak) and the three spatially overlapping emission components were extracted.  The emission line intensity, line profile centroid velocity and velocity dispersion for the Gaussian fit to the blueshifted Component 1 is shown in the top panels, Component 2 I, v$_c$ and $\sigma_v$ is shown at middle, and redshifted Component 3 intensity, velocity and dispersion is in the lower panels.  Blue contours trace the $\sim$10, 40, 70, and 90\% emission levels of optical continuum flux, and black contours in the velocity and dispersion maps trace the 10, 30, 55 and 70\% contours of the integrated [S~II] map from Figure~\ref{fig.opt_morph}c.   \label{fig.sii_components}}
\end{figure}

Figure~\ref{fig.sii_components} shows the three Gaussian component fit intensity and kinematic maps for spatially resolved line fits to the [S~II] (6716\AA ) emission profile.  The intensity map is the sum image across the wavelengths of the Gaussian fit to the [S~II] emission component.  The velocity and dispersion images present the line centroid velocity and the dispersion FWHM for the Gaussian fit to the component feature in the line profile, as demonstrated in Figure~\ref{fig.sii_profile}.  Component 1 traces the most strongly blueshifted emission seen in the environment of T Tau, from the blueshifted western jet.  We find that component 1 of the [S~II] emission profile arises from the position of T Tau N and extends to the west at profile velocity centroids of -100 to -140km/s.  The blue arrow in the component 1 intensity map shows the position and angle of this blueshifted collimated western jet from T Tau N.  The inner jet is $\sim$40\% brighter than the wider angle outflow emission.  Lower velocity blue-shifted emission with higher dispersion encompasses this fast blueshifted flow, with decreasing velocity with wider outflow opening angle. This component is tracing the inner emission from the long known HH~155 outflow that extends to the west toward Hinds variable nebula and NGC~1555.    The brightest [S~II] emission seen in the component 1 intensity map is seen near to the inner regions and to the south of T Tau N.  This may represent an inner component of lower velocity shocked emission with higher dispersion.  

The brightest [S~II] emission from the T Tau outflows is seen in Component 2.  This mid-velocity emission fills the full spatial field around T Tau (Figure~\ref{fig.sii_components}, middle panels).  This emission exhibits slightly blue-shifted velocity to the south ($\sim$-10km/s), and slightly redshifted to the north ($\sim$+7km/s), with a low overall velocity dispersion.  The structure of this inner emission is consistent with the maps of extended emission [S~II] to the north and south measured by \cite{solf99}, though at a lower overall kinematic magnitude ($\pm\sim$10~km/s versus their measured $\pm\sim$45~km/s).  Our kinematic measurements of Component 2 agree with their interpretation that the bulk of the measured emission arises from the southeast to northwest flow from T Tau S, representing the innermost regions of the extended Herbig-Haro flow catalogued as HH~255.    Additionally, the radio source "T Tau R" identified and measured by \cite{ray97} had the same orientation and kinematics ($\sim$10~km/s), and so could likely have been a knot of emission in this northwestern flow from T Tau S.  The south-easterrn outflow has a wider opening angle and poorer collimation.  The bulk of the outflow motion may be in the tangential direction for these flows.   The red arc highlighted in the Component 2 intensity map designates the brightest region of [S~II] emission seen in a similar location to a bright arc of [S~II] measured by \cite{solf99}, and is spatially coincident with an arc of H$_2$ emission seen in the imaging maps of \cite{herb07}.  The position of this bright region of shocked emission in [S~II] and H$_2$ has been nearly static for over the decade spanning these respective observations \citep{solf99,herb07}.

The red arrow in the Component 3 intensity map of Figure~\ref{fig.sii_profile} (lower panels) shows a small-scale ($\sim$2$\arcsec$) inner redshifted collimated [S~II] jet that is traced to within $\sim$0.$\arcsec$4 of the position of T Tau S.  The upper knot in this jet is designated 'NW 2'; it is also seen in H$_2$ emission and discussed in detail in the following section.  The lower knot is similarly labeled "NW 1" and also discussed in its context with the H$_2$.  This collimated jet has a mean redshifted velocity of $\sim$20~km/s and it seems to be a more collimated and strongly shifted outflow component that is encompassed by the lower velocity wide-angle redshifted emission mapped by Component 2.  This redshifted inner jet feature has a different intensity and kinematic profile than seen in the other emission traced in the Component 3 fit.  Particularly, the lower level emission to the east of this jet has more strongly redshifted emission, with velocities to $\sim$100km/s.   This redshifted emission has lower dispersion closer to the stars, and decreasing velocity centroid with higher dispersion in the northern-most emission.    A weak knot of v$_{c}\sim$100km/s redshifted emission is seen $\sim$1.$\arcsec$5 due east of T Tau N.  This might be an inner eastern counterpart to the strongly blueshifted HH~155 (Component 1), but the general assumption is that the majority of the corresponding redshifted flow of HH~155 is obscured by the optically thick disk of T Tau N \citep{solf99}.

\subsection{Infrared Outflow Diagnostics: [Fe~II] and H$_2$}\label{sec:outflows.ir}

\begin{figure}
\plotone{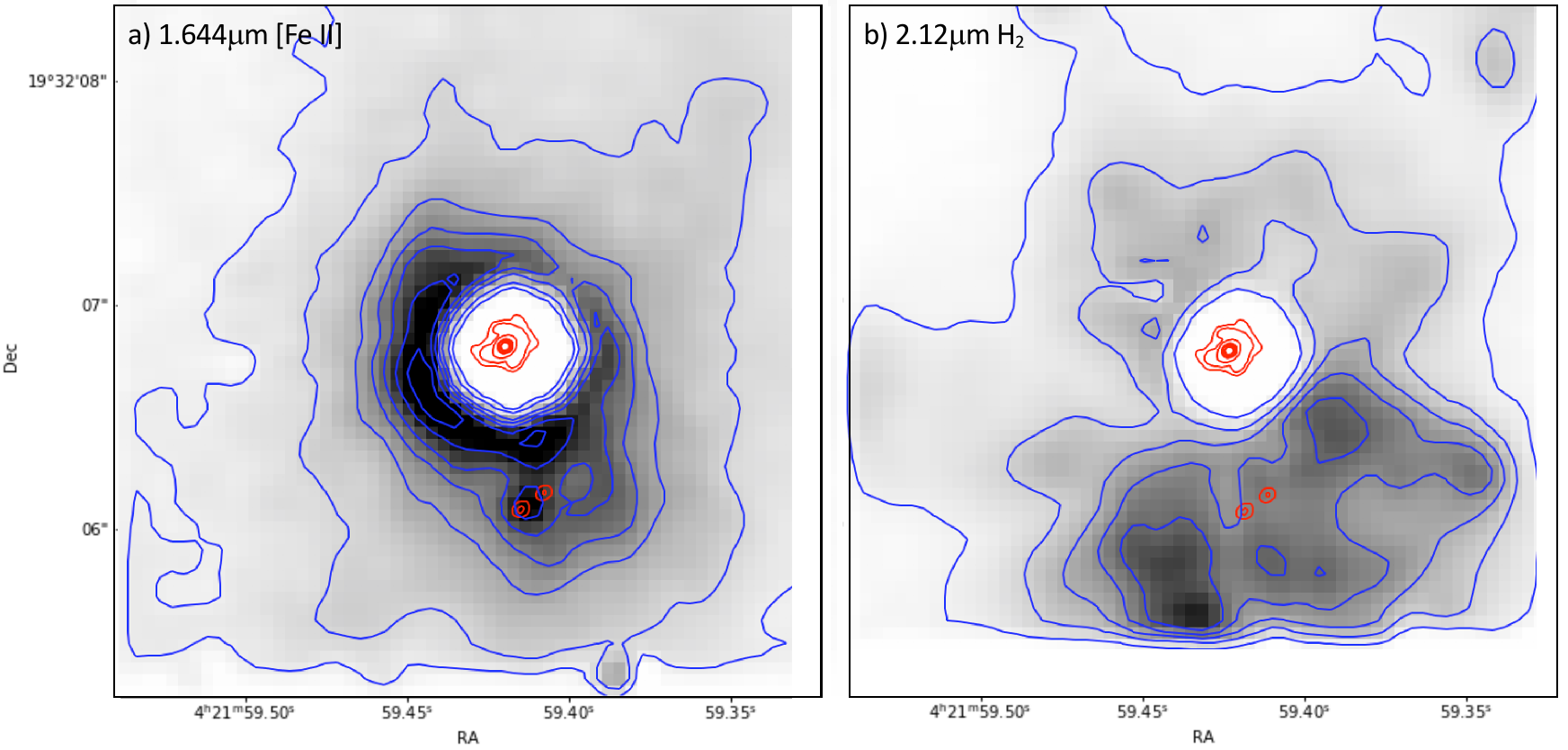}
\caption{Integrated maps of extended near infrared emission line species of [Fe~II] (1.644$\mu$m) and H$_2$ (2.12$\mu$m) from December 2006.  The maps have been subtracted for the continuum emission and the inner 0.$\arcsec$0.25 radius has been masked to designate the position of the 0.$\arcsec$5 diameter occulting spot used for the observations.   The images were scaled linearly from 0.0 to 90\% of the peak line flux.  Overplotted in red are contours of the continuum flux (K-band) from the near contemporaneous high resolution AO observations at levels of 5, 10, 20, 60, 70, 80, 90\% of the peak emission (from T Tau N).  Blue contours trace the $\sim$10, 15, 25, 40, 70, 80 and 90\% emission line levels in the [Fe~II] maps and $\sim$10, 40, 70, and 90\% emission line levels in H$_2$. \label{fig.ir_morph}}
\end{figure}

Our near-infrared integral field spectroscopic observations reveal the extended line emission in the {\it v}=1-0 S(1) 2.122$\mu$m ro-vibrational H$_2$ emission and 1.644$\mu$m [Fe~II] forbidden emission.   Figure~\ref{fig.ir_morph} shows a view of the inner 3$\arcsec \times$ 3$\arcsec$ regions for maps of the  [Fe II] and H$_2$ from the 2006 observation epoch.  The 0.$\arcsec$5 diameter occulting spot was used for these observations to enhance sensitivity to low level extended line emission.   In both cases, the spatially extended [Fe~II] and H$_2$ emission fills more than 80\% of the field of view, but the two lines show different morphological structure.  The strongest [Fe~II] flux comes from the T Tau N system, with significant emission encompassing the position of the occulting spot.  The strongest [Fe~II] from the T Tau S environment appears to arise from T Tau Sa.  These detections suggest that a moderate amount of the [Fe~II] emission arises from the heated gas in the inner circumstellar disks of T Tau N and Sa.  The [Fe~II] emission elsewhere in the field is diffuse and does not follow the same 'knots and arcs' morphological structure as seen in the H$_2$ map.  No strong peak of H$_2$ emission is seen at the position of either T Tau Sa or Sb.  The strongest H$_2$ rises from the south east of T Tau S in a bright knot of apparent outflow emission.  The extensive H$_2$ knot structures encompassing T Tau S are not measured in the [Fe~II] transition, which traces gas at similar densities but higher excitation temperature in outflows from young stars \citep{taka06}. 

\cite{herb07},  \cite{gust10} and \cite{beck08} have published high spatial resolution ro-vibrational 2.12$\mu$m H$_2$ emission maps from 2002, 2004 and 2005, respectively.  As seen in Figure~\ref{fig.ir_morph}b, these images also showed extensive knots of H$_2$ emission surrounding the T Tau Sa+Sb binary.  Analysis of emission line ratios suggest LTE gas with a shock excitation origin, probably from outflowing gas \citep{gust10, beck08}.  \cite{herb07} labeled four of the measured H$_2$ emission knots as C1 - C4 and discussed their potential origin in shocked emission from the outflows from the T Tau Sa+Sb system.  Particularly, they highlighted the C1 knot and its surrounding emission as having a bow-shock shape, possibly emanating from T Tau Sa and driving a western outflow component.  \cite{gust10} compared the positions of the C1 - C4 four knots in their second epoch of H$_2$ spectral imaging data acquired two years later, and found strongly varying knot morphology.  In \cite{gust10}, only the southern C3 and C4 knots showed possible tangential motion at an average $\sim$0.$\arcsec$09/year level, indicating outward flowing gas.

\begin{figure}
\includegraphics[scale=0.67,angle=0]{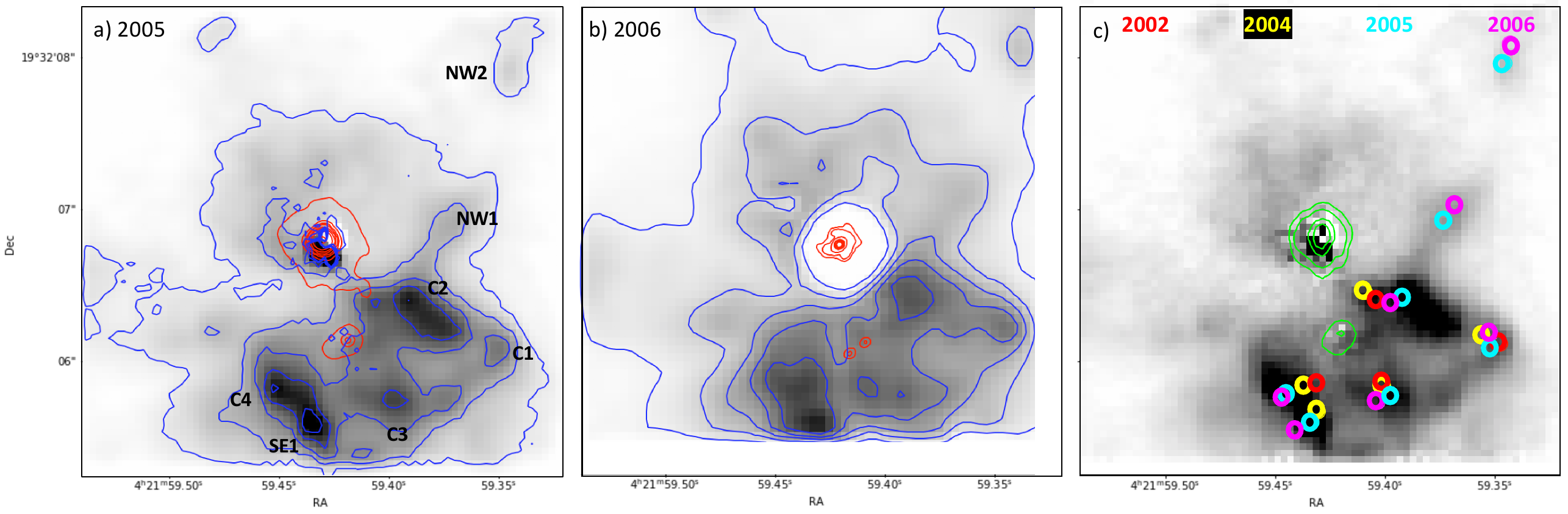}
\caption{The near IR 2.12$\mu$m H$_2$ images from October 2005 (a) and December 2006 (b).  The images are scaled from 0 to 90\% of the peak H$_2$ flux (at knot SE1 in both cases).  Blue contours trace the 10, 40, 70, and 90\% emission contours of the H$_2$.  Overplotted in red are contours of the continuum flux (K-band) from the near contemporaneous high resolution AO observations at levels of 5, 10, 20, 60, 70, 80, 90\% of the peak emission (from T Tau N).  Panel (a) also has a red continuum contour at 1\% included to outline the extended flux around T Tau Sa, which is extremely faint at the 2005 epoch.  H$_2$ knots C1 - C4 (from \cite{herb07}) and SE1, NW1 and NW2 are designated in (a).  Panel (c) shows the 2005 H$_2$ image with overlaid colored circles that show the positions of the seven emission knots from the 2002 to 2006 observing epochs, from data presented and discussed in Appendix 1. \label{fig.h2_pm}}
 \end{figure}

Figure~\ref{fig.h2_pm}a and b present the epoch 2005 and 2006 H$_2$ line emission maps side-by-side.  In Appendix 1 we measure the knot positions of the C1 - C4 knots, as well as three additional knots labeled SE1, NW1 and NW2 (identified in Figure~\ref{fig.h2_pm}a).  The NW1 and NW2 knots are also clearly seen in the [SII] maps discussed in the previous section.  Figure~\ref{fig.h2_pm}c shows the four epochs of H$_2$ knot positions (from Appendix 1) over-plotted onto the epoch 2005 H$_2$ map; red shows the 2002 measurement, yellow is 2004, cyan is 2005 and magenta is 2006.  All knot locations are corrected to the 2006 epoch for the motion of the position of T Tau Sa; this correction is shown in Appendix 1.   We find that the C1 and C2 knots are not outflowing from the position of T Tau Sa+Sb, the position of these knots varies by over 0.$\arcsec$1 in a random manner and appears to be dominated by fluctuations in the morphology of the underlying gas structures.  Particularly, the bow-shock type morphology of the C1 knot was used by \cite{herb07} to conclude that the east-west flow is emanating from T Tau S, but this knot is roughly stationary in average position and does not appear to be outflowing.  Hence its origin appears to be somehow associated with excitation of the static material in the circumbinary ring.  The location of this bow-shock shaped arc of emission in knot C1 is $\sim$1$\arcsec$, or $\sim$140~AU, approximately to the west of T Tau Sa.  Circumbinary rings around young binary star systems have an inner cavity of dynamically cleared material, the extent of which depends on the system separation and the orbital eccentricity of the binary \citep{arty94}.  It seems plausible that instead of tracing an inner outflow shock, this arc C1 instead measures static H$_2$ emission excited on the inner wall of the circumbinary ring around T Tau Sa+Sb.  However at the $\sim$140~AU distance to arc C1, this would imply that the inner cavity of the ring is cleared to over 10 times the $\sim$12.5~AU binary semi-major axis.

From Figure~\ref{fig.h2_pm}c and Appendix 1, we find that Knot C3 does seem to be flowing away from the position of T Tau Sa+Sb.  However, the derived tangential velocity of C3 is significantly lower than the other flowing knots.  This is a likely result of the strong variation in the brightness and morphology of this knot, which results in difficulty identifying the dominant gas flow.  Knots C4, SE1, NW1 and NW3 are all flowing away from the position of T Tau Sa+Sb.   Table~\ref{tab.h2} presents the tangential velocities measured in the outflowing knots, the radial velocities measured from the H$_2$ emission profile from the 2006 IFU spectral imaging data (v$_{LSR}$), and the resulting 3-D space velocity.  The uncertainty in the tangential velocity is $\pm$9~km/s and is dominated by the morphological variations in the knots; the absolute radial velocity measurement uncertainty is $\sim12$~km/s defined by the spectral resolution.  The southern knots C4 and SE1 are blue shifted, and the northwest knots NW1 and NW2 are slightly red shifted.  Knots C1, C2 and C3 are not included in Table~\ref{tab.h2} because they do not have a measurable velocity or have significant morphological variation that limits detection of the gas flow.

\begin{deluxetable*}{c|ccc}
\tablecaption{H$_2$ Knot Velocities\label{tab.h2}}
\tablewidth{0pt}
\tablehead{
\colhead{Knot Designation} & \colhead{Tangential Velocity} & \colhead{Radial Velocity} &\colhead{3-D Velocity} \\
\colhead{} & \colhead{(km/s)}  & \colhead{(km/s)} &\colhead{(km/s)} }
\startdata
C4 & 41 & -14  & 43 \\
SE1 & 36 & -14 & 39  \\
NW1 & 34 &  +2 & 34 \\
NW2 & 41 & +4 &  41   \\
\enddata
\end{deluxetable*}

The average tangential motion of the outflowing knots from T Tau Sa+Sb is 38$\pm$9~km/s, and the magnitude of the overall average 3-D space motion is 40$\pm$15~km/s.  The outflow motion is predominantly in the tangential direction, and the derived average inclination of the SE-NW outflow is 4$\pm$6$^{\circ}$ assuming bi-polarity.  The southern outflow appears to have a wider opening angle than the more collimated emission seen in the NW1 and NW2 knots.  This is evident both in the arc-like morphological structure in the C4 and SE1 knots and the stronger magnitude of the blue shifted radial velocity.  The N1 and NW2 knots are also seen in the [S~II] emission presented and discussed in \S5.1.  The [S~II] emission knots are isolated and measured in the component 3 intensity map in Figure~\ref{fig.sii_components}.  The morphology of the linear, collimated emission in the redshfited jet that ends in the NW2 knot is similar in [S~II] and H$_2$, though the [S~II] is more strongly redshifted ($\sim$20~km/s versus $<$5~km/s in H$_2$).  The morphology and kinematics are consistent with an interpretation that the northwestern outflow from T Tau S consists of knots with a faster atomic component mapped in [S~II] and associated with slower encompassing H$_2$ emission \citep{taka06}.  

\section{Southern Arcs in Burnham's Nebula: Outflow Ejection Triggered by Binary Periastron Passage?}\label{sec:outflows.binary}

\begin{figure}
\plotone{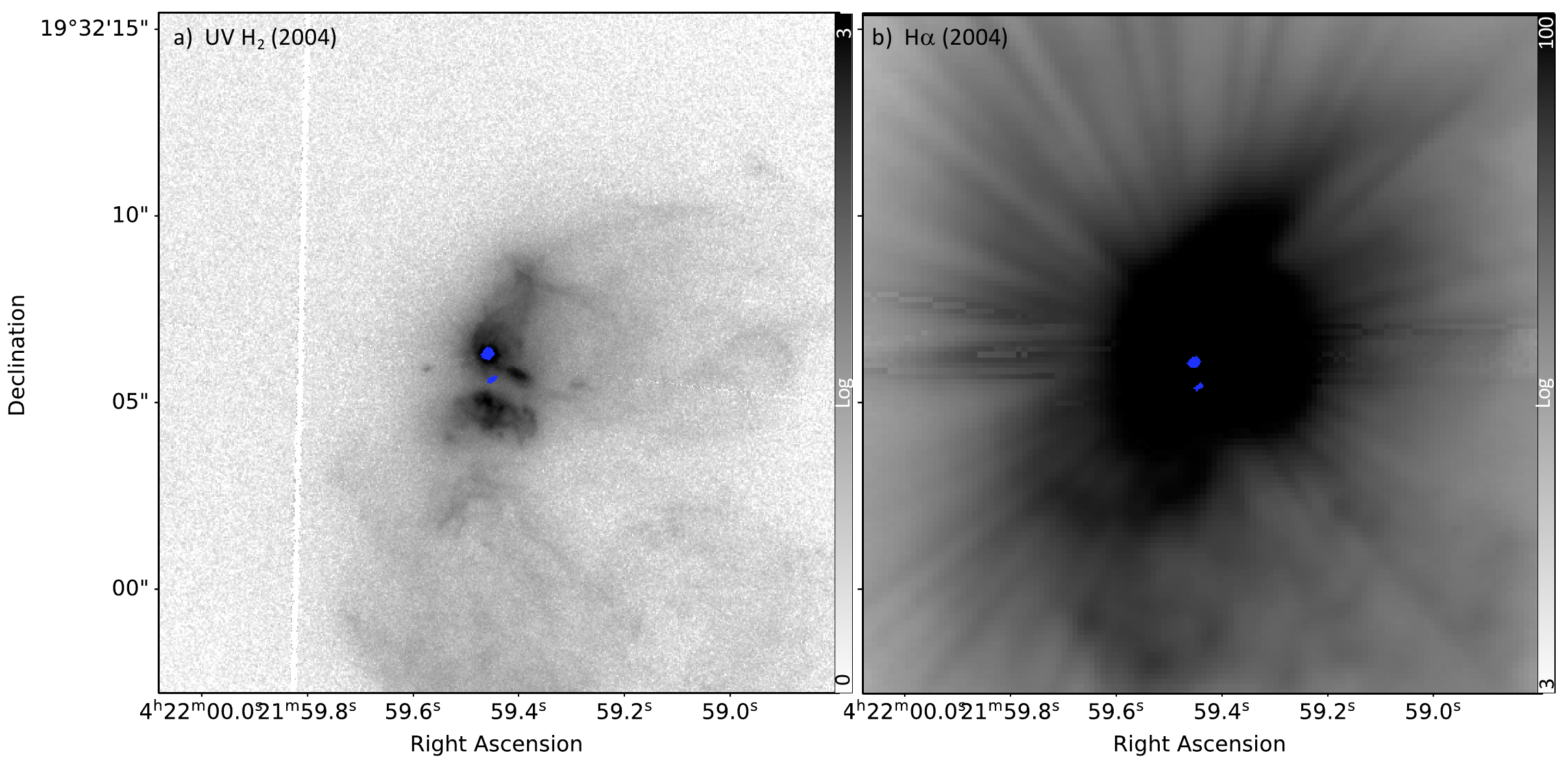}
\caption{The Ultraviolet H$_2$ image showing extended low density outflow material in the inner 18$\arcsec\times$18$\arcsec$ region of T Tau (a), and a view of the H$\alpha$ outflow arcs of emission on the same spatial scale (b).  Blue contours trace the stars and show the positions of T Tau N and S.  T Tau Sa+Sb are blended in this wide view.
\label{fig.h2_halpha}}
\end{figure}

\begin{figure}
\includegraphics[scale=0.65,angle=0]{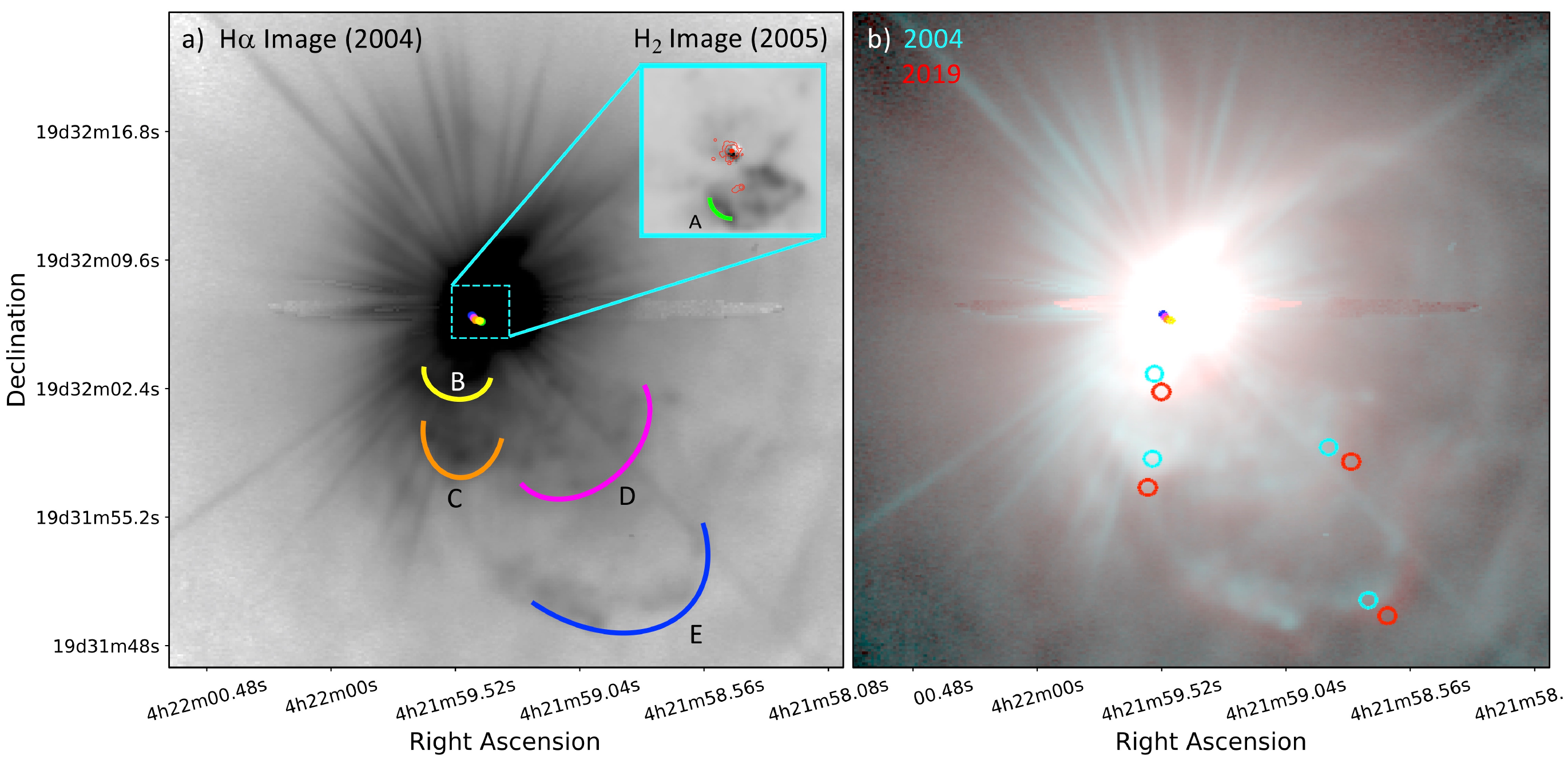}
\caption{ (a) The log-scaled H$\alpha$ image with the southern outflow loops identified as colored arcs (labeled A through E). The zoomed view at right shows the near-IR H$_2$ image from 2005 (\S5), revealing the inner outflowing arc, designated A.  The zoomed view to the left highlights the colored asterisks from green through blue that map that the positions of T Tau Sa relative to N, at the time of the five last periastron passages of T Tau Sb.  (b) A two color image of the 2004 epoch extended H$\alpha$ emission (cyan; log scaled 3\% to $\sim$50\% of the peak flux) and the 2019 epoch H$\alpha$ (red; log scaled 2\% to 20\% of the peak flux).  Positions for the "B" through "E" arcs that were used to derive tangential velocities are overplotted as cyan and red circles, respectively.  Colored inner circles map the position of T Tau Sa, as also seen in (a).  \label{fig.halpha_arc_ID}}
\end{figure}

Figure~\ref{fig.h2_halpha} shows the inner 15$\arcsec\times$15$\arcsec$ view of the spatially extended Ultraviolet H$_2$ emission from T Tau (a), side-by-side with a zoomed view of the same H$\alpha$ emission map (b).  At distances greater than $\sim$1$\arcsec$ from the central stars, the ultraviolet H$_2$ emission likely traces low density H$_2$ gas in outflow shocks.  The H$\alpha$ emission also traces the shock interfaces in the Herbig-Haro flows.  However, the UV H$_2$ and H$\alpha$ line emission maps show different morphologies.  In Figure~\ref{fig.h2_halpha}a, the bright lobes of H$_2$ emission surrounding the UV-dark circumbinary ring are seen, and low level UV H$_2$ extends to the south and north of T Tau.  Figure~\ref{fig.h2_halpha}b shows that the inner $\sim$3$\arcsec$ region has extensive and strong H$\alpha$ emission with an arc extending $\sim$3.$\arcsec$2 to the southeast.  This southeastern arc is measured to be slightly blue shifted by$\sim$14~km/s in the spectral images analyzed in \S5; this arc comprises the blue shifted southern emission in the "Component 2" [S~II] profile outflow that arises from T Tau S.  Additional arcs of southern H$\alpha$ emission are seen in Figure~\ref{fig.h2_halpha}b and Figure~\ref{fig.halpha_arc_ID} at increasing distance and decreasing brightness levels.  These extended H$\alpha$ outflow arcs are not seen clearly in the UV H$_2$.

\cite{gust10} first postulated that the inner outflow from T Tau S seen in near-IR H$_2$ might have been launched close to the time of the last periastron passage of T Tau Sb around T Tau Sa.  From Figure~\ref{fig.h2_halpha}b, we see arcs of H$\alpha$ emission that appear to be somewhat regular and periodic.  It seems plausible that these arcs might have been launched close to the time of past periastron passages.  To investigate this, we use information derived and presented by \cite{scha20} on the positions of T Tau Sa and Sb during historical orbital epochs, coupled with the location of these seemingly periodic arcs and motion from two epochs of H$\alpha$ imaging.  The details of this analysis are presented in Appendix 2.  

In Figure~\ref{fig.halpha_arc_ID}a we show a wider view of the H$\alpha$ image with an inner 3$\arcsec\times$3$\arcsec$ inset of the epoch 2005 near-IR H$_2$ emission.  Colored traces highlight the approximate locations of the leading shock-front emissions associated with each of the five arcs.  The inner Arc A is traced in green (in the inset), Arc B in yellow,  Arc C is in orange, Arc D is in magenta and the outermost Arc E is blue.  Overplotted in the image center in corresponding colors are the estimated positions of T Tau Sa with respect to T Tau N at the time of the past five orbital periastron passages of the T Tau S binary.   The positions of the star and arcs are included in Appendix 2.   In Figure~\ref{fig.halpha_arc_ID}b we show in cyan the same epoch 2004 H$\alpha$ image presented in (a), with the epoch 2019 image included in red.  Cyan and red circles highlight locations where the kinematics of the outflowing H$\alpha$ arcs were measured over the almost 15 year baseline between the two observations (see Appendix 2).   Table~\ref{tab.halpha} presents the four arcs measured in H$\alpha$, their distances, and tangential velocities.  The arcs measured in H$\alpha$ emission are moving faster than the inner H$_2$ knots, which is consistent with on-axis atomic flows expanding at a greater rate than slower molecular components \citep{taka06}.  We also saw this effect in the radial velocities of the NW1 and NW2 knots measured in [S~II] and H$_2$ (\S5.2).  Although the uncertainties in the kinematics are appreciable (see Appendix 2), the different H$\alpha$ arcs show evidence for variations in expansion velocities, particularly in the B vs. C-E arcs.  This might be expected if the flows are precessing.   
 
\begin{deluxetable*}{c|cc}
\tablecaption{Analysis of H$\alpha$ Arc Positions \label{tab.halpha}}
\tablewidth{0pt}
\tablehead{
\colhead{Arc} & \colhead{H$\alpha$ Arc}  &  \colhead{Tangential}  \\ 
\colhead{Designation} &  \colhead{Distance}  &  \colhead{Velocity} }
\startdata
A & 0.$\arcsec$62$\pm$0.09  &  36$\pm$9  \\
B &   3.$\arcsec$2$\pm$0.3 &  56$\pm$12 \\
C &  7.$\arcsec$7$\pm$0.4 &  78$\pm$16  \\
D & 11.$\arcsec$3$\pm$0.6   &  74$\pm$20  \\
E &  18.$\arcsec$8$\pm$0.5  &  73$\pm$16 \\
\enddata
\end{deluxetable*}

\begin{figure}
\plottwo{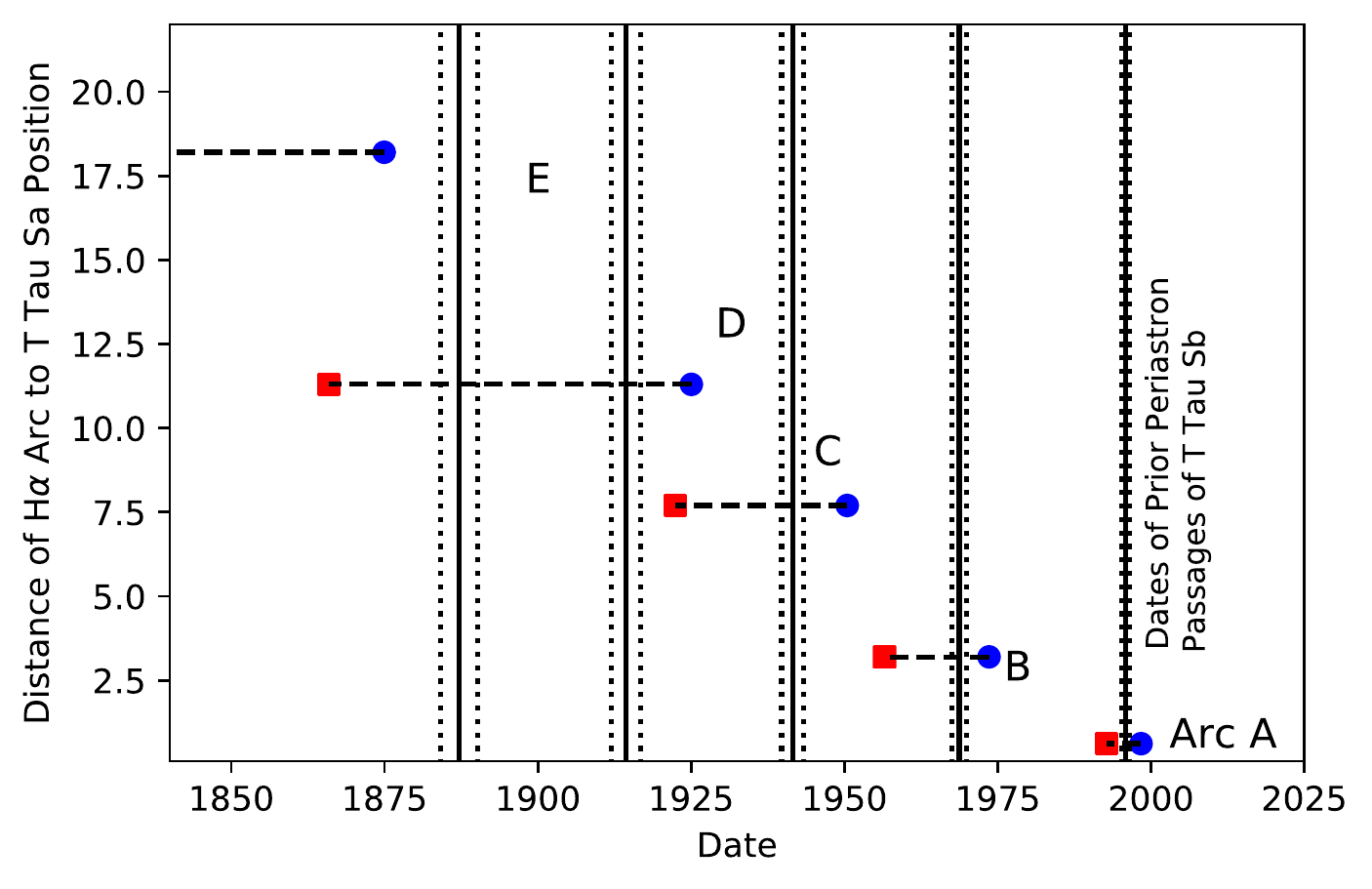}{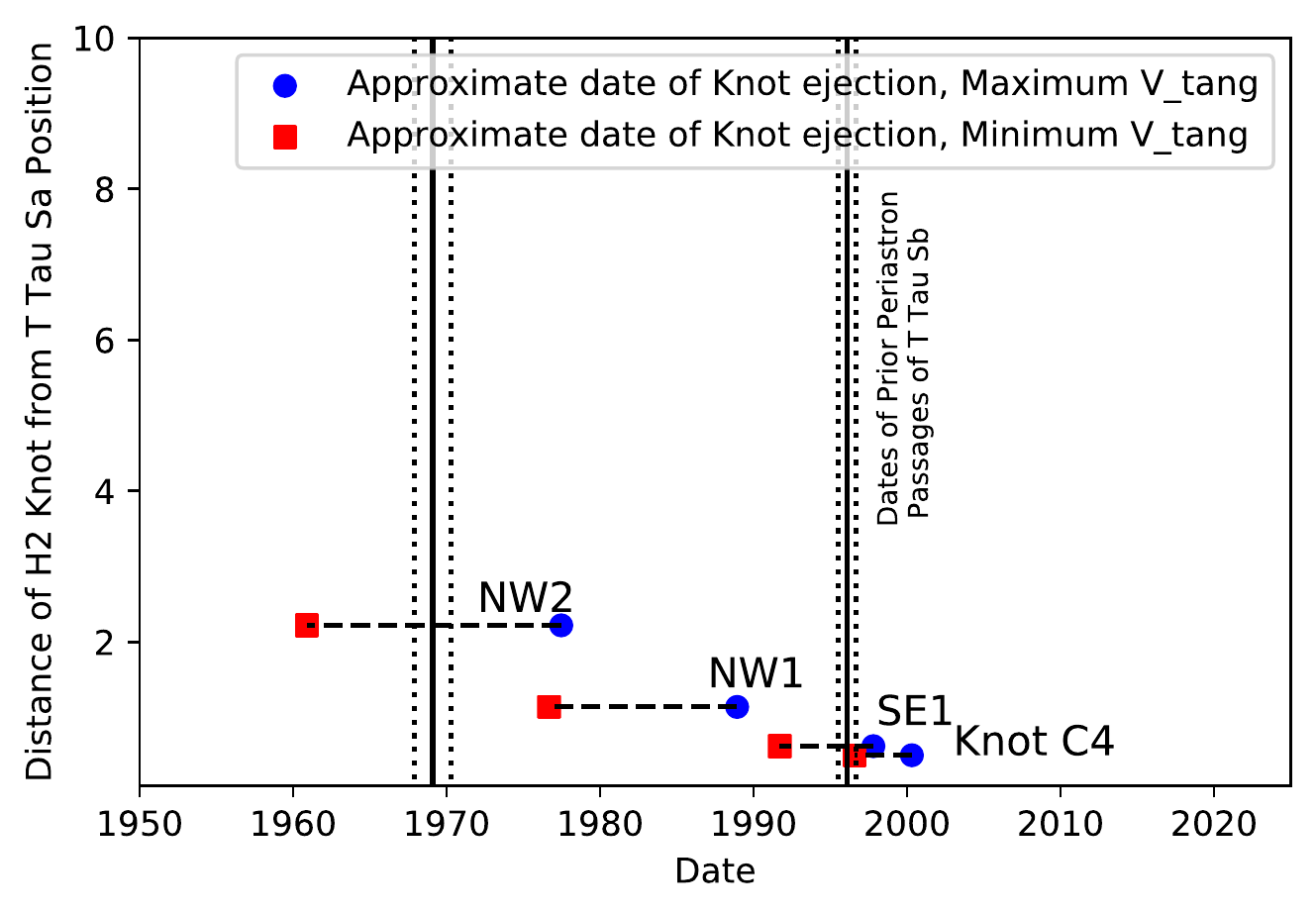}
\caption{{\it Left:} A comparison between the periastron passage dates (vertical solid lines) and the extrapolated outflow ejection dates for the A through E arcs of H$\alpha$ emission assuming two different tangential flow velocities, bounded by the red and blue points.  {\it Right: } A comparison between the periastron passage dates (vertical solid lines) and the extrapolated outflow ejection dates for the inner H$_2$ emission knots designated C4 through NW2 and discussed in detail in \S5.2.  The maximum and minimum tangential velocities used to bound the approximate ejection dates are from the measurements made for each knot and presented in Table~\ref{tab.h2}.
\label{fig.ejections}}
\end{figure}

Figure~\ref{fig.ejections}a presents an investigation of the historical periastron passage dates for the T Tau S binary (vertical lines) from the 27.2$\pm$0.7 year orbit model \citep{scha20} and their relation to best estimates for outflow arc ejection dates.   The extrapolated bounding dates for outflow ejection of the A through E arcs of emission were derived using the measured arc distances and the measured tangential velocities (Appendix 1 and 2) .  The timing of the A through D arc ejections span the dates of past periastron passages of T Tau Sb, with uncertainties increasing back in time.  In fact, the correspondence for launch of the most recent Arc A is within less than 1 month of the last Sb periastron in 1996. This provides intriguing evidence that the ejection of the southern outflowing arcs from T Tau S are contemporaneous with, and triggered by, the close passage of T Tau Sb around T Tau Sa.   The uncertainties in this analysis are dominated by the accuracies of the position and tangential motion of the arcs.  Adopting a lower mass for T Tau N \citep{flor20} in the historical orbital model is a small secondary effect in this result.   We find that the uncertainties on the arc D kinematics are large enough that it could have been launched either 4 or 5 periastron passages ago.  The arc E tangential velocity does not result in a launch timing overlap with the orbit periastron from five passages ago, this arc is moving slower than necessary to align well with the position of T Tau Sa in that launch time frame.  It may be that as the arcs move away from the system, their kinematic launch signatures are dampened by interaction with ambient material that slows the flow, or arc E might be from a prior periastron passage.  These results can be improved with future epochs of optical imaging and particularly 3-D kinematic mapping that can disentangle the spatially overlapping outflow arcs. 

The approximate outflow ejection dates of the H$_2$ knots are also derived and shown in Figure~\ref{fig.ejections}b.  This was carried out using the distance of the H$_2$ knots from the approximate position of Sa in the 2006 epoch measurement, coupled with the derived tangential velocities for each knot as presented in Appendix 1.  Figure~\ref{fig.ejections}b shows that the H$_2$ knots designated SE1 and NW2 have outflow ejection dates that bound the past two periastron passages of Sb around Sa from 1996 and in 1969.  The knot of H$_2$ emission designated C4 seems to have been ejected after the last periastron passage.  Interestingly, the NW1 knot is traced back to the position of T Tau Sa in a timeframe that corresponds to the apastron of the orbit of T Tau Sb.  Hence, while four of the loops and arcs of H$\alpha$ emission seem to be ejected at dates consistent with periastron passage (Figure~\ref{fig.ejections}a), the inner H$_2$ emission shows evidence that outflow ejection events also may occur at half phases of the orbital period.

The apparent position angle of the A through E outflow arcs has varied by over 80 degrees, from 130$^{\circ}$ measured for the A arc, to $\sim$220$^{\circ}$ for Arc D.  Hence, this flow from T Tau Sa+Sb appears to be precessing significantly.  In fact, the variation in position angle from the Arc C to Arc D flow was greater than 40$^{\circ}$, implying a high rate of jet precession assuming these two arcs map consecutive periastron ejections separated by the $\sim$27.2 year orbital period.  Interestingly, the Arc D and Arc E flows are nearly aligned with the current orientation of the circumbinary ring around T Tau S (\S4). 

\section{Discussion}\label{sec:disc}

Ever since its nature as a young triple star system was revealed, T Tau has served as a demonstration laboratory for spatially resolved measurement of young multiple star systems \cite{kore97,reip00}.  The apparent relative configuration of the stars on the sky classify T Tau as a non-hierarchical triple system, where the separation of the tight binary is less than $\sim$10$\times$ the extent of the wider system \citep{reip00, reip14}.  Non-hierarchical multiples are inherently gravitationally unstable.  The evolving orbital interactions of the stars in these young systems can lead to circumstellar (or circumbinary) disks that are mis-aligned relative to one another, or mis-aligned and interacting outflows.  Numerical simulations suggest that the majority of high order multiple systems in young star forming regions are likely gravitationally unstable and dynamically evolving, either into a more stable configuration or toward ejection of one star into an extremely wide orbit or gravitationally unbound state \citep{reip12, reip14}.  High angular resolution measurements of mm dust in young triple star systems continue to provide evidence for mis-aligned disks and dynamical complexity \citep{kurt18}.  The T Tau system provides an important case where disks and outflows can be spatially resolved, but also the unique situation where the orbit of the tight binary can be monitored and mapped \citep{scha14, kohl16, scha20}.

\begin{figure}
\includegraphics[scale=0.67,angle=0]{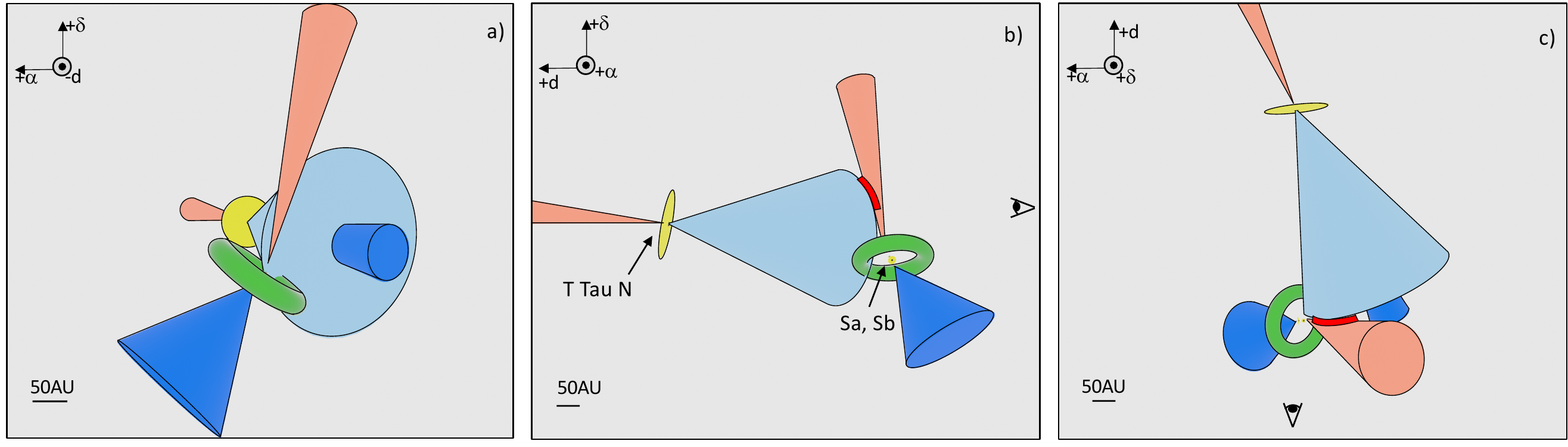}
\caption{A simple 3-D geometrical model of the stars, circumstellar disks, the circumbinary ring and the known outflows from the T Tau system.  The panels present different views down the geometrical axes (shown at the upper right):  (a) presents the view looking toward the T Tau system down the distance axis, which is our sky view, (b) shows the system geometry looking down the +RA axis (from due East), and (c) presents the top view of T Tau looking down the +$\delta$ Declination axis.  The circumstellar disks are in yellow, the circumbinary ring around T Tau S is in green, outflows from T Tau N and S are shown in red and blue depicting their radial velocities.  The outflow from T Tau N is presented as a fast on-axis flow in darker blue, surrounded by a slower wide angle outflow in light blue.  An estimate of the line-of-sight distance between T Tau N and S is discussed in \S6.  The red arc in panels (b) and (c) show where the two outflows could be interacting and exciting the brightest [S~II] emission viewed in the system (\S5.1).  The 50AU key is presented to show the different scale in the three views.
\label{fig.geom_3D}}
\end{figure}

Our multi-wavelength investigation of the disks and outflows in the T Tau system allows for a representative 3-dimensional (3-D) geometrical model of the stars, disks and outflows in the system to be constructed.   Figure~\ref{fig.geom_3D} shows three 3-D geometrical views of the observed star/disk/outflow components in the T Tau triple based on the accumulated data presented in this study.  The circumstellar disks are shown in yellow.  The blue shifted outflow from T Tau N is seen as a fast on axis component in dark blue surrounded by a lower velocity wider angle flow seen in a lighter shade of blue.  The circumbinary disk encircling T Tau S is presented in green; its geometry is constrained by the direct mm detection of material and the nebulosity seen in silhouette in UV and optical emission line species (\S4).  Hence the circumbinary disk is viewed nearly edge on, tilted at a PA$\sim$60$^{\circ}$ east of north and inclined into the line of sight by 5$^{\circ}$.  The geometries of the red and blue shifted outflows from T Tau S are derived from the analysis of [S~II] and H$_2$ presented in \S5.  The collimated redshifted outflow from T Tau S extends to the north-northwest from the system at a PA of 350$^{\circ}$.  The blueshifted flow from T Tau S has a wider opening angle and extends to the southeast of the system.  The geometry shown here is consistent with the PA and orientation of the inner "Arc A" knot measured in H$_2$ emission (\S5).  Our measurements of H$_2$ suggest that the inner outflows from T Tau S are oriented nearly in the plane of the sky with an inclination of just $\pm$5-10$^{\circ}$ from edge-on.  The radial velocity results from [S~II] emission are consistent with this interpretation.   The redshifted eastern flow from T Tau N is presented in Figure~\ref{fig.geom_3D} only for geometrical completeness, detection of this outflow emission is uncertain (\S5.1).

Figure~\ref{fig.geom_3D} shows the blueshifted outflow from T Tau N as a fast on axis jet (dark blue) surrounded by a lower velocity wider angle flow (light blue), consistent with the kinematics and morphology revealed in the blueshifted "Component 1" emission from the [S~II] emission profile (\S5.1).  This blueshifted outflow from T Tau N has similar structure to the well-studied blueshifted jet from the DG Tauri system \citep{pyo03, maur14, whit14}, where the fast, collimated central jet is encompassed by a wide-angle slower flow.  DG Tau has been used as a prime demonstration case for this outflow structure, consistent with MHD models of central collimated poloidal flows encompassed by a wind launched at larger radii from the disk \citep{ferr06, fend06}. In the case of T Tau N, the blueshifted brighter central flow is oriented nearly due west at a PA$\sim$260$^{\circ}$ East of North, with increasing velocity with distance from the star.  However, the strong velocity magnitude of greater than 140~km/s for the blueshifted central flow coupled with the 28$^{\circ}$ inclined view of the disk from T Tau N reveal that we are nearly looking down the central axis of this jet.  

In Figure~\ref{fig.geom_3D}, we included a red arc in panels b) and c) that traces the location where the blue-shifted nearly pole-on flow from T Tau N appears to be colliding with the northwestern redshifted flow from T Tau S.  Interestingly, this red arc is consistent with the position of the brightest region of [S~II] measured in the outflow maps of \S5.1 (also highlighted with a red arc in Figure~\ref{fig.sii_components}.  A bright region of shocked [S~II] emission was detected near this location in the past long-slit maps of \cite{solf99}.  Additionally, this location is also co-spatial with the bright arc of H$_2$ emission, designated "NW Arc", presented by \cite{herb07}.   Hence, we propose that the blue-shifted western flow from T Tau N may be colliding with the northwestern redshifted flow from T Tau S and creating this shock emission that has persisted in approximate position for more than $\sim$10 years.   Under this interpretation, a simple geometrical analysis allows for a better determination of the relative position of the stars in 3-D.  The location of the brightest region of [S~II] emission that we measure is 1.$\arcsec$9 from the position of T Tau N at an angle $\sim$330$^{\circ}$ East of North, and 2.$\arcsec$4 from T Tau Sa at a $\sim$355$^{\circ}$ orientation.  Using these apparent relative positions of the bright [S~II] arc to the stars, coupled with the knowledge that the blueshifted flow from T Tau N is oriented $\sim$28$^{\circ}$ from pole-on (perpendicular to the disk) and the northwestern flow from T Tau S is $\sim$5$^{\circ}$ into the plane of the sky, allows us to derive that T Tau N is located $\sim$620~AU behind the T Tau Sa+Sb binary in this geometry.  This placement of T Tau N relative to the Sa+Sb binary is used to define the positions of the stars in Figure~\ref{fig.geom_3D}.  

\begin{figure}
\includegraphics[scale=0.67,angle=0]{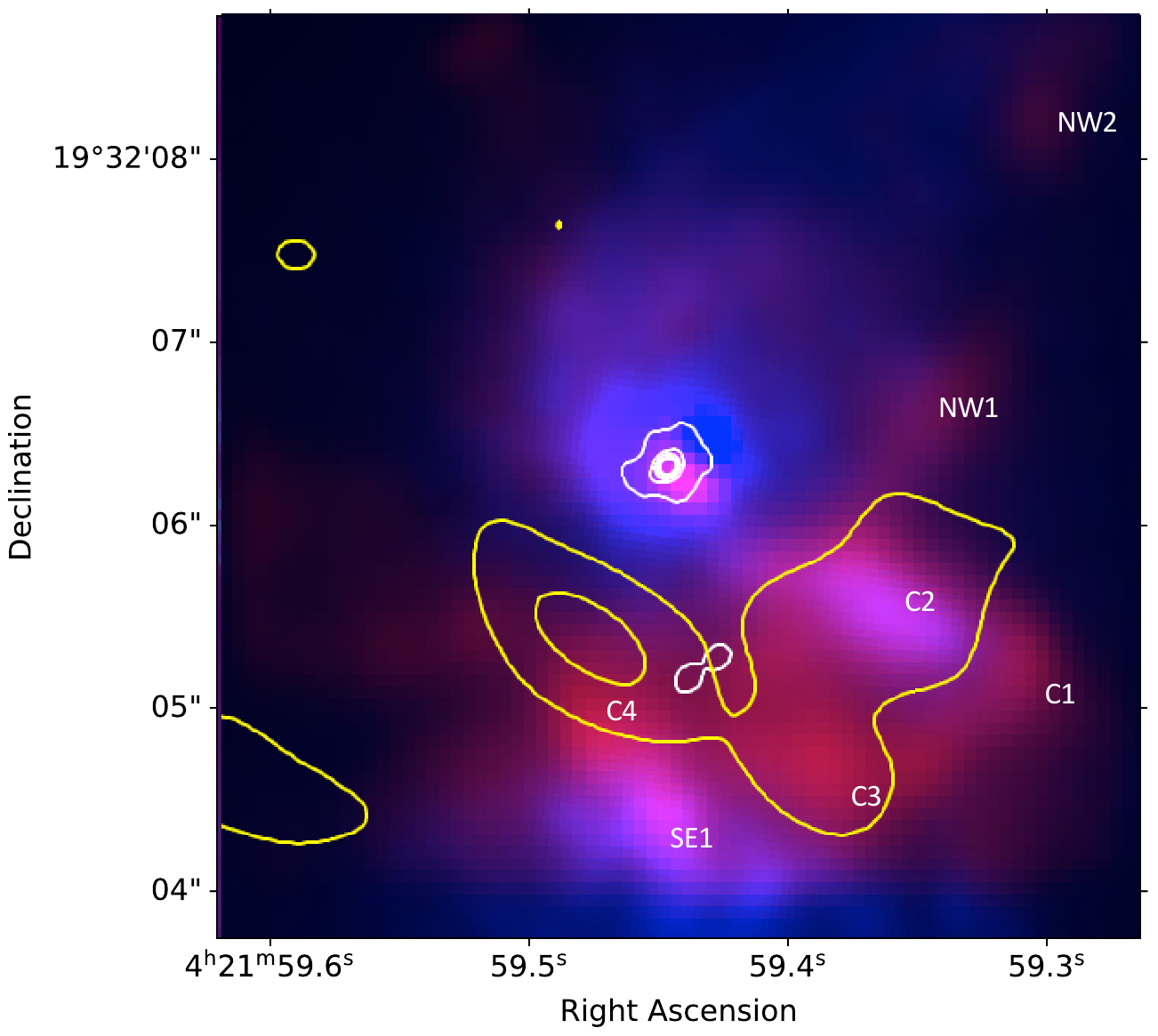}
\caption{(a) The UV Image (blue) which traces electronic UV H$_2$ emission shown with the {\it v}=1-0 S(1) near IR line map (red).  Overlaid are contours of the near-IR continuum (white) tracing the stars and the mm dust residual emission (yellow; \S5).  The near-IR H$_2$ knots discussed in in \S5.2 are labeled. 
\label{fig.geom_sa_wind_outflow}}
\end{figure}

Figure~\ref{fig.geom_sa_wind_outflow} presents a comparison of the UV electronic H$_2$ emission measured in 2004 with the Hubble ACS F140LP imaging (blue) and and the near-IR 2.12$\mu$m ro-vibrational H$_2$ from the 2005 observation epoch (red).  Overplotted contours trace emission from the three stars (white; 2.1$\mu$m continuum) and the detected dust in the circumbinary ring (yellow; 1.3mm residuals).   The UV H$_2$ from the inner environment around T Tau S is seen primarily at the locations designated C2 and SE1 locations in the IR images; the dark lane of obscuring material in the circumbinary ring attenuates UV emission from the other regions.  We found in \S5 that the near-IR H$_2$ emission at both the C1 and C2 locations seemed to be variable in morphology, but not comprised of outflowing gas.  Comparison with the UV image suggests that the C2 region of H$_2$ emission is tracing a static distribution of low A$_v$ gas at the northwestern surface of the circumbinary ring (Figure~\ref{fig.geom_sa_wind_outflow}).  The C2 knot measured in the UV and near IR corresponds to the northern lobe of the dark lane seen in silhouette, and knot C1 is at the westernmost edge of the distribution of near IR H$_2$ around the binary.   Conversely, the SE1 knot is outflowing and is tracing blue-shifted gas in the southern outflow.  The coincidence of UV and near-IR H$_2$ emission at the SE1 location means that this gas also lies along a line of sight with low dust attenuation.  Hence, the H$_2$ emission from the environment of T Tau S seems to be comprised of both static and outflowing distributions of gas with highly varying levels of line-of-sight extinction.  

In \S5.2 we discussed the arc-shaped morphology of the H$_2$ emission knot designated C1 \citep{herb07} as arising from excited gas at the inner edge of material the circumbinary ring around T Tau Sa+Sb.  This seems like a plausible interpretation under the knowledge that this arc of emission appears to be nearly static in position and not outflowing from the system.  However, this requires that the inner dynamically cleared cavity of the circumbinary ring is extremely large, over 10 times the orbital semi-major axis \citep{arty94}.  Simulations by \cite{fran19} suggest that circumbinary disks that are misaligned in eccentric binaries should have smaller not larger inner circumbinary disk radii.  From our geometrical model and disk parameters from the ALMA mm model data fits (Table~\ref{tab.cs_disks}), we find that the inner circumstellar disk around T Tau Sa is within $\sim$20 degrees of being perpendicular to the circumbinary ring.  If T Tau Sa has a poorly collimated wide angle wind, as suggested by the existence of appreciable 1.08$\mu$m He~I emission in the environment (Figure~\ref{fig.cbring}d), the flow orientation could be near to the plane of the circumbinary ring.  The apparent orientation of the T Tau Sa star+disk system could cause its outflowing gas to collide with the inner edge of the circumbinary ring, exciting the static H$_2$ shocks at the C1 and C2 locations (Figure~\ref{fig.geom_sa_wind_outflow}).  Additionally, interaction of a wind from T Tau Sa with the circumbinary ring might play an extra role in increasing the diameter of the dynamically cleared ring cavity, as gas and dust are slowly eroded by the wind.  High spatial resolution, high contrast near-IR maps of T Tau from \cite{yang18} revealed evidence for small dust grains that appeared to be streaming toward the west of T Tau S.  Particularly, we postulate that the "N3" and "N4" arcs of H-band polarimetric emission identified by \cite{yang18} could be evidence of small dust grains streaming westward from the circumbinary ring after being eroded by a wide angle wind from T Tau Sa.

From the [S~II] and H$_2$ spectral imaging of T Tau (\S5), we find that knots in the northwestern outflow from T Tau S are more collimated than the wider angle southeastern flow.  This type of outflow asymmetry can be caused by inner disk warps or structure in the quadrupole magnetic field of the launching star \citep{love10, fend13, dyda15, fend18}.  Additionally, the NW2 knot from the northwestern outflow from T Tau S seen in the 'Component 3' [S~II] fit (from \S5.1) has a stronger kinematic redshift versus H$_2$ or the encompassing lower velocity [S~II] (e.g., 'Component 2').   This implies that the fast, atomic on-axis jet in the [S~II] redshifted flow from T Tau S does have a slower wide-angle wind or low velocity flow component.  Investigations of outflow structure from young stars have revealed that many systems with fast, collimated central atomic jets often have wider angle winds traced in molecular species \citep{pyo03, taka06, angl07, taka07, maur14, whit14, louv18}.   Increasingly sophisticated theoretical models have been developed to explain the observed varieties of outflow structures, including asymmetric and mono-polar outflows caused by inner disk warps, or uneven mass accretion and outflow through complex quadrupolar magnetic field structures \citep{ferr06, fend06, fend09, love10, fend13, dyda15}. 

In a single star system, mass outflows remove a fraction of the angular momentum from the system through the jets and outflows.   The magnetohydrodynamical processes that link mass accretion from the inner disk onto the star regulate this process and allow for the accretion and outflows to proceed smoothly.  The mass outflow rate is correlated with mass accretion rate, with a typical mass accretion to outflow ratio of $\sim$10\% \citep{hart95, nisi18}.  Although rare, mass outflows triggered by binary orbital motion have been postulated to explain regular and repeatable outflow ejections \citep{angl07, mund10, louv18}, and the wavy shape of inner jets \citep{hao16}.  The monopolar outflow from the HH~30 young star system has served as a demonstration laboratory for outflow ejections triggered by binary orbital passage \citep{angl07}.   A close periastron passage could disturb inner disks into enhanced mass accretion onto the star, causing subsequent mass outflow as material is ejected from the inner system to balance the excess angular momentum \citep{angl07, mund10, louv18}.  Modeling studies have also considered how orbital motion and dynamical interaction between disks in a young binary system might affect mass outflow and jet structure \citep{raga09, shei15, shei18}.  \cite{shei18} found that the disk+orbit inclination and mass ratio of a binary can both affect mass outflow in a jet from one of the stars in the system, and high inclinations between the disk and binary orbit can inhibit outflow activity or cause jet precession.  The models show that the mass ratio of the binary can result in asymmetries in the lobes of outflows.  Our analysis temporally links the outflow ejecta with the orbital periastron passage of T Tau Sb around Sa.  The T Tau Sa+Sb binary provides the unique opportunity to measure multi-wavelength stellar brightness and mass accretion properties, while resolving the binary through periastron passage in anticipation of the next mass ejection.  In the outflow from T Tau S, we find evidence for precession and a more collimated northwestern jet compared to the wide-angle southern flow.  Additionally, the orbit of the Sa+Sb binary with respect to the disk of T Tau Sa is inclined by more than the $\sim$30$^{\circ}$ considered as an extreme inclination in the models of \cite{shei18}.   Hence, the T Tau S system may provide a specific geometry with defined observable characteristics to further develop MHD models of jets and outflow launching from interacting young binary star systems. 

The three dimensional geometrical model of T Tau presented in Figure~\ref{fig.geom_3D} is consistent with our current knowledge of the locations and orientations of the disks and outflows in this complex triple system.  The geometry of the circumstellar disk around T Tau Sb is not known, but is presented in Figures~\ref{fig.geom_3D} as being perpendicular to the northwestern red and southeastern blueshifted flows from T Tau S.   Knowledge of the position of the disk around Sa is derived from the mm measurements in \S4 (also \cite{mana19}), and is consistent with the mid-IR interferometric measurement from \cite{ratz09}.  This disk orientation for Sa is nearly perpendicular to the measured geometry of the outflows from T Tau S revealed in [S~II].  Hence, if the outflows from T Tau S originated from Sa, then the magnetic outflow pole geometry would have to be oriented less than $\sim$30$^{\circ}$ from the circumstellar disk plane.  This seems unlikely, and as a result we postulate that the outflows from T Tau S may originate from the lower mass T Tau Sb component.  \cite{bohm94} and \cite{solf99} find a stronger historical magnitude of blue and redshifted [S~II] radial velocities associated with the southeastern to northwestern flows from T Tau S than we measure.  These stronger radial velocities measured in the past, coupled with the variation in the outflow position angles in the H$\alpha$ maps, support the idea that the axis of the outflow from T Tau S is precessing significantly.  As T Tau Sb orbits around T Tau Sa, it is possible that both its disk and outflow axis are precessing \citep{raga09, shei18}.  However, the near-IR flux of T Tau Sa (Figure~\ref{fig.ir_var}) varies wildly in a manner that is consistent with rampant and rapid accretion variations \citep{vanb10}, which might in-turn drive a strong bi-polar outflow.  Hence, it is not clear which star, Sa or Sb, originates the northwestern-southeastern outflow.  High resolution measurement of the inner disks of T Tau Sa and Sb with ALMA over the course of the 27 year orbit can measure the disk precession around the stars directly, correlate it with outflow diagnostics, and help determine which of the T Tau S stars actually drives this mass outflow.

\subsection{T Tau: A Historical Perspective}\label{sec:disc.hist}
  
\begin{figure}
\includegraphics[scale=1.0,angle=0]{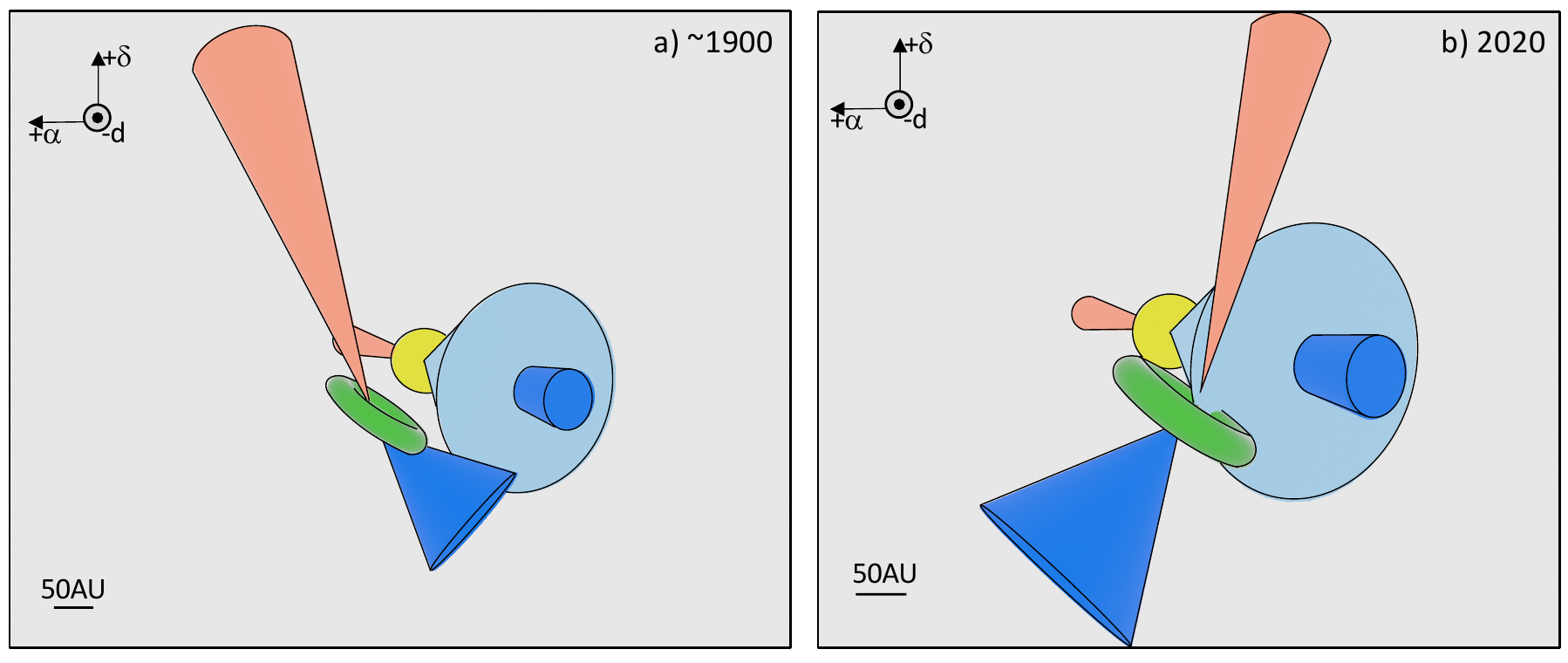}
\caption{A historical view of the T Tau system circa $\sim$1900 (a) with T Tau N, its disk (yellow) and the circumbinary ring around T Tau S (green).  For comparison, our view of the T Tau system as it is in 2020 is presented in (b); the same view as Figure~\ref{fig.geom_3D}a.  The red and blueshifted outflows are included in both panels.  Panel (a) presents the same star, disk and outflow components as in (b), but the positions of T Tau Sa, Sb, their disks and the circumbinary ring have been moved to their estimated location from the early 1900s.  The orientation of the outflows from T Tau S has been altered for consistency with the position angle of the "Arc E" flow, which would have been newly ejected from the last periastron passage.  The 50AU key is presented to show the slightly different scale in the two views.
\label{fig.geom_history}}
\end{figure}

Figure~\ref{fig.geom_3D} presents our geometrical interpretation of the multi-wavelength observations of the T Tau triple star and the associated disks and outflows, as discussed in detail in the previous section.  In Figure~\ref{fig.geom_history}, we show a similar geometrical view that has been tracked backward in time.  Particularly, the position of T Tau Sa+Sb with respect to T Tau N has been traced through the past positions in the wide orbit of the N-S system.  We have dialed the geometry of the stars, disks and outflows back to their potential locations at the time of periastron passage of Sb around Sa four orbits ago in the early 1900s.   The positions of the stars have been approximated based on our current knowledge of the wide orbit of the N-S system (Appendix 2).   The circumstellar disks are presented with the same geometry as today.  The blueshifted western outflow from T Tau N is also shown with the geometry we see today, because of a lack of information on how the flow might have changed during the past $\sim$100 years.  However, the geometry of the southeastern+northwestern outflow from T Tau S assumes bi-polarity and has been tilted to accurately reflect the position angle of the blue-shifted "E" arc, which would have been the innermost measurable arc at this epoch.    

Interestingly, one of the obvious things that we see from this putative historical geometrical configuration for the T Tau N and Sa+Sb system is that the redshifted outflow from T Tau S would have been to the east of our line of sight in the past.  Today, the redshifted outflow from T Tau S is westward of our line of sight to T Tau N.   At some point in the past $\sim$100 years, the northern outflow from T Tau S must have moved through our line of sight toward T Tau N.  An alignment of this sort would cause blobs of material from the redshifted outflow to randomly occult T Tau N, resulting in a flickering of the star and random dimming of its optical brightness.  Of course, this random flux variability is precisely what was observed in T Tau N prior to 1930 (Figure~\ref{fig.figaavso}).  Moreover, an abrupt cessation of the random brightness variations could be explained by the edge of the northern redshifted flow orbiting past the alignment with T Tau N.  The time duration that T Tau N experienced these random fluctuations in optical brightness would be defined by the width of the northern outflow from T Tau S along the line of sight, plus a combination of the tangential velocity of the wide orbit and precessional speed of the outflow.  In fact, if the position angle of the northern outflow from T Tau S is indeed precessing, as suggested by the variation in the southern outflow PAs seen in the H$\alpha$ image analysis of \S6, then a future alteration of the tilt of the flow might once again align it with our line of sight to T Tau N.  Hence, it seems feasible that the optical brightness of T Tau N could also vary in this manner in the future.  Derivation of the 3-D outflow kinematics could clarify exactly how the southern outflow has been historically precessing, which would lead to a better prediction for the future position angle of the northern flow, assuming bipolarity.  The proposed geometry of T Tau that we present can explain a number of unique characteristics that are seen in the system, and it offers a solid interpretation for the historical rampant optical variability of T Tau N and its abrupt cessation.  
	
\subsection{T Tau: What to Expect in the Future}\label{sec:disc.fut}

T Tau Sb will orbit through periastron around T Tau Sa in early 2023 \citep{scha20}.  Based on the results of the circumstellar disk fits to the ALMA archival data (\S5, \cite{mana19}), the disk around T Tau Sa is nearly perpendicular to the orbital plane of the Sa+Sb stars.   The orbital periastron distance between Sa and Sb is only $\sim$5AU, and the disk around T Tau Sa has a 3.5~AU radius.  Hence, tidal interaction between the stars and disks at the close passage distance is certainly expected.   ALMA can potentially spatially resolve the disk structure and any morphological variations during periastron close approach, monitoring observations may reveal our first direct view of young binary disk-disk interactions during a close periastron passage.  Tidal interaction and disk stripping has been seen on a much wider and more massive scale in the RW Aur system \citep{cabr06}, following the last orbital periastron passage of RW Aur B around A.   Moreover, monitoring changes in the radio emission of the T Tau S binary as it passes through periastron could provide direct evidence for interacting magnetic fields during the closest approach of Sa and Sb.  Relating the observed wide-scale radio emission characteristics to the binary orbit might reveal a clearer understanding of the non-thermal and polarized radio emission components measured previously \citep{phil93, skin94, ray97, john03, smit03, loin07}.  Radio wavelengths might also provide the first method to directly detect extended emission in a new outflow ejection launched from the T Tau S binary.

The light curves of T Tau Sa+Sb (Figure~\ref{fig.ir_var}) show IR flux variability linked with the Sa-Sb orbit.  Numerical simulations of pulsed accretion onto eccentric binaries suggests that accretion outbursts can occur with modulation defined by the orbital period, and that one of the stars can accrete at a much higher rate than the other \citep{muze13,muno16}.   To date, the strongest evidence for pulsed accretion associated with the T Tau orbital motion has come from the IR brightness fluctuations attributed to T Tau Sa \citep{vanb10}.  \cite{ghez91} reported a grey $\sim$2 magnitude flare from 2-10$\mu$m in observations acquired in 1990, six years prior to the last orbital periastron passage.  Interestingly, at the corresponding time frame in the current orbit, the IR flux of Sa increased to a comparable level as measured by \cite{ghez91}, and to its brightest level seen since the early 2000's (Figure~\ref{fig.ir_var2}).  If the brightness of T Tau Sa modulates repeatably over the binary orbit then we may expect a continued decrease in flux leading up to periastron, with a significant peak in brightness occurring $\sim$2-3 years after periastron passage in 2025 to 2026.  To date, the magnitude and mechanism of the mass accretion onto T Tau Sa and Sb during their close approach is unclear.  However, continued resolved observations that can disentangle accretion and outflow tracers will be important to understand the origin of these pulse events and how they are triggered by the binary orbital motion.

In the distant future, the T Tau S binary will continue its wider relative orbital motion around T Tau N (as shown in Figure~6 of \cite{scha20}).   There are two interesting possibilities to speculate about the future configuration for the system.  First, as T Tau Sa+Sb continues its wider orbit around the N - S center of mass, the apparent motion appears to have the binary moving east-ward relative to T Tau N; and thus entering the blue-shifted outflow channel of the HH~155 flow from T Tau N.  We postulated that the northwestern outflow from T Tau S is already interacting with the blueshifted flow from T Tau N and causing a strong shock in [S~II] (Figure~\ref{fig.sii_components}).  If the T Tau S binary is in fact entering the blueshifted outflow channel of HH~155, then the fast on-axis flow from T Tau N will shock and erode material in the circumbinary ring around T Tau S.  Strong emission from the interacting outflows would be expected, possibly with blueshifted acceleration in the kinematics of the outflows from T Tau S.  Second, results from \S4 show that the low level 1.3mm residual emission from the circumbinary ring around T Tau S extends to $>$2$\arcsec$ to the northeast of the binary.  It seems plausible that as the T Tau S binary continues its wide orbital motion in the N - S system, a northern extension of the circumbinary ring could occult T Tau N and cause the optical flux of the star to dim.  This might be especially true if there is any amount of low density material trailing the circumbinary ring, such as in tidally stripped trailing streams of material, as seen in the RW Aur system \citep{cabr06}.  \cite{flor20} has analyzed the relative motion of the stars and also postulated that T Tau N may dim in flux in the future as it is obscured by material associated with T Tau S.  While speculative, both of these future scenarios we discuss will result in easily testable predictions:  1) there will be evidence for interaction between the blueshifted western flow from T Tau N with the circumbinary ring or outflowing material from T Tau S, and 2) a future dimming of the combined system flux at all wavelengths within the next $\sim$100 years would be expected as the relative orbital motion of the stars takes T Tau N behind the circumbinary ring material encircling T Tau S.

\facilities{AAVSO, Gemini:Gillett (GMOS, Michelle, NIRI, NIFS), HST (ACS), CXO, ALMA}

\software{  \\
 Gemini IRAF and DRAGONS:  \url{https://www.gemini.edu/sciops/data-and-results/processing-software} \\
 Python and Astropy v. 3.8 \url{https://www.astropy.org}  \\
 GILDAS software package: http://www.iram.fr/IRAMFR/GILDAS \\
 Blender v. 2.8 3-D Data Visualization and Animation package: \url{https://www.blender.org }
  }
 
\section{Acknowledgements}

Most of the 'historical firsts' of the T Tau system described in \S1.1 came directly from TLB's pleasant dinnertime conversations at the Mauna Kea Hale Pohaku basecamp in the late 1990's with fellow T Tau enthusiast, George Herbig.  Thank you and RIP, George.  We are grateful to P. C. Schneider for information on the Chandra ACIS 0th order HETG image of the T Tau system, and for sharing with us the super-sampled X-ray image of T Tau presented in Figure 4.  We thank our anonymous referee for providing comments which improved our manuscript and aided in identification and correction of an error in our initial analysis.  TLB posthumously thanks Peter McGregor, friend, mentor and NIFS instrument PI, for his demonstration leadership on all observations using Gemini NIFS.  We thank Lisa Prato for providing comments on an early draft of this manuscript.   GHS acknowledges support from NASA Keck PI Data Awards administered by the NASA Exoplanet Science Institute.  We acknowledge with thanks the variable star observations from the {\it AAVSO International Database} contributed by observers worldwide and used in this research.  We are grateful to R. S. Fisher and T. Kerr for their assistance setting up and observing the mid-IR images and spectra of T Tau acquired during system verification of the Michelle instrument on Gemini North.  Based on observations obtained at the international Gemini Observatory which is managed by the Association of Universities for Research in Astronomy (AURA) under a cooperative agreement with the National Science Foundation on behalf of the Gemini Observatory partnership: the National Science Foundation (United States), National Research Council (Canada), Agencia Nacional de Investigaci—n y Desarrollo (Chile), Ministerio de Ciencia, Tecnolog'a e Innovaci—n (Argentina), MinistŽrio da Cincia, Tecnologia, Inova›es e Comunica›es (Brazil), and Korea Astronomy and Space Science Institute (Republic of Korea).  Observations for this project acquired at the Gemini North Observatory include data from program IDs:  GN-2003B-SV-81, GN-2004B-Q-80 and GN-2006B-DD-06.  

\section{Appendix 1:  Tangential Motions from Four Epochs of H$_2$ Spectral Imaging}\label{sec:abs}

We analyzed four epochs of 2.12$\mu$m ro-vibrational H$_2$ maps from the environment of T Tau.  For this, we used the maps of \cite{herb07} from 2002, \cite{gust10} from 2004, \cite{beck08} from 2005 and unpublished measurements from 2006  (described in Table 1).  Figure~\ref{fig.h2_pm}a and~\ref{fig.h2_pm}b present the 2.12$\mu$m H$_2$ emission maps from October 2005 and December 2006.  The central $\pm$0.$\arcsec$25 region around T Tau N was masked out where the 0.$\arcsec$5 diameter occulting spot was used for the 2006 observation (Table~\ref{tab.obs}).  The C1 - C4 knots identified by \cite{herb07} are shown in Figure~\ref{fig.h2_pm}a, and three additional knots labeled SE1, NW1 and NW2 are also designated.  For consistency, the positions of each of the knots were re-estimated for at all four measurement epochs.  This was done for the H$_2$ maps of \cite{herb07} and \cite{gust10} by digitizing the published images onto a regular measurement grid and estimating their locations.  The position angles and distances of the knots with respect to the position of T Tau Sa are shown in Table~\ref{tab.a1_1} for the four epochs from 2002 to 2006.  Knot SE1 is not clearly measured in \cite{herb07}, and the NW1 and NW2 knots could be estimated only in the epoch 2005 and 2006 measurements shown in Figure~\ref{fig.h2_pm}.

As discovered by \cite{gust10}, the morphologies of the H$_2$ emission knots in the environment of T Tau S are highly variable.  In the images from 2005 to 2006 shown in Figure~\ref{fig.h2_pm}a and ~\ref{fig.h2_pm}b, the morphology and relative brightness of the knots is seen to change appreciably on this $\sim$1 year timescale.  The double-lobed emission structure seen in knot C2 in 2005 merged together in 2006, the C2 knot decreased in relative brightness compared to other knots, and it developed an extension to the southeast.  The relative brightness of two apparent knots making up the C3 complex changed; the brighter knot in the west in 2005 became the fainter knot in 2006.  The C4 knot developed an arc-shaped morphology in 2006.  Measurement of detailed structure in the SE1 knot in 2006 was affected by the fact that it lay near the edge of the southern field of view.  The position angles and distances presented in Table~\ref{tab.a1_1} are the best locations of peaks in the contour maps for each of the measurement epochs with respect to the position of T Tau Sa at that epoch.  Table~\ref{tab.a1_2} shows the positions of T Tau Sa with respect to its location in the 2006 epoch, these values are used to correct the position for the wide orbit of T Tau S around N in the distance measurements.  The accuracy of the values in Table~\ref{tab.a1_1} by far are dominated by the variable knot morphology variations and difficulty clearly defining the emission that should be consistently compared from year to year.  Overall, we estimate the accuracy of the position angle and distance measurements in Table~\ref{tab.a1_1} to be $\pm$3$^{\circ}$ and $\pm$0.$\arcsec$04.  Within these measurement uncertainties, our remeasured knot positions for the 2002 and 2004 epochs are consistent with the values presented by \cite{gust10}.

\begin{deluxetable*}{c|cc|cc|cc|cc|cc|cc|cc}
\tablecaption{Four Epoch Positions of H$_2$ Emission Knots (2002 - 2006) \label{tab.a1_1}}
\tablewidth{0pt}
\tablehead{
\colhead{Date}  & \multicolumn2c{Knot C1} &  \multicolumn2c{Knot C2} &  \multicolumn2c{Knot C3} & \multicolumn2c{Knot C4} &
\multicolumn2c{Knot SE1} & \multicolumn2c{Knot NW1} & \multicolumn2c{Knot NW2} \\
\colhead{}  & \colhead{PA} & \colhead{distance} & \colhead{PA}& \colhead{distance} & \colhead{PA} & \colhead{distance} & \colhead{PA}& \colhead{distance} & \colhead{PA}& \colhead{distance} & \colhead{PA}& \colhead{distance} & \colhead{PA} & \colhead{distance} }
\startdata
2002 & 270$^{\circ}$ & 1.$\arcsec$05 & 313$^{\circ}$ & 0.$\arcsec$41 & 215$^{\circ}$ & 0.$\arcsec$42 & 110$^{\circ}$ & 0.$\arcsec$26 & -- & -- & --  & -- & -- & -- \\
2004 & 272$^{\circ}$ & 0.$\arcsec$93 & 305$^{\circ}$ & 0.$\arcsec$38 & 229$^{\circ}$ & 0.$\arcsec$45 & 121$^{\circ}$ & 0.$\arcsec$34 & 168$^{\circ}$ & 0.$\arcsec$46 & -- & -- & -- & --  \\
2005 & 267$^{\circ}$ & 0.$\arcsec$98 & 331$^{\circ}$ & 0.$\arcsec$54 & 227$^{\circ}$ & 0.$\arcsec$52 & 130$^{\circ}$ & 0.$\arcsec$48 & 166$^{\circ}$ & 0.$\arcsec$55 & 313$^{\circ}$ & 1.$\arcsec$04 & 301$^{\circ}$ & 2.$\arcsec$04 \\
2006 & 272$^{\circ}$ & 1.$\arcsec$00 & 331$^{\circ}$ & 0.$\arcsec$46 & 217$^{\circ}$ & 0.$\arcsec$51 & 130$^{\circ}$ & 0.$\arcsec$50 & 157$^{\circ}$ & 0.$\arcsec$62 & 307$^{\circ}$ & 1.$\arcsec$14 & 301$^{\circ}$ & 2.$\arcsec$16  \\
\enddata
\tablecomments{All knot position measurements are with respect to the epoch position of T Tau Sa.  Knot designations are overlaid on the H$_2$ image shown in Figure~\ref{fig.h2_pm}a. The knot designations C1-C4 and epoch 2002 H$_2$ positions are remeasured from \cite{herb07}, and 2004 knot positions are derived from \cite{gust10}.}
\end{deluxetable*}

\begin{deluxetable*}{c|cc}
\tablecaption{Positions of T Tau Sa (2002 - 2006) \label{tab.a1_2}}
\tablewidth{0pt}
\tablehead{
\colhead{Date} & \multicolumn2c{T Tau Sa Position}  \\
\colhead{} & \colhead{$\Delta$X} &\colhead{$\Delta$Y}  }
\startdata
2002 & -0.$\arcsec$002 & -0.$\arcsec$049  \\
2004 & -0.$\arcsec$005 & -0.$\arcsec$022  \\
2005 & -0.$\arcsec$003 & -0.$\arcsec$011  \\
2006 & 0.0 & 0.0    \\
\enddata
\end{deluxetable*}

\section{Appendix 2:  Tangential Motions from Two Epochs of H$\alpha$ Spectral Imaging}\label{sec:abs2}

\begin{figure}
\plotone{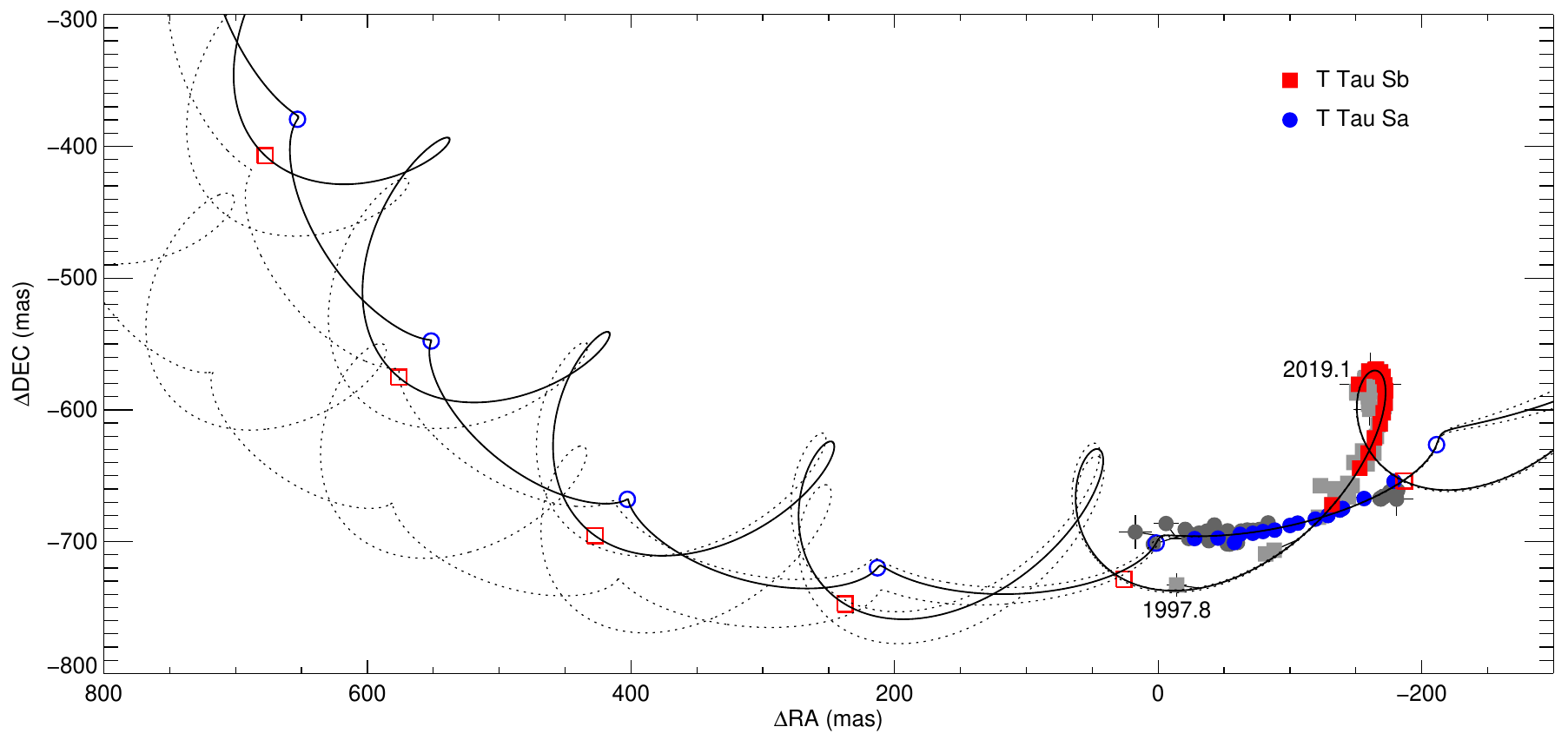}
\caption{The filled symbols show the measured positions of T Tau Sa and Sb with respect to T Tau N for all resolved measurements of the three stars.  The solid lines trace the motion of Sa and Sb back in time through their modeled wide orbit around T Tau N, from \cite{scha20}.  The positions at the times of the last five orbital periastron passages are marked by the open blue circles for T Tau Sa and the open red squares for T Tau Sb.    The dotted lines show the of the motion of Sa+Sb expected from different models for the wide orbit around T Tau N based on its mass (e.g., dotted orbits in Figure 6 of \cite{scha20}).  
\label{fig.historical_orbits}}
\end{figure}

Here we derive the motions of the southern outflow arcs with respect to the historical stellar positions of T Tau S to estimate if launch timing is contemporaneous with characteristics of the binary.  Figure 6 of \cite{scha20} shows the wide orbit of the combined T Tau S system (blue curve) around the position of T Tau N.  In Figure~\ref{fig.historical_orbits}, the historical positions of T Tau Sa and Sb are mapped back through this wide orbit in a zoomed view of the binary motion.  The filled symbols show the prior positions of T Tau Sa and Sb with the same color coding as in Figure 3 of \cite{scha20}.  The open blue circles and the open red squares show the locations of T Tau Sa and Sb, respectively, during the past five periastron passages of Sb around Sa.   \cite{scha20} and \cite{flor20} ponder the possibility that T Tau N is a less massive $\sim$1M$_{\odot}$ T Tauri star, but this changes the historical positions of T Tau Sa and Sb in prior orbital periastron passages by less than 0.$\arcsec$1. 
 
Table~\ref{tab.halpha_ap1} shows the H$\alpha$ B through E arcs with their distance and position angle from the positions of T Tau Sa at the time of the last four periastron passages. Also included in Table~\ref{tab.halpha_ap1} is the measured change in distance of the arc between the epoch 2004 and 2019 optical imaging observations (see also Figure~\ref{fig.halpha_arc_ID}b).  The distance differences between the 2004 and 2019 arc positions were estimated independently at three locations; the reported value is the average, and the uncertainty is half of the range in the three independent measurements.  Included in Table~\ref{tab.halpha_ap1} are the positions of Sa with respect to T Tau N at the time of the prior five periastron passages of the T Tau S orbit (colored circles in Figure~\ref{fig.halpha_arc_ID}). Taken in combination, the values presented in this table were used to derive the historical launch time frames shown in Figure~\ref{fig.ejections}a by referencing the 2004 epoch arc measurement to the location of T Tau Sa at the time of historical periastron passages of the binary, and using the angular change in arc position from 2004 to 2019 (14.989 year baseline) to derive the tangential velocity. These velocities are presented in the last column of Table~\ref{tab.halpha}. The uncertainties in the arc positions range from 0.$\arcsec$3 to 0.$\arcsec$5-0.$\arcsec$6 and increase with distance from T Tau.  In the epoch 2019 image the knots in the extended C and D arcs are less distinct than in 2004, resulting in moderate errors on their measured positions and movement.  As a result, kinematic uncertainties are quite large (up to 20km/s; Table~\ref{tab.halpha}).  

\begin{deluxetable*}{c|ccclcc}
\tablecaption{Analysis of H$\alpha$ Arc Positions \label{tab.halpha_ap1}}
\tablewidth{0pt}
\tablehead{
\colhead{Arc} & \colhead{H$\alpha$ Arc}  & \colhead{Position} & \colhead{2004-2019} & \multicolumn2c{Periastron Position of Sa w/r/t N} \\  
\colhead{Designation} &  \colhead{Distance}  & \colhead{Angle} & \colhead{$\Delta$Distance} &  \colhead{$\Delta$RA} & \colhead{$\Delta$Dec} } 
\startdata
B &   3.$\arcsec$2$\pm$0.3 & 135$^{\circ}$ & 1.$\arcsec$28$\pm$0.28  & -0.$\arcsec$212 & -0.$\arcsec$719 \\
C &  7.$\arcsec$7$\pm$0.4 & 175$^{\circ}$ & 1.$\arcsec$77$\pm$0.35 & -0.$\arcsec$402 & -0.$\arcsec$667 \\
D & 11.$\arcsec$3$\pm$0.6  &220$^{\circ}$ &  1.$\arcsec$68$\pm$0.40 &  -0.$\arcsec$551 & -0.$\arcsec$547 \\
E &  18.$\arcsec$8$\pm$0.5  & 210$^{\circ}$ & 1.$\arcsec$64$\pm$0.33  & -0.$\arcsec$653 & -0.$\arcsec$379 \\
\enddata
\end{deluxetable*}

\bibliographystyle{aasjournal}
\bibliography{references}

\end{document}